\documentclass[12pt,a4paper]{article}
\usepackage[utf8]{inputenc}
\usepackage[lmargin=1in, rmargin=1in, tmargin=1in, bmargin=1in]{geometry}
\usepackage{setspace}
\usepackage{titling}
\usepackage{lipsum}
\usepackage{comment}
\usepackage{physics}

\allowdisplaybreaks
\providecommand{\keywords}[1]
{
  \small	
  \textbf{\textit{Keywords---}} #1
}

\RequirePackage{amsthm,amsmath,amsfonts,amssymb}
\RequirePackage{natbib}
\RequirePackage[colorlinks,citecolor=blue,urlcolor=blue]{hyperref}
\RequirePackage{graphicx}

\usepackage{enumerate}
\usepackage{placeins}
\usepackage{multirow}
\usepackage[font=small]{caption}

\usepackage{enumitem} 

\usepackage{longtable} 

\usepackage{tikz}
\usetikzlibrary{calc,matrix,decorations.pathmorphing,shapes,arrows.meta,backgrounds,automata}
\usepackage[small]{subfigure}
\usepackage{float}

\usepackage{bbold} 
\usepackage{mathrsfs}

\usepackage{mathtools, nccmath}

\usepackage{makecell}

\newcommand{\indep}{\rotatebox[origin=c]{90}{$\models$}}  

\newcommand{\Cov}{\operatorname{Cov}}

\newcommand{\E}{\operatorname{E}}

\theoremstyle{plain}
\newtheorem{prop}{Theorem}

\theoremstyle{remark}

\newtheorem{assu}{Assumption}

\newtheorem{subassu}{Assumption}[assu] 

\def\X{{\boldsymbol X}}
\def\x{{\boldsymbol x}}
\def\Q{{\boldsymbol Q}}
\def\q{{\boldsymbol q}}

\def\C{{\boldsymbol C}}

\DeclareMathOperator{\Pro}{Pr}
\def\one{\mathbb{1}}
\def\P{\mathbb{P}}
\newcommand{\defeq}{\vcentcolon=}

\def\pheight{\frac{w(\X,g')}{\E_{\mathcal{P}_t}(w(\X,g') \mid \Q,g')}}    

\newcommand{\bigslash}{\mathrel{\smash{\raisebox{4.2ex}{\rotatebox{160}{\rule{0.5pt}{8ex}}}}}}

\title{\Large Nonparametric Causal Decomposition of Group Disparities}
\author{Ang Yu\thanks{Department of Sociology, University of Wisconsin-Madison. Email: ayu33@wisc.edu}, Felix Elwert\thanks{Department of Sociology, Department of Biostatistics and Medical Informatics, and Department of Population Health Sciences, University of Wisconsin-Madison}}
\date{December 14, 2024}

\begin{document}

\maketitle

\begin{abstract}
We introduce a new nonparametric causal decomposition approach that identifies the mechanisms by which a treatment variable contributes to a group-based outcome disparity. Our approach distinguishes three mechanisms: group differences in 1) treatment prevalence, 2) average treatment effects, and 3) selection into treatment based on individual-level treatment effects. Our approach reformulates classic Kitagawa-Blinder-Oaxaca decompositions in causal and nonparametric terms, complements causal mediation analysis by explaining group disparities instead of group effects, and isolates conceptually distinct mechanisms conflated in recent random equalization decompositions. In contrast to all prior approaches, our framework uniquely identifies differential selection into treatment as a novel disparity-generating mechanism. Our approach can be used for both the retrospective causal explanation of disparities and the prospective planning of interventions to change disparities. We present both an unconditional and a conditional decomposition, where the latter quantifies the contributions of the treatment within levels of certain covariates. We develop nonparametric estimators that are $\sqrt{n}$-consistent, asymptotically normal, semiparametrically efficient, and multiply robust. We apply our approach to analyze the mechanisms by which college graduation causally contributes to intergenerational income persistence (the disparity in adult income between the children of high- vs low-income parents). Empirically, we demonstrate a previously undiscovered role played by the new selection component in intergenerational income persistence.
\end{abstract}


\section{Introduction}
Social and health scientists often seek to decompose an outcome disparity between groups in terms of the contributions of an intermediate treatment variable. 
For example, how much, and in what ways, do racial differences in medical care contribute to racial disparities in health \citep{howe_african_2014}? How does childbearing contribute to the gender wage gap \citep{cha_is_2023}? And what are the roles of education in the association between parents' and their children's social class (social mobility) \citep{ishida_class_1995}? 
The common structure of these questions is that they seek to quantify the mechanisms by which a treatment variable \emph{causally} explains an observed \emph{descriptive} group disparity. 

Prior research has addressed such questions using three approaches, none of which is fully appropriate for the task of causally decomposing descriptive disparities. First, popular Kitagawa-Blinder-Oaxaca (KBO) decompositions \citep{kitagawa_components_1955, blinder_wage_1973, oaxaca_male-female_1973} are defined in terms of parametric regression coefficients and do not answer any causal question by design (\citeauthor{fortin_decomposition_2011}, \citeyear{fortin_decomposition_2011}, p.13; \citeauthor{lundberg_what_2021}, \citeyear{lundberg_what_2021}, p.542). Second, causal mediation analysis (CMA) \citep{vanderweele_explanation_2015}, though formulated in causal terms, decomposes the causal effects of group membership rather than observed group disparities. Third, recently developed random-equalization decompositions \citep{vanderweele_causal_2014, jackson_decomposition_2018, jackson_meaningful_2021, lundberg_gap-closing_2024} conflate distinct disparity-generating mechanisms, which limits their interpretability, explanatory power, and usefulness for policy analysis. Importantly, all prior approaches neglect that outcome disparities can in part be explained by group-differential selection of individuals into treatment as a function of their individual treatment effects. 

In this article, we propose a nonparametric causal decomposition approach that remedies these limitations. In contrast to KBO decompositions, our decompositions are formulated as model-free causal estimands with interventional interpretations. In contrast to CMA, we decompose observed group disparities rather than possibly ill-defined causal effects of group membership. In contrast to random equalization decompositions, our approach divides the outcome disparity into conceptually unambiguous components. In contrast to all prior approaches, we reveal a novel disparity-generating mechanism, termed the ``selection component,’’ that has not previously been incorporated in decomposition analysis. 

Our framework distinguishes three causal mechanisms through which an intermediate treatment variable can produce a group disparity in an outcome. First, mean outcomes may differ across groups because the groups receive the treatment at different rates (differential prevalence). Second, even if both groups have the same treatment prevalence, mean outcomes may differ across groups because the treatment has different average effects across groups (differential effects). Third, even if both groups have the same treatment prevalence and the same average treatment effects, mean outcomes will nonetheless differ across groups if members of one group select into treatment more strongly on their treatment effects than members of the other group (differential selection); for example, if the treatment is randomly distributed to members of one group, but specifically given to only those members who will benefit the most from the treatment in the other group. 
Conceptually, prior decompositions were limited to considering differential prevalence and differential effects, leading to the belief that these two are the only possible mechanisms \citep[e.g.,][]{ward_how_2019, diderichsen_differential_2019}. Isolating differential selection into treatment as a source of group disparities is the central conceptual contribution of our approach. 

We introduce an unconditional and a conditional decomposition. The unconditional decomposition is useful for investigating the overall (marginal) contributions of a treatment to an outcome disparity. The conditional decomposition is useful for investigating the contributions of the treatment within levels of one or more pre-treatment covariates. For example, in the social sciences, the unconditional decomposition might quantify the overall contributions of racial differences in incarceration rates to racial income disparities; and the conditional decomposition could identify the contributions of racial differences in mortgage receipt to racial disparities in home ownership conditional on credit scores, i.e., considering only the part of the racial differences in mortgage receipt that remains after accounting for racial differences in credit scores. In medical research, the unconditional decomposition might inform the overall contributions of vaccinations to racial health disparities; and the conditional decomposition can quantify the contributions of receiving a medical treatment to health disparities conditional on indications for that treatment.  

We develop nonparametric estimators for our causal decompositions using efficient influence functions (EIF) under the assumption of conditional ignorability of the treatment. Our estimators can be implemented via data-adaptive methods such as machine learning (ML) and accommodate high-dimensional confounders. We  derive the conditions under which the estimators are $\sqrt{n}$-consistent, asymptotically normal, and semiparametrically efficient. The estimators are also doubly or even quadruply robust in terms of consistency. The estimators are implemented in the R package cdgd \citep{yu_2023}, available from CRAN.

As an empirical application, we study the causal contributions of college graduation to intergenerational income persistence (the complement to income mobility), defined as the disparity in income attainment across parental income groups. This application contributes to multiple literatures in sociology and economics. Policy-wise, it provides insights into how interventions on college education may alter intergenerational income persistence. R Code for the empirical application is available at \url{https://github.com/ang-yu/causal_decomposition_case_study}. 

Our paper proceeds as follows. Section 2 introduces our causal decompositions and their interventional interpretations. We explicate the contributions of our framework by formally contrasting it with KBO decompositions, CMA, and random equalization decompositions. In section 3, we introduce the estimators and their asymptotic theory. Section 4 presents the empirical application. Section 5 concludes with extensions. All proofs are collected in the supplementary appendices. 

\section{Estimands}
\subsection{Unconditional decomposition}
We consider a binary treatment variable $D_i \in \left\lbrace 0,1 \right\rbrace$ for each individual $i$. Let $Y_{i}^0$ and $Y_{i}^1$ be the potential outcomes \citep{rubin_estimating_1974} of $Y_i$ under the hypothetical intervention to set $D_i=0$ and $D_i=1$, respectively. Let $\tau_i \coloneqq Y_{i}^1 - Y_{i}^0$ denote the individual-level treatment effect. For expositional convenience, we assume that higher values of $Y_i$ are better. Suppose that the population contains two disjoint groups, $G_i=g \in \left\lbrace a,b \right\rbrace$, where $a$ denotes the advantaged group and $b$ denotes the disadvantaged group. We use subscript $g$ to indicate group-specific quantities, for example, $\E_g(Y_i) \coloneqq \E(Y_i \mid g)$. Henceforth, we suppress subscript $i$ to ease notation.
In our empirical application, $G$ is parental income group, $Y$ is adult income, and $D$ is college graduation. 
 
\begin{assu}[SUTVA]
     $Y=(1-D) Y^0 + D Y^1$. \label{assu1}
\end{assu} 
Assuming only the stable unit treatment value assumption (SUTVA) on the relationship between the treatment and the outcome \citep{rubin_randomization_1980,rubin_comment_1986}, the observed outcome disparity between group $a$ and $b$ can be decomposed into four components: 
\begin{align}
&\phantom{{}={}} \E_a(Y)-\E_b(Y) \label{eqt1}  \\
&= \underbrace{ \E_a \left(Y^0 \right)-\E_b \left(Y^0 \right) }_{\text{\small Baseline}}
+ \underbrace{\E_b(\tau) [\E_a(D)-\E_b(D)]}_{\text{\small Prevalence}} + \underbrace{\E_a(D)[ \E_a(\tau) - \E_b(\tau) ]}_{\text{\small Effect}} 
+ \underbrace{\Cov_a(D, \tau) -  \Cov_b(D, \tau)}_{\text{\small  Selection}}.  \nonumber  
\end{align} 
First, the ``baseline'' component reflects the difference in mean baseline potential outcomes, $Y^0$, between groups, i.e., the outcome disparity in the complete absence of the treatment.\footnote{The baseline component is identical to the ``counterfactual disparity measure'' proposed by \citet{naimi_mediation_2016}. In contrast to the two-way decomposition of \citet{naimi_mediation_2016}, our four-way decomposition distinguishes more mechanisms.} In our application, the baseline component is the part of the income disparity in adulthood that is not attributable to college graduation in any way. Second, the ``prevalence'' component indicates how much of the group disparity is due to differential prevalence of the treatment. For example, it indicates the extent to which the difference in college graduation rates between parental income groups contributes to the outcome disparity. Third, the ``effect'' component reflects the difference in average treatment effects (ATE) between groups. Thus, it reveals the contribution of group-differential ATEs of college graduation to the adult income disparity. 

Fourth, the ``selection'' component captures the contribution of differential selection into treatment based on the individual-level treatment effects. Selection into treatment within each group is captured by $\Cov_g(D,\tau)$. This covariance is positive if group members who would benefit more from the treatment are more likely to receive the treatment. In our example, differential selection will increase the income disparity in adulthood if selection into college graduation is more positive in the higher parental income group than in the lower parental income group. 

Both the effect component and the selection component account for the contribution of effect heterogeneity to group disparities. Whereas the effect component captures the contribution of \emph{between}-group effect heterogeneity, the selection component captures the contribution of \emph{within}-group effect heterogeneity. To our knowledge, no prior decomposition has captured the role of differential selection into treatment in the generation of group disparities. 

To further explicate our novel selection component, we provide two interpretations for the covariance between the treatment and treatment effect, $\Cov(D, \tau)$. First, when the receipt of the treatment is mainly based on self-selection (e.g, college graduation), the covariance may indicate the extent to which the choice to take up the treatment is rational with respect to returns to the treatment. This interpretation has been extensively studied in economics and sociology \citep{heckman_structural_2005, brand_who_2010, heckman_returns_2018}, where a quantity closely related to $\Cov(D, \tau)$ has been labeled ``sorting on gains''. We formally explicate the connection to ``sorting on gains'' in Appendix F. Second, if treatment assignment is mainly administered by external decision-makers (e.g., drug prescription), the covariance indicates how effectively the treatment is assigned to individuals (see Appendix B). Thus, the selection component could also be called the sorting component or the effectiveness component.

Finally, the sum of the prevalence, effect, and selection components can be thought of as the unconditional total contribution of the treatment. Thus, equation (\ref{eqt1}) decomposes the total contribution of a treatment into three distinct causal mechanisms.

\subsection{Interventional interpretation} \label{sec:intervention}
We expect that most researchers will use our decomposition for retrospective causal explanation of 
existing disparities. However, since our decomposition  is formulated in counterfactual terms, it is also prescriptive for future interventions to change disparities. 
Here, we explicate a three-step sequential intervention on the treatment, which successively eliminates the selection, prevalence, and effect components, in this order. The baseline component is the remaining disparity after the intervention.

We express this three-step sequential intervention using the randomized intervention notation  \citep{didelez_direct_2006, geneletti_identifying_2007}, where $R(D \mid g)$ represents a randomly drawn value of the treatment $D$ from group $g$. Then, $\E_g \left(Y^{R(D \mid g') } \right)$ denotes the post-intervention mean potential outcome for group $g$ after each member of group $g$ has received a random draw of the treatment from group $g'$. When $g=g'$, the intervention amounts to a random redistribution of the treatment within the group. Using its definition, we can rewrite the post-intervention mean potential outcome:
\begin{align}
    \E_g \left(Y^{R(D \mid g') } \right) = \E_g (Y^0) + \E_{g'}(D)\E_g(\tau). \label{eqt2}
\end{align}

Then the components of the unconditional decomposition can be re-written as follows:
\begin{align*}
   \text{Selection} &= \underbrace{\E_a(Y) - \E_b(Y) - \left[ \E_a \left(Y^{R(D \mid a)} \right) - \E_b \left(Y^{R(D \mid b)}\right) \right]}_{\text{\small Change in disparity in the first step}} \\
   \text{Prevalence} &= \underbrace{\E_a \left(Y^{R(D \mid a)} \right) - \E_b \left(Y^{R(D \mid b)}\right) - \left[\E_a \left(Y^{R(D \mid a)} \right)-\E_b \left(Y^{R(D \mid a)} \right)\right]}_{\text{\small Change in disparity in the second step}} \\
   \text{Effect} &= \underbrace{\E_a \left(Y^{R(D \mid a)} \right)-\E_b \left(Y^{R(D \mid a)} \right) - \left[ \E_a \left(Y^0 \right) - \E_b \left(Y^0 \right) \right]}_{\text{\small Change in disparity in the third step}}  \\
   \text{Baseline} &= \underbrace{\E_a \left(Y^0 \right) - \E_b \left(Y^0 \right)}_{\text{\small Remaining disparity}}.
\end{align*}
The first step of the intervention internally randomizes the treatment within each group. In this step, the pre-intervention disparity is $\E_a(Y) - \E_b(Y)$, and the post-intervention disparity is $\E_a \left(Y^{R(D \mid a)} \right) - \E_b \left(Y^{R(D \mid b)}\right)$. Since randomizing the treatment within each group sets $\Cov_a(D,\tau)=\Cov_b(D,\tau)=0$ and removes differential selection between groups, the change in disparity resulting from this step equals the selection component.

The second step of the intervention equalizes treatment prevalence by giving members of group $b$ random draws of the treatment from group $a$. In this step, the pre-intervention disparity is $\E_a \left(Y^{R(D \mid a)} \right) - \E_b \left(Y^{R(D \mid b)}\right)$, and the post-intervention disparity is $\E_a \left(Y^{R(D \mid a)} \right)-\E_b \left(Y^{R(D \mid a)} \right)$. Thus, the prevalence component is the change in disparity resulting from this equalization step.\footnote{This step also justifies the scaling factors in the prevalence and effect components. Intuitively, the prevalence component is scaled by $\E_b(\tau)$, because randomly changing treatment prevalence in group $b$ affects the outcome disparity only if the treatment has an effect in group $b$ on average. The scaling factor in the effect component follows algebraically to complete the decomposition. Different interventions would lead to different scaling factors. For example, if group $a$ is instead intervened to have the treatment prevalence of group $b$, the prevalence component would be scaled by $\E_a(\tau)$.} 

The third step of the intervention sets the treatment to zero for all individuals in both groups. In the third step, the pre-intervention disparity is $\E_a \left(Y^{R(D \mid a)} \right)-\E_b \left(Y^{R(D \mid a)} \right)$, and the post-intervention disparity is $\E_a \left(Y^0 \right) - \E_b \left(Y^0 \right)$. When nobody receives the treatment, the influence of differential ATEs across groups is deactivated, which is why the change in disparity in this step equals the effect component.\footnote{The effect component also corresponds to the change in disparity brought about by giving group $a$ the ATE of group $b$. Although the notion of interventions on effects or structural relations often appears in the literature \citep[e.g.,][]{malinsky_intervening_2018, diderichsen_differential_2019, brady_rethinking_2017}, these interventions cannot be expressed in the potential outcomes notation. In contrast, the third step of our sequential intervention fits in the potential outcomes framework, as it eliminates the effect component by intervening on the treatment variable.\label{fn: intervene_effect}}

Finally, at the end of the three-step sequential intervention, the remaining disparity is the baseline component, which is the part of the disparity not attributable to the treatment and cannot be affected by the intervention on the treatment.\footnote{The three-step interventional interpretation also elucidates the role of the treatment coding scheme. If the labels of treatment ($D=1$) and control ($D=0$) are swapped, the selection and prevalence components will remain unchanged, as their corresponding intervention steps are agnostic to treatment labels. However, the effect and baseline components will change, because the third intervention step assigns everyone to what is labeled as control. Therefore, the choice of treatment labels is a substantive decision to the extent that the researcher seeks to interpret the baseline and effect components.} We present a visualization of the unconditional decomposition in Appendix B.

Importantly, each step of the sequential intervention removes one single component without affecting any other components. Conversely, each component of the decomposition isolates a distinct policy lever that can be sequentially intervened upon. Therefore, estimates of our unconditional decomposition directly enable policy makers to choose between implementing just the first step, the first two steps, or all three steps together.
Such freedom of choice is useful, for example, when the first step is effective for reducing the disparity, but the later two steps would either empirically increase the disparity or be infeasible to implement (e.g., due to cost concerns).
In our empirical example, if the selection component is positive, then a low-cost information intervention might ameliorate the income disparity by preventing suboptimal selection into college among lower-income students, even if the prevalence and effect components do not provide viable disparity-reducing policies.

\subsection{Comparison with prior work}
We compare our new approach to three prior decomposition frameworks. We highlight that no prior decomposition contains a selection component; furthermore, prior approaches require strong assumptions to recover any of our other components. 

\subsubsection{Comparison with the KBO decomposition}
Disparities research in the social and health sciences traditionally employs KBO decompositions.  
The KBO decomposition that most closely resembles our approach decomposes the outcome disparity between groups into four components with respect to treatment $D$ and pre-treatment covariates $\X$:\footnote{In practice, many KBO decompositions are farther removed from our approach. In particular, research often does not separate $D$ from $\X$ or heed the temporal order of variables.}  
\begin{align}
&\phantom{{}={}} \E_a(Y)-\E_b(Y) \label{kbo} \\ 
&= \underbrace{\alpha_a-\alpha_b }_{\text{\small Intercept}}
+ \underbrace{\beta_b [\E_a(D)-\E_b(D)]}_{\text{\small Endowment }}
+ \underbrace{\E_a(D)[\beta_a - \beta_b ]}_{\text{\small Slope}} +  \underbrace{\boldsymbol{\gamma}_b^\intercal [\E_a(\X)-\E_b(\X)] + \E_a(\X)^\intercal (\boldsymbol{\gamma}_a-\boldsymbol{\gamma}_b)}_{\text{\small Residual}}, \nonumber
\end{align}
where $\alpha_g$, $\beta_g$, and $\boldsymbol{\gamma}_g$ are coefficients from the group-specific linear regression that contains only main effects:
$$Y=\alpha_g+\beta_g D + \boldsymbol{\gamma}_g^\intercal \X + \epsilon.$$

This decomposition attains a causal interpretation under (i) the causal assumption of conditional ignorability of the treatment, $Y^d \indep D \mid g, \x$, $\forall d, g, \x$, and (ii) the parametric assumption that the group-specific linear regressions are correctly specified. If and only if both assumptions are satisfied, the endowment and slope components in the KBO decomposition are equivalent to our prevalence and effect components, respectively; and the sum of the intercept and residual components equals our baseline component.\footnote{Prior work has offered alternative causal interpretations for KBO decompositions under various assumptions. For example, a prominent literature shows that KBO decompositions can estimate the average treatment effect on the treated \citep{fortin_decomposition_2011, kline_oaxaca-blinder_2011, yamaguchi_decomposition_2015}. Similarly, \citet{chernozhukov_sorted_2018} show that a KBO decomposition can estimate the partial treatment effect. And \citet{huber_causal_2015} discusses using a KBO decomposition for estimating the natural indirect effect \citep{pearl_direct_2001}. None of these interpretations accommodates a descriptive group variable, with one important exception: \citet{jackson_decomposition_2018} establish a connection between the KBO decomposition (\ref{kbo}) and the unconditional random equalization decomposition that is discussed in Section \ref{sec:URED}.}

Our decomposition (\ref{eqt1}) differs from the KBO decomposition in three respects. First, our decomposition is inherently causal, because it is directly formulated as estimands in potential outcomes notation. In contrast, the KBO decomposition requires additional assumptions to support a causal interpretation. Such assumptions are rarely stated in practice. Second, our decomposition is nonparametric, whereas the KBO decomposition is model-based and hence relies on a particular functional form. Third, the KBO decomposition does not contain a selection component, because the assumed functional form imposes effect homogeneity within each group. By contrast, our nonparametric decomposition does not impose any effect homogeneity and hence contains a selection component as a distinctive conceptual contribution.

\subsubsection{Comparison with causal mediation analysis}
CMA decomposes a total effect of an exposure on an outcome into components in terms of an intermediate mediator. When CMA is used to understand a group disparity, our group variable is treated as the exposure, and our treatment variable becomes the mediator. The CMA literature is vast and contains many different decompositions \citep{vanderweele_explanation_2015}. We focus on a three-way CMA decomposition \citep{vanderweele_attributing_2014,vanderweele_unification_2014} that is useful for illustrating similarities and differences with our decomposition (\ref{eqt1}):   

\begin{align}
    &\phantom{{}={}} \underbrace{\E \left(Y^a \right)-\E (Y^b)}_{\text{\small Total effect of exposure $G$}} \label{eqt:CMA} \\
    &=
    \underbrace{\E \left( Y^{a, 0} - Y^{b, 0} \right)}_{\text{\small Controlled direct effect}} +
    \underbrace{\E \left[ \left( Y^{b,1}-Y^{b,0} \right) \left(D^a-D^b \right) \right]}_{\text{\small Pure indirect effect}} + \underbrace{\E \left[D^a  \left(Y^{a, 1}-Y^{a, 0} - Y^{b, 1} + Y^{b, 0} \right) \right]}_{\text{\small Portion attributable to interaction}}, \nonumber 
\end{align}
where $Y^g$ and $D^g$ are, respectively, the potential outcomes of $Y$ and $D$ when assigned group $g$, and $Y^{g,d}$ is the potential outcome of $Y$ when jointly assigned both group $g$ and treatment $d$. 

Certain equivalences between equation (\ref{eqt:CMA}) and our unconditional decomposition can be established under very strong assumptions: two unconditional ignorability assumptions for $G$, i.e., $Y^{g,d} \indep G$, $\forall d,g$, and $D^g \indep G$, $\forall g$; two SUTVA-type assumptions, $\E_g \left(Y^{g,d} \right)=\E_g \left(Y^d \right)$ and $\E_g \left(D^g \right)=\E_g (D), \forall d,g$; and a cross-world independence assumption \citep{pearl_direct_2001}, $Y^{g,d} \indep D^{g'}, \forall d,g,g'$. Under these assumptions, the conditional direct effect (CDE) equals our baseline component; the pure indirect effect (PIE) equals our prevalence component; and the portion attributable to interaction (PAI) equals our effect component. These equivalences are intuitive: both the CDE and the baseline component capture a group-based outcome difference when the intermediate variable, $D$, is held at $0$; both the PIE and the prevalence component address the role of the prevalence of $D$ in the relationship between the group and the outcome; finally, the PAI and the effect component both reflect how the effect of $D$ interacts with group membership.

However, all existing CMA decompositions, to our knowledge, differ from our unconditional decomposition in three crucial respects. First, CMA decomposes a different quantity: a total effect of group membership, rather than the observed, descriptive, group disparity. Our focus on decomposing descriptive disparities is useful, because descriptive disparities between groups are often the object of interest in their own right in the social and health sciences and the focus of popular and policy concerns. For example, income disparities in adulthood between the children of rich and poor parents are often viewed as concerning, regardless of whether these disparities originate from the causal effect of parental income or from confounding factors such as parents’ education and race.
Furthermore, some group variables, such as race and gender, may be immutable attributes on which it is hard to define an intervention \citep{rubin_estimating_1974, holland_statistics_1986}. As a result, CMA estimands, when encoding a hypothetical intervention on group membership, may not even be well-defined.

Second, the identification assumptions of CMA are much stronger. Whereas, as we show in Section \ref{estimation_sec}, our approach requires only causal assumptions on $D$, CMA requires assumptions on both $D$ and $G$ \citep{vanderweele_explanation_2015}. This is why, above, we need assumptions on $G$ to establish equivalences between equation (\ref{eqt:CMA}) and our unconditional decomposition. 

Third, there is no selection component in CMA. Although it is possible to define an analogous selection component for CMA in terms of exposure-induced selection into the mediator based on the net effect of the mediator on the outcome, i.e., $\Cov \left(D^a, Y^{a, 1}-Y^{a, 0} \right)-\Cov \left(D^b, Y^{b, 1}-Y^{b, 0} \right)$, such a component, to our knowledge, has not appeared in existing CMA decompositions.\footnote{In a companion paper, we discuss in more detail the analogous selection component for CMA in relation to the difference between the total effect and its randomized interventional analogue \citep{yu2024naturalmediationeffectsdiffer}. Also, \citet{zhou_attendance_2024} presents a decomposition that contains the first half of CMA's selection component, $\Cov (D^a, Y^{a, 1}-Y^{a, 0} )$.}

\subsubsection{Comparison with the unconditional random equalization decomposition} \label{sec:URED}
The newest entry into decomposition methodology is the random equalization decompositions \citep{vanderweele_causal_2014, jackson_decomposition_2018, sudharsanan_educational_2021, lundberg_gap-closing_2024}, which exist in unconditional (URED) and conditional versions (CRED). 

The URED is defined in terms of a one-step intervention that equalizes treatment prevalence by assigning random draws of the treatment from the advantaged group, $R(D \mid a)$, to the disadvantaged group \citep{jackson_decomposition_2018}:
\begin{gather}
\E_a(Y)-\E_b(Y)=\underbrace{\E_b \left(Y^{R(D \mid a)} \right)-\E_b(Y)}_{\text{\small Change in disparity}} + \underbrace{\E_a(Y)-\E_b \left(Y^{R(D \mid a)} \right)}_{\text{\small Remaining disparity}} .  \nonumber 
\end{gather}
Thus, the outcome disparity is decomposed into the change in disparity resulting from the intervention and the remaining disparity after the intervention.

The URED shares several similarities with our approach. It decomposes the observed disparity, and it is formulated as causal estimands with interventional interpretations. Furthermore, when there is no selection into treatment, the change in disparity of the URED equals our prevalence component, and the remaining disparity equals the sum of our baseline and effect components. 

However, the URED also differs from our unconditional decomposition in important respects. First, as a two-way decomposition, the URED contains less information than our four-way decomposition. Correspondingly, the URED only informs a one-step intervention, which is less flexible than the three-step sequential intervention our decomposition provides. 

Second, when selection is present, the URED does not isolate any of the mechanisms identified in our decomposition (\ref{eqt1}). Notably, the URED does not have a component isolating the contribution of differential treatment prevalence across groups.  
Formally, the change in disparity of the URED equals $\E_b(\tau)[\E_a(D)-\E_b(D)]-\Cov_b(D, \tau)$, which mixes our prevalence component with selection into treatment in the disadvantaged group. 
Intuitively, by assigning the disadvantaged group random draws of the treatment, the URED's one-step intervention simultaneously equalizes treatment prevalence across groups and randomizes treatment assignment in the disadvantaged group.
By contrast, as noted in Section \ref{sec:intervention}, the three-step sequential intervention corresponding to our unconditional decomposition neatly separates equalization from randomization. Consequently, unlike our prevalence component, the URED's change in disparity is generally non-zero even if there is no difference in treatment prevalence, $\E_a(D)=\E_b(D)$.

By the same token, the URED does not contain a selection component, because the contribution of differential selection to the disparity is split between the URED's two components. The change in disparity contains the negative of selection into treatment in the disadvantaged group, and the remaining disparity contains selection into treatment in the advantaged group alongside our baseline and effect components, $\E_a(Y^0)-\E_b(Y^0)+ \E_a(D)[\E_a(\tau)-\E_b(\tau)] + \Cov_a(D,\tau)$.\footnote{\citet{lundberg_gap-closing_2024} introduces a variant of the URED whose one-step intervention equalizes treatment prevalence across groups by assigning random draws from the marginal treatment distribution in the population, $R(D)$, to both groups. The resulting change in disparity is  $\E_a(Y)-\E_b(Y)-\left[\E_a \left(Y^{R(D)} \right)-\E_b \left(Y^{R(D)} \right) \right]$. Rewriting this as
$\E(\tau)[\E_a(D)-\E_b(D)] + \Cov_a(D, \tau) - \Cov_b(D, \tau) - [\Pro(G=a)-\Pro(G=b)][\E_a(D) - \E_b(D)][\E_a(\tau)-\E_b(\tau)] $ shows that this URED variant also fails to separate the contributions of differential prevalence and differential selection.}

\subsection{Conditional decomposition}
Our conditional decomposition evaluates the contributions of the treatment to the outcome disparity conditional on a vector of pre-treatment covariates $\Q$. 
Thus, the conditional decomposition quantifies how the disparity would change if differences in treatment  selection, prevalence, and effect were removed only between members of the advantaged and disadvantaged groups who share the same values of $\Q$. The conditional decomposition is useful in practice because it informs interventions that may be normatively more desirable or practically more feasible than their unconditional counterparts \citep[see][]{jackson_meaningful_2021}.\footnote{\citet{jackson_meaningful_2021} calls $\Q$ ``treatment-allowable covariates'' in the sense that the analyst allows group differences in treatment assignment that are associated with $\Q$ to remain and does not investigate their remediation.} 

In our application, $\Q$ is academic achievement in high school. The conditional decomposition thus informs interventions on college graduation that are conducted only within levels of prior achievement, preserving the relationship between prior achievement and college degree attainment. 

\begin{assu}[Common support]
    $\text{supp}_a(\Q) = \text{supp}_b(\Q)$. \label{assu2}
\end{assu} 
In order to introduce the conditional decomposition, we assume common support on $\Q$, which rules out the scenario where certain $\Q$ values only exist in one group. Under Assumptions \ref{assu1} and \ref{assu2}, we then obtain the following conditional decomposition:
\begin{align}
    \E_a(Y)-\E_b(Y) &= \underbrace{\E_a \left(Y^0 \right)-\E_b \left(Y^0 \right)}_{\text{\small Baseline}}  + \underbrace{\int [\E_a(D \mid \q)-\E_b(D \mid \q)]\E_b(\tau \mid \q) f_b(\q) \dd \q}_{\text{\small Conditional prevalence}}  \label{eqt5} \\
    &\phantom{{}={}} + \underbrace{\int [\E_a(\tau \mid \q)-\E_b(\tau \mid \q)] \E_a(D \mid \q) f_a(\q) \dd \q}_{\text{\small Conditional effect}}  + \underbrace{\E_a[\Cov_a(D, \tau \mid \Q)] - \E_b[\Cov_b(D, \tau \mid \Q)}_{\text{\small Conditional selection}}] \nonumber \\
    &\phantom{{}={}} + \underbrace{\int \E_a(D \mid \q) \E_b(\tau \mid \q) [f_a(\q)-f_b(\q)] \dd \q}_{\text{\small $\Q$-distribution}}. \nonumber
\end{align}
The baseline component in the conditional decomposition is the same as the baseline component in the unconditional decomposition: it is the level of the disparity had nobody received the treatment. 
The conditional prevalence component represents the part of the disparity that is due to differences in treatment prevalence across groups within levels of $\Q$. In our application, it indicates the extent to which the difference in college graduation rates between equally achieving advantaged and disadvantaged students contributes to the income disparity in adulthood. The conditional effect component reflects the contribution of the difference in the group-specific conditional average treatment effect (CATE) given $\Q$. Thus, it reveals the contribution of group-differential CATEs of college graduation given prior academic achievement. The conditional selection component is the contribution of group-differential selection into treatment net of $\Q$, for example, the contribution of differential sorting into college graduation after accounting for achievement in high school.  

Analogous to the unconditional total contribution of the treatment, the conditional total contribution of the treatment is thus the sum of the conditional prevalence, effect, and selection components. The $\Q$-distribution component, which appears only in the conditional decomposition, equals the difference between the unconditional and conditional total contributions of the treatment to the outcome disparity. Hence, the $\Q$-distribution component is the between-$\Q$, as opposed to within-$\Q$, part of the unconditional total contribution of the treatment.\footnote{Furthermore, the $\Q$-distribution component identifies the amount of the disparity associated with (i) the relationship between $G$ and $\Q$ and (ii) the within-group relationship between $\{D,\tau\}$ and $\Q$. Clearly, if $G \indep \Q$ or $\{D,\tau\} \indep \Q \mid g$, the $\Q$-distribution component will be zero. In our application, if there is no group difference in prior achievement or if prior achievement is not associated with college graduation and the effect of college graduation on adult income, then the $\Q$-distribution component will be zero. \label{fn: Q_interpretation}}

\subsubsection{Interventional interpretation} \label{sec: con_intervention}
Similar to the unconditional case, our conditional decomposition has an interventional interpretation that maps a three-step sequential intervention to the decomposition components.  
Let $\E_g \left(Y^{R(D \mid g',\Q)} \right)$ be the mean potential outcome of group $g$ when its members with certain $\Q$ values are given treatments randomly drawn from those members of group $g'$ who have the same $\Q$ values.
We can rewrite this mean potential outcome as:
\begin{align}
    \E_g \left(Y^{R(D \mid g',\Q)} \right) = \E_g \left(Y^0 \right) + \int \E_g(\tau \mid \q) \E_{g'}(D \mid \q) f_g(\q) \dd \q. \label{eqt6}
\end{align}
This allows us to express the components of the conditional decomposition as:
\begin{align}
    &\text{Conditional selection} = \underbrace{\E_a(Y)-\E_b(Y)- \left[\E_a \left(Y^{R(D \mid a,\Q)} \right)-\E_b \left(Y^{R(D \mid b,\Q)} \right) \right]}_{\text{\small Change in disparity in the first step}} \label{eqt7} \\
    &\text{Conditional prevalence} = \underbrace{\E_a \left(Y^{R(D \mid a,\Q)} \right)-\E_b \left(Y^{R(D \mid b,\Q)} \right) - \left[\E_a \left(Y^{R(D \mid a,\Q)} \right) - \E_b \left(Y^{R(D \mid a,\Q)} \right) \right]}_{\text{\small Change in disparity in the second step}} \nonumber \\
    &\text{Conditional effect} + \text{$\Q$-distribution} = \underbrace{\E_a \left(Y^{R(D \mid a,\Q)} \right) - \E_b \left(Y^{R(D \mid a,\Q)} \right) -\left[\E_a \left(Y^0 \right) - \E_b \left(Y^0 \right) \right]}_{\text{\small Change in disparity in the third step}} \nonumber  \\
    &\text{Baseline} = \underbrace{\E_a \left(Y^0 \right) - \E_b \left(Y^0 \right)}_{\text{\small Remaining disparity}}. \nonumber 
\end{align}

Therefore, the first step of the intervention randomizes the treatment within groups and within $\Q$ levels. In this step, the pre-intervention disparity is $\E_a(Y)-\E_b(Y)$, the post-intervention disparity is $\E_a \left(Y^{R(D \mid a,\Q)} \right)-\E_b \left(Y^{R(D \mid b,\Q)} \right)$, and the resulting change in disparity equals the conditional selection component. 
The second step equalizes treatment prevalence across groups but within $\Q$ levels. In this step, the pre-intervention disparity is $\E_a \left(Y^{R(D \mid a,\Q)} \right)-\E_b \left(Y^{R(D \mid b,\Q)} \right)$, the post-intervention disparity is $\E_a \left(Y^{R(D \mid a,\Q)} \right) - \E_b \left(Y^{R(D \mid a,\Q)} \right)$, and the change in disparity equals the conditional prevalence component. The third step sets the treatment to zero for all individuals. Thus, this final step retains only the baseline component as the post-intervention disparity, while the change in disparity is the sum of the conditional effect and $\Q$-distribution components.\footnote{Different from the effect component in the unconditional decomposition, there does not exist an obvious interventional step that isolates the conditional effect component itself. Setting $D=0$ within each level of $\Q$ in analogy to the third step of the unconditional decomposition sets $D=0$ marginally and hence eliminates both the conditional effect and $\Q$-distribution components. Equalizing the CATE given 
$\Q$ across groups would eliminate the conditional effects component alone, but, as noted in Footnote \ref{fn: intervene_effect}, interventions on effects are unconventional within the potential outcomes framework.} \footnote{For equation (\ref{eqt6}) to hold, we only require $\text{supp}_g(\Q) \subseteq \text{supp}_{g'}(\Q)$. This implies that the three-step intervention is well-defined as long as $\text{supp}_b(\Q) \subseteq \text{supp}_{a}(\Q)$. Intuitively, at each level of $\Q$ in group $b$, we must be able to find members of group $a$ with the same $\Q$ values in order to conduct the equalization intervention (the second step). Note that $\text{supp}_b(\Q) \subseteq \text{supp}_{a}(\Q)$ is a weaker condition than Assumption \ref{assu2}. However, under the weaker condition, only the pre- and post-intervention disparities are well-defined, not all components in equation (\ref{eqt5}).} 

\subsubsection{Comparison with path-specific effects in causal mediation analysis}
Since there is no apparent counterpart of our  conditional decomposition in the KBO framework, we proceed to comparison with CMA. 
As a subfield of CMA, path-specific effects provide fine-grained decompositions of the total effect of an exposure when there is more than one mediator \citep[][Section 5.5]{avin_identiability_2005, vanderweele_explanation_2015}. Treating the group variable $G$ as the exposure and all variables in $\{\Q, D\}$ as mediators, such that $\Q$ is temporally after $G$ but before $D$, we have the following path-specific decomposition of the total effect of the exposure:
\begin{align*}
    &\phantom{{}={}} \underbrace{\E \left(Y^a \right)-\E (Y^b)}_{\text{\small Total effect of exposure $G$}} \\
    &=
    \underbrace{\E \left( Y^{a, \Q^a, D^{a, \Q^a}} - Y^{b, \Q^a, D^{a, \Q^a}} \right)}_{\text{\small Paths involving neither $\Q$ nor $D$}} +
    \underbrace{\E \left( Y^{b, \Q^a, D^{a, \Q^a}} - Y^{b, \Q^b, D^{a, \Q^b}} \right)}_{\text{\small Paths involving $\Q$}} + \underbrace{\left( Y^{b, \Q^b, D^{a, \Q^b}} - Y^{b, \Q^b, D^{b, \Q^b}} \right)}_{\text{\small Paths involving $D$ but not $\Q$}},
\end{align*}
where $Y^{g, \Q^{g'}, D^{g'', \Q^{g'''}}}$ is a potential outcome with ``nested counterfactuals'' indicating the assignment of $G$, $\Q$, and $D$.
The total effect of $G$ is decomposed into the paths involving neither $\Q$ nor $D$ (i.e., $A \rightarrow Y$), the paths involving $\Q$ (i.e., the combination of $G \rightarrow \Q \rightarrow D \rightarrow Y$ and $G \rightarrow \Q \rightarrow Y$), and the paths involving $D$ but not $\Q$ (i.e., $G \rightarrow D \rightarrow Y$). Figure A1 in Appendix A.8. illustrates these paths.

The equivalence between the paths involving $D$ but not $\Q$ and the conditional prevalence component can be established under some very strong assumptions detailed in Appendix A.8. This equivalence is intuitive, as our conditional prevalence component captures the contribution of differential prevalence of the treatment $D$ after accounting for $\Q$. However, none of other components of our conditional decomposition is isolated in the path-specific decomposition under these assumptions. Moreover, unlike the path-specific decomposition, our conditional decomposition does not impose any temporal order between $G$ and $\Q$ or require ignorability assumptions on $G$ or $\Q$.

\subsubsection{Comparison with the conditional random equalization decomposition}
Analogous to the URED in Section \ref{sec:URED}, the CRED \citep{jackson_meaningful_2021} is defined in terms of a one-step intervention that equalizes treatment prevalence within levels of $\Q$ by assigning members of the disadvantaged group who have certain $\Q$ values to random draws of the treatment from advantaged group members who have the same $\Q$ values. The CRED thus decomposes the outcome disparity into the change in disparity resulting from this intervention and the corresponding remaining disparity: 
\begin{gather}
\E_a(Y)-\E_b(Y)=\underbrace{\E_b \left(Y^{R(D \mid a, \Q)} \right)-\E_b(Y)}_{\text{\small Change in disparity}} + \underbrace{\E_a(Y)-\E_b \left(Y^{R(D \mid a, \Q)} \right)}_{\text{\small Remaining disparity}} .  \nonumber 
\end{gather}

As before, the key difference between our conditional decomposition and the CRED is that the CRED is a two-way decomposition that does not isolate any of the mechanisms identified in our conditional decomposition. Notably, the CRED does not have a conditional prevalence component. Formally, the CRED's change in disparity equals 
\begin{gather}
\int  [\E_a(D \mid \q) - \E_b(D \mid \q)] \E_b(\tau \mid \q) f_b(\q) \dd \q - \E_b[\Cov_b(D, \tau \mid \Q)], 
\end{gather}
which mixes our conditional prevalence component with the disadvantaged group's conditional selection.
Intuitively, within $\Q$ levels, the CRED's intervention simultaneously equalizes treatment across groups and randomizes treatment in the disadvantaged group. Consequently, unlike our conditional prevalence component, the URED’s change in disparity is generally non-zero even if there is no difference in treatment prevalence given any $\Q$ value, $\E_a(D \mid \q) = \E_b(D \mid \q), \forall \q$. The CRED does not contain a conditional selection component either, because the CRED splits the contribution of differential conditional selection between its two components.\footnote{\citet{lundberg_gap-closing_2024} also proposes a variant of the CRED, whose change in disparity can be written as $\E_a[\Cov_a(D,\tau \mid \Q)] - \E_b[\Cov_b(D,\tau \mid \Q)] + \int[\E_a(D \mid \q) - \E_b(D \mid \q)][\E_a(\tau \mid \q)f_a(\q) p_b(\q) + \E_b(\tau \mid \q)f_b(\q) p_a(\q)] \dd \q$, where $p_g(\q) = \Pro(G=g \mid \q)$. Hence, this CRED variant also does not separate the contributions of differential conditional prevalence and differential conditional selection.}

\section{Identification, estimation, and inference} \label{estimation_sec}
We identify our unconditional and conditional decompositions using the standard assumptions of conditional ignorability and overlap. Without loss of generality, let $\Q \subseteq \X$.
\begin{assu}[Conditional ignorability]
    $Y^d \indep D \mid \x, g$, $\forall d, \x, g$. \label{assu3}
\end{assu}
\begin{assu}[Overlap]
    $0 < \E(D \mid \x, g) <1$, $\forall \x, g$. \label{assu4}
\end{assu}

We develop nonparametric and efficient estimators for our decompositions.
These estimators are ``one-step'' estimators based on the EIFs of the decomposition components, which remove the bias from naive substitution estimators \citep{bickel_efficient_1998, van_der_vaart_asymptotic_2000,hines_demystifying_2022}. The estimators contain some nuisance functions, which can be estimated using flexible ML methods coupled with cross-fitting. Under conditions specified below, our estimators are $\sqrt{n}$-consistent, asymptotically normal, and semiparametrically efficient. Thus, we are able to construct asymptotically accurate Wald-type confidence intervals and hypothesis tests. Our estimators also have double or quadruple robustness properties. 

To unburden notation, we define the following functions of the observed data: $\mu(d,\X, g)=\E(Y \mid d, \X, g)$, $\pi(d,\X, g) = \Pro(D=d \mid \X, g)$, and $\omega(d, \Q, g)=\E[\mu(d,\X,g) \mid \Q,g]$. Also recall that $p_g = \Pro(G=g)$, and $p_g(\Q) = \Pro(G=g \mid \Q)$. We use circumflexes to denote estimated quantities.

\subsection{Unconditional decomposition}
All components of the unconditional decomposition can be expressed as linear combinations of the total disparity and two generic functions of the potential outcomes evaluated at appropriate values of $d$, $g$, and $g'$: $\xi_{dg} \defeq \E \left(Y^d \mid g  \right)$ and $\xi_{dgg'} \defeq \E \left(Y^d \mid g \right)\E \left(D \mid g' \right)$. The relationship between the components of the unconditional decomposition and the generic functions is as follows:
\begin{align*}
    \text{Baseline} &= \xi_{0a}-\xi_{0b}  \\
    \text{Prevalence} &= \xi_{1ba}-\xi_{0ba}-\xi_{1bb}+\xi_{0bb} \\
    \text{Effect} &= \xi_{1aa}-\xi_{0aa} - \xi_{1ba}+\xi_{0ba} \\
    \text{Selection} &= \E_a(Y)-\E_b(Y) - \xi_{0a}+\xi_{0b} - \xi_{1aa}+\xi_{0aa} + \xi_{1bb}- \xi_{0bb}.
\end{align*}
Hence, the EIFs and one-step estimators for the decomposition components directly follow from those for $\xi_{dg}$ and $\xi_{dgg'}$.\footnote{These generic functions also provide a basis for the estimation of Jackson and VanderWeele's (\citeyear{jackson_decomposition_2018}) version of the URED, since its change in disparity can be represented as $\xi_{0b} + \xi_{1ba}-\xi_{0ba}-\E_b(Y)$.}
Under Assumptions \ref{assu1}, \ref{assu3}, and \ref{assu4}, $\xi_{dg}$ and $\xi_{dgg'}$ can be identified as the following nonparametric functionals:
\begin{align*}
    \xi_{dg} &= \E \left[\mu(d,\X,g) \mid g \right] \\
    \xi_{dgg'} &= \E \left[\mu(d,\X,g) \mid g \right] \E \left[D \mid g' \right].
\end{align*}
These identification results then enable the derivation of the  EIFs for $\xi_{dg}$ and $\xi_{dgg'}$.\footnote{When the conditional ignorability assumption (\ref{assu3}) is not credible, researchers may evaluate the robustness of their causal conclusions by performing a
sensitivity analysis, as we do in Appendix H. Alternatively, they may abandon the causal interpretation in favor of a descriptive interpretation. The nonparametric descriptive quantity $\E[\mu(1,\X,G)-\mu(0,\X,G)]$ has been labeled the average controlled difference (ACD) between treatment and control given $\X$ and $G$ \citep{li_propensity_2013, li_balancing_2018}. In the same spirit, one can also descriptively interpret the identified nonparametric functionals for our decomposition components. We define $\E[\mu(1,\X,g)-\mu(0,\X,g) \mid g]$ as the group specific ACD, $\mu(1,\X,G)-\mu(0,\X,G)$ as the controlled outcome difference (COD), and $\E(D \mid \X,G)$ as the controlled treatment prevalence (CTP). Then, the functional for the prevalence component is the group difference in treatment prevalence scaled by group $b$'s ACD, the functional for the effect component is the group difference in ACD scaled by group $a$'s treatment prevalence, and the functional for the selection component is the group difference in the covariance between the COD and the CTP. Formally, 
    \begin{align*}
        \text{Prevalence} &: \E[\mu(1,\X,b)-\mu(0,\X,b) \mid b][\E(D \mid a)-\E(D \mid b)] \\
        \text{Effect} &: \E(D \mid a)\{ \E[\mu(1,\X,a)-\mu(0,\X,a) \mid a] - \E[\mu(1,\X,b)-\mu(0,\X,b) \mid b] \} \\
        \text{Selection} &: \Cov[\mu(1,\X,a)-\mu(0,\X,a), \E(D \mid \X,a) \mid a] - \Cov[\mu(1,\X,b)-\mu(0,\X,b), \E(D \mid \X,b) \mid b].
    \end{align*}} 
\begin{prop}[EIF, unconditional decomposition]
Under Assumptions \ref{assu1}, \ref{assu3}, and \ref{assu4}, the EIF of $\xi_{dg}$ is
$$\phi_{dg}(Y,D,\X,G) \defeq \frac{\one(G=g)}{p_g} \left\{ \frac{\one(D=d)}{\pi(d, \X,g)} [Y-\mu(d,\X,g)] + \mu(d,\X,g) - \xi_{dg} \right\},$$ 
and the EIF of $\xi_{dgg'}$ is
\begin{align*}
    \phi_{dgg'}(Y,D,\X,G) &\defeq \frac{\one(G=g)}{p_g}  \left\{\frac{\one(D=d)}{\pi(d,\X,g)}[Y-\mu(d,\X,g)] + \mu(d,\X,g) \right\} \E \left(D \mid g' \right) \\
    &\phantom{{}={}} + \frac{\one(G=g')}{p_{g'}} \xi_{dg} \left[D - \E \left(D \mid g' \right) \right] - \frac{\one(G=g)}{p_g} \xi_{dgg'}.
\end{align*}
In Appendix C, we also derive the general EIFs with survey weights.\footnote{Related EIFs have appeared in prior work. \citet{park_groupwise_2024} give the EIF for $\xi_{1g}-\xi_{0g}$. The EIF for $\xi_{0,g}+\xi_{1gg'}-\xi_{0gg'}$ coincides with the EIF for the quantity denoted as $\theta_c$ in \citet{diaz_nonparametric_2021}, when pre-treatment confounders in the latter are omitted. Neither of these prior works accommodates survey weights.} 
\end{prop}

We use the EIFs as estimating equations, i.e., set their sample averages to zero and solve for $\xi_{dg}$ and $\xi_{dgg'}$. The one-step estimators of $\xi_{dg}$ and $\xi_{dgg'}$ thus are
\begin{align*}
    \hat{\xi}_{dg} &\defeq \frac{1}{n} \sum \frac{\one(G=g)}{\hat{p}_g} \left\{ \frac{\one(D=d)}{\hat{\pi}(d, \X,g)} [Y-\hat{\mu}(d,\X,g)] + \hat{\mu}(d,\X,g)\right\} \\
    \hat{\xi}_{dgg'} &\defeq \frac{1}{n} \sum \frac{\one(G=g)}{\hat{p}_g}  \left\{\frac{\one(D=d)}{\hat{\pi}(d,\X,g)}[Y-\hat{\mu}(d,\X,g)] + \hat{\mu}(d,\X,g) \right\} \hat{\E} \left(D \mid g' \right).
\end{align*}

Each estimator contains two nuisance functions, $\pi(d,\X,g)$ and $\mu(d,\X,g)$. The estimators are consistent as long as either one of the two nuisance functions is consistently estimated. Although $p_g$ and $\E(D \mid g)$ are technically also nuisance functions, their consistent estimation is left implicit hereafter.

\begin{prop}[Double robustness in consistency, unconditional decomposition]
Under Assumptions \ref{assu1}, \ref{assu3}, and \ref{assu4}, either consistent estimation of $\mu(d,\X,g)$ or of $\pi(d,\X,g)$ is sufficient for the consistency of $\hat{\xi}_{dg}$ and $\hat{\xi}_{dgg'}$.\footnote{The estimator of our unconditional decomposition is doubly robust with respect to the same two nuisance functions as the classic augmented-inverse-probability-of-treatment-weighting (AIPW) estimator of the ATE \citep{robins_estimation_1994}.} 
\end{prop}

The nuisance functions $\pi(d,\X,g)$ and $\mu(d,\X,g)$ can be estimated using various methods. We focus on a nonparametric approach where the nuisance functions are estimated using flexible ML models with cross-fitting. The use of cross-fitting allows for weaker conditions on the estimation of the nuisance functions \citep{chernozhukov_double/debiased_2018,kennedy_semiparametric_2022}. In practice, we implement cross-fitting by randomly splitting the data into two disjoint subsamples, fitting the ML models in each subsample and plugging in values of $\X$ from the other sample to obtain the nuisance function estimates. To improve the finite-sample performance of the estimators, we stabilize the weight, $\one(D=d)/\hat{\pi}(d,\X,g)$, by dividing it by its sample average.

To study the asymptotic distribution of the cross-fitted one-step estimators for the unconditional decomposition, we invoke three additional assumptions about the nuisance functions, which are the same as the assumptions required for the double ML estimator of the ATE \citep{chernozhukov_double/debiased_2018,kennedy_semiparametric_2022}. We let $\| \cdot \|$ denote the $L_2$-norm. 
\refstepcounter{assu} 
\label{assu5} 
\begin{subassu}[Boundedness]
    With probability 1, $\hat{\pi}(d,\X,g) \geq \eta$, $\pi(d,\X,g) \geq \eta$, and $|Y-\hat{\mu}(d,\X,g)| \leq \zeta$, for some $\eta>0$ and some $\zeta < \infty$, $\forall d, g$. \label{assu5a}
\end{subassu}

\begin{subassu}[Consistency]
$\| \hat{\mu}(d,\X,g) - \mu(d,\X,g) \| =o_p(1)$, and $\| \hat{\pi}(d,\X,g) - \pi(d,\X,g) \| =o_p(1)$, $\forall d, g$. \label{assu5b}
\end{subassu}

\begin{subassu}[Convergence rate]
$\|\hat{\pi}(d,\X,g)-\pi(d,\X,g)\| \|\hat{\mu}(d,\X,g)-\mu(d,\X,g)\|=o_p(n^{-1/2})$, $\forall d, g$. \label{assu5c}
\end{subassu}

\begin{prop}[Asymptotic distributions, unconditional decomposition] \label{prop3}
Under Assumptions \ref{assu1}, \ref{assu3}, \ref{assu4}, \ref{assu5a}, \ref{assu5b}, and \ref{assu5c}, the cross-fitted one-step estimators for $\hat{\xi}_{dg}$ and $\hat{\xi}_{dgg'}$ are $\sqrt{n}$-consistent, asymptotically normal, and semiparametrically efficient, i.e., $\sqrt{n} \left(\hat{\xi}_{dg} - \xi_{dg} \right) \xrightarrow{d} \mathcal{N} \left(0, \sigma^2_{dg} \right)$, and $\sqrt{n}\left(\hat{\xi}_{dgg'} - \xi_{dgg'} \right) \xrightarrow{d} \mathcal{N} \left(0, \sigma^2_{dgg'} \right)$, where $\sigma^2_{dg} \defeq \E[\phi_{dg}(Y,D,\X,G)^2]$ and $\sigma^2_{dgg'} \defeq \E[\phi_{dgg'}(Y,D,\X,G)^2]$ are the respective semiparametric efficiency bounds.
\end{prop}

We consistently estimate $\sigma^2_{dg}$ and $\sigma^2_{dgg'}$ using the averages of the squared estimated EIFs. The asymptotic distributions can then be used to construct hypothesis tests and confidence intervals. Since the unconditional decomposition components are simple additive functions of the observed disparity, $\xi_{dg}$, and $\xi_{dgg'}$, all properties established for the estimators of $\xi_{dg}$, and $\xi_{dgg'}$ (double robustness, $\sqrt{n}$-consistency, asymptotic normality, and semiparametric efficiency) carry over to the final estimators of the decomposition components.

\subsection{Conditional decomposition}
\label{sec:cond_estimation}
Relative to the unconditional decomposition, the estimation of the conditional decomposition requires consideration of one additional generic function:
$$\xi_{dgg'g''} \defeq \E \left[\E \left(Y^d   \mid \Q, g \right) \E \left(D \mid \Q, g' \right) \mid g'' \right],$$
where $(d, g, g', g'')$ denotes any one of eight combinations of treatment status and group memberships. The relationship between components of the conditional decomposition and the generic functions is as follows:
\begin{align*}
    \text{Baseline} &= \xi_{0a}-\xi_{0b}  \\
    \text{Conditional prevalence} &= \xi_{1bab}-\xi_{0bab}-\xi_{1bbb}+\xi_{0bbb} \\
    \text{Conditional effect} &= \xi_{1aaa}-\xi_{0aaa} - \xi_{1baa}+\xi_{0baa} \\
    \Q\text{-distribution} &= \xi_{1baa}-\xi_{0baa} - \xi_{1bab}+\xi_{0bab} \\
    \text{Conditional selection} &= \E_a(Y)-\E_b(Y) - \xi_{0a}+\xi_{0b} - \xi_{1aaa} + \xi_{0aaa} + \xi_{1bbb}-\xi_{0bbb}.
\end{align*}
Since the estimation of $\xi_{dg}$ was discussed in the previous subsection, we now focus on $\xi_{dgg'g''}$. The EIFs, one-step estimators, and their asymptotic distributions for the components of the conditional decomposition will then follow.\footnote{We thereby also provide efficient and nonparametric estimation for the change in disparity in the CRED of \citet{jackson_meaningful_2021}, which can be represented as $\xi_{0b}+\xi_{1bab}-\xi_{0bab}-\E_b(Y)$.}
Under Assumptions  \ref{assu1}, \ref{assu2}, \ref{assu3}, and \ref{assu4}, we identify $\xi_{dgg'g''}$ as $$\xi_{dgg'g''}=\E \left[\omega(d,\Q,g) \E \left(D \mid \Q, g' \right) \mid g'' \right].$$

\begin{prop}[EIF, conditional decomposition]
Under Assumptions \ref{assu1}, \ref{assu2}, \ref{assu3}, and \ref{assu4}, the EIF of $\xi_{dgg'g''}$ is 
\begin{align*}
     &\phantom{{}={}} \phi_{dgg'g''}(Y,D,\X,G) \\
    &=\frac{\one(G=g'')}{p_{g''}} \left[ \omega(d,\Q,g) \E \left(D \mid \Q, g' \right) - \xi_{dgg'g''} \right] + \frac{\one(G=g')p_{g''}(\Q)}{p_{g'}(\Q)p_{g''}} \left[ D-\E \left(D \mid \Q, g' \right) \right] \omega(d,\Q,g) \\
    &\phantom{{}={}} + \frac{\one(G=g) p_{g''}(\Q)}{p_g(\Q)p_{g''}} \left\{ \frac{\one(D=d)}{\pi(d,\X,g)} [Y-\mu(d,\X,g)]+\mu(d,\X,g) - \omega(d,\Q,g) \right\} \E \left(D \mid \Q,g' \right).
\end{align*}
In Appendix C, we also derive the general EIF with survey weights. 
\end{prop}
We again construct the one-step estimator by using the EIF as an estimating equation. 
\begin{align*}
    &\phantom{{}={}} \hat{\xi}_{dgg'g''} \\
    &=\frac{1}{n} \sum \frac{\one(G=g'')}{ \hat{p}_{g''}} \hat{\omega}(d,\Q,g) \hat{\E} \left(D \mid \Q, g' \right) + \frac{\one(G=g') \hat{p}_{g''}(\Q)}{ \hat{p}_{g'}(\Q) \hat{p}_{g''}} \left[ D- \hat{\E} \left(D \mid \Q, g' \right) \right] \hat{\omega}(d,\Q,g) \\
    &\phantom{{}={}} + \frac{\one(G=g) \hat{p}_{g''}(\Q)}{\hat{p}_g(\Q) \hat{p}_{g''}} \left\{ \frac{\one(D=d)}{\hat{\pi}(d,\X,g)} [Y-\hat{\mu}(d,\X,g)]+ \hat{\mu}(d,\X,g) - \hat{\omega}(d,\Q,g) \right\} \hat{\E} \left(D \mid \Q,g' \right).
\end{align*}
This estimator contains five nuisance functions: $p_g(\Q)$, $\pi(d,\X,g)$, $\mu(d,\X,g)$, $\E(D \mid \Q, g)$, and $\omega(d,\Q,g)$. As is the case for the unconditional decomposition, consistent estimation of the conditional decomposition does not require all nuisance functions to be consistently estimated. In particular, $\hat{\xi}_{dgg'g''}$ is quadruply robust to inconsistently estimated nuisance functions. 

\begin{prop}[Quadruple robustness in consistency, conditional decomposition]
Under Assumptions \ref{assu1}, \ref{assu2}, \ref{assu3}, and \ref{assu4}, $\hat{\xi}_{dgg'g''}$ is consistent if one of four minimal conditions holds, as summarized in Table 1.
\end{prop}

\begin{table}[t]
\centering
\caption{Quadruple robustness of $\hat{\xi}_{dgg'g''}$. For each scenario defined by $g,g'$ and $g''$, $\hat{\xi}_{dgg'g''}$ is consistent if any of four minimal sets of nuisance functions indicated by check marks is consistently estimated. The first three panels concern the conditions for the consistent estimation of the conditional prevalence ($g=g''$), conditional effect ($g'=g''$), and conditional selection ($g=g'=g''$) components, respectively. Since all conditions require that either $\pi(d,\X,g)$ or $\mu(d,\X,g)$ be consistently estimated, the baseline component is always consistently estimated. The bottom panel shows the four minimal combinations of nuisance functions that must be consistently estimated so that all components of the conditional decomposition are consistently estimated simultaneously.} 
\begin{tabular}{ccccc}
 $\mu(d,\X,g)$ & $\pi(d,\X,g)$ & $\omega(d,\Q,g)$ & $p_g(\Q)$ & $\E(D \mid \Q,g)$  \\
 \hline 
\multicolumn{5}{c} {$g=g''$ (conditional prevalence)} \\
 \hline 
\checkmark & & \checkmark & \checkmark & \\
& \checkmark & \checkmark & \checkmark & \\
\checkmark & & & & \checkmark \\
& \checkmark & & & \checkmark \\
\hline
\multicolumn{5}{c} {$g'=g''$ (conditional effect)} \\
\hline 
\checkmark & & \checkmark & & \\
& \checkmark & \checkmark & & \\
\checkmark & & & \checkmark & \checkmark \\
& \checkmark & & \checkmark & \checkmark \\
\hline
\multicolumn{5}{c} {$g=g'=g''$ (conditional selection)} \\
\hline 
\checkmark & & \checkmark & & \\
& \checkmark & \checkmark & & \\
\checkmark & & & & \checkmark \\
& \checkmark & & & \checkmark \\
\hline
\multicolumn{5}{c} {All components} \\
\hline 
\checkmark & & \checkmark & \checkmark & \\
& \checkmark & \checkmark & \checkmark & \\
\checkmark & & & \checkmark & \checkmark \\
& \checkmark & & \checkmark & \checkmark \\
\hline
\end{tabular}
\end{table}

As before, we estimate the nuisance functions nonparametrically using ML with cross-fitting. 
To improve finite-sample performance, $\hat{\xi}_{dgg'g''}$ can be stabilized by dividing $\one(D=d)/\hat{\pi}(d,\X,g)$, $\one(G=g')\hat{p}_{g''}(\Q)/\hat{p}_{g'}(\Q)\hat{p}_{g''}$, and $\one(G=g)\hat{p}_{g''}(\Q)/\hat{p}_g(\Q)p_{g''}$ by their respective sample averages. 

The estimation of $\omega(d,\Q,g)$ deserves particular attention, because its cross-fitting is nonstandard, and it can be doubly robust itself. 
Specifically, we adopt a pseudo-outcome approach \citep[e.g.,][]{van_der_laan_statistical_2006,semenova_debiased_2021}, where the pseudo outcome for each $d$ is defined as 
$$ \delta_d \left(Y, D, \X, G \right) \defeq \frac{\one(D=d)}{\pi(d, \X,G)} [Y-\mu(d,\X,G)] + \mu(d,\X,G),$$
which is motivated by the fact that $\omega(d,\Q,g)=\E \left[\delta_d \left(Y, D, \X, G \right) \mid \Q,g \right]$.
We first randomly draw two disjoint subsamples from the data. Then we estimate $\pi(d, \X,G)$ and $\mu(d,\X,G)$ in each subsample \emph{without} cross-fitting and obtain the estimated pseudo outcome $\hat{\delta}_d \left(Y, D, \X, G \right)$. Finally, we obtain estimates of $\omega(d,\Q,g)$ using cross-fitting, i.e., we fit $\E\left[\hat{\delta}_d \left(Y, D, \X, G \right) \mid \Q, g \right]$ separately in each subsample and plug in values of $\Q$ from the respective other subsample. Using this procedure, we ensure that the fitting of $\omega(d,\Q,g)$, which relies on estimating the pseudo outcome, is done separately in each subsample. Provided that $\E\left[\hat{\delta}_d \left(Y, D, \X, G \right) \mid \Q, g \right]$ can be consistently estimated, this approach enables consistent estimation of $\omega(d,\Q,g)$ if either $\mu(d,\X,g)$ or $\pi(d,\X,g)$ is consistently estimated.

To establish the asymptotic distributions of the cross-fitted one-step estimators for the conditional decomposition, we invoke Assumptions \ref{assu6}, which augments Assumptions \ref{assu5} with respect to the additional nuisance functions needed for the conditional decomposition. 
\refstepcounter{assu} 
\label{assu6} 
\begin{subassu}[Boundedness]
    With probability 1,  $\hat{\pi}(d,\X,g) \geq \eta$, $\pi(d,\X,g) \geq \eta$, $\hat{p}_g(\Q) \geq \eta$, $p_g(\Q) \geq \eta$, 
$|Y-\hat{\mu}(d,\X,g)| \leq \zeta$, 
$|Y-\mu(d,\X,g)| \leq \zeta$, 
$|\mu(d,\X,g)| \leq \zeta$,
$|\omega(d,\Q,g)| \leq \zeta$, and $\left| \frac{\hat{p}_g(\Q)}{\hat{p}_{g'}(\Q)} \right| \leq \zeta$, for some $\eta>0$ and $\zeta < \infty$, $\forall d,g,g'$. \label{assu6a}
\end{subassu}

\begin{subassu}[Consistency]
$\| \hat{\mu}(d,\X,g) - \mu(d,\X,g) \| =o_p(1)$, $\| \hat{\pi}(d,\X,g) - \pi(d,\X,g) \| =o_p(1)$, $\left\| \hat{\omega}(d,\Q,g) - \omega(d,\Q,g) \right\| =o_p(1)$, $\left\| \hat{\E}(D \mid \Q,g) - \E(D \mid \Q,g) \right\| =o_p(1)$, and $\left\| \hat{p}_g(\Q) -p_g(\Q) \right\|=o_p(1)$, $\forall d,g$. \label{assu6b}
\end{subassu}

\begin{subassu}[Convergence rate]
First, we require $\left\|\hat{\pi}(d,\X,g)-\pi(d,\X,g) \right\| \|\hat{\mu}(d,\X,g)-\mu(d,\X,g)\|=o_p(n^{-1/2}), \forall d,g$. Second, depending on the specific combination of $g,g'$, and $g''$ in a $\xi_{dgg'g''}$, we require $\left\| \hat{\E}(D \mid \Q, g) - \E(D \mid \Q, g)  \right\| = o_p(n^{-1/2}), \forall g$, when $g=g''$; $\left\| \hat{\omega}(d,\Q,g) - \omega(d,\Q,g)  \right\|= o_p(n^{-1/2}), \forall d,g$, when $g'=g''$; and \\ $\left\| \hat{\omega}(d,\Q,g) - \omega(d,\Q,g)  \right\| \left\| \hat{\E}(D \mid \Q, g) - \E(D \mid \Q, g)  \right\| = o_p(n^{-1/2}), \forall d,g$, when $g=g'=g''$. \label{assu6c}
\end{subassu}

\begin{prop}[Asymptotic distribution, conditional decomposition] \label{prop6}
Under Assumptions \ref{assu1}, \ref{assu2}, \ref{assu3}, \ref{assu4}, \ref{assu6a}, \ref{assu6b}, and \ref{assu6c}, the cross-fitted one-step estimator of $\xi_{dgg'g''}$ is $\sqrt{n}$-consistent, asymptotically normal, and semiparametrically efficient, i.e., $\sqrt{n} \left( \hat{\xi}_{dgg'g''} - \xi_{dgg'g''} \right) \xrightarrow{d} \mathcal{N} \left(0, \sigma^2_{dgg'g''} \right)$, where $\sigma^2_{dgg'g''} \defeq \E \left[\phi_{dgg'g''}(Y,D,\X,G)^2 \right]$ is the semiparametric efficiency bound.
\end{prop}

Since the conditional decomposition components are additive functions of the observed disparity, $\xi_{dg}$, and $\xi_{dgg'g''}$, double robustness, $\sqrt{n}$-consistency, asymptotic normality, and semiparametric efficiency all carry over to the final estimators of the decomposition components. We conduct hypothesis tests and construct confidence intervals analogously to the unconditional decomposition. 

Finally, for ML-based estimation of the conditional decomposition, we note a tension between asymptotic normality and semiparametric efficiency on one hand, and consistency on the other. ML does not typically satisfy the convergence rate conditions for establishing asymptotic normality and semiparametric efficiency for three components of the conditional decomposition: conditional prevalence, conditional effect, and $\Q$-distribution. However, we still prefer ML for all components because it achieves consistent estimation without imposing parametric assumptions. Note that this tension does not exist for the baseline and conditional selection components, because their convergence rate conditions can reasonably be achieved by ML.

\section{Application}
\subsection{Overview}

Our application decomposes the contributions of college graduation to the perpetuation of income inequality across generations, which is also known as intergenerational income persistence, the complement to income mobility.
Groups ($G$) are defined by parental income; the outcome ($Y$) is offspring's adult income; and the treatment ($D$) is college graduation. Although previous research has indirectly touched upon all four components of our unconditional decomposition to some extent, this analysis is the first to present a direct and unified decomposition.  Our findings reveal that differential selection into college completion by income origin reduces intergenerational income persistence and increases mobility. This result has not previously been discovered and introduces a new mechanism into the classic Origin-Education-Destination triangle \citep[e.g.,][]{breen_social_2004} in social stratification research.

The baseline component of our decompositions represents the part of the disparity in adult income that is unaccounted for by college graduation. Prior research enumerates multiple sources of intergenerational income persistence that may operate independently of college graduation. For example, parental income is associated with a variety of pre-college characteristics that may directly influence income attainment, such as cognitive skills and noncognitive traits in adolescence \citep{heckman_effects_2006}. Moreover, people from more privileged backgrounds likely benefit from their parents' human, social, and financial capitals regardless of their own formal educational attainment. Interventions on college graduation would not eliminate these channels of income persistence. 

Conceptually speaking to the prevalence component in our unconditional decomposition, social scientists have long regarded education as a mediator in the intergenerational reproduction of socioeconomic inequalities (\citealp[chapter 4 \& 5]{blau_american_1978}; \citealp[p.255-9]{featherman_opportunity_1978}; \citealp{ishida_class_1995}). Specifically, research documents large differences in college graduation rates across parental income groups \citep{ziol-guest_parent_2016, bailey_gains_2011}, and simulations suggest that rising educational inequalities have strengthened intergenerational income persistence over time \citep{bloome_educational_2018}. Nonetheless, prior work is descriptive in nature.

Related to our effect component, there is an active literature on heterogeneity in the effects of college graduation on adult income, although results vary.  Brand et al. (\citeyear{brand_uncovering_2021}) and Cheng et al. (\citeyear{cheng_heterogeneous_2021}) find larger effects of college graduation on the income of people from more disadvantaged backgrounds. By contrast, \cite{zhou_equalization_2019}, \cite{fiel_great_2020}, and \cite{yu_leveraging_2021} find no statistically significant heterogeneity in the effects of college completion on income across parental income groups. However, none of these works evaluates the extent to which income disparities can be attributed to groupwise differential effects of college.

Finally, some prior work has addressed selection into college as a function of college effects on income, i.e. $\Cov(D, \tau)$. Here, too, results are mixed. \citet{brand_who_2010} and \citet{brand_uncovering_2021} find negative selection, i.e., those who are least likely to attend college would benefit most from it. By contrast, the instrumental variable analysis of \citet{heckman_returns_2018} finds positive selection into college. However, prior work estimates selection in the pooled population rather than within parental income groups, thereby missing the link between the difference in group-specific selection and the group-based outcome disparity that our approach identifies. (Appendix F clarifies and synthesizes related notions of ``selection into college'' in the social science literature.)

\subsection{Data, variables and estimation}
We analyze the National Longitudinal Survey of Youth 1979, a nationally representative U.S. cohort study of individuals born between 1957 and 1964.
We restrict the sample to respondents who were between 14 to 17 years old at baseline in 1979 to ensure that income origin is measured prior to respondents' college graduation. We also limit the analysis to respondents who graduated from high school by age 29. The sample size of our complete-case analysis is $N=2,008$. Missingness mostly occurs in the outcome variable due to loss to follow-up (22\%), with less missingness in other variables (<6\%). Appendix Table A1 presents associations between outcome missingness and baseline covariates.

We contrast parental income-origin groups ($G$), defined as the top 40\% and bottom 40\% of family income averaged over the first three waves of the survey (1979, 1980, and 1981, when respondents were 14 to 20 years old) and divided by the square root of the family size to adjust for need \citep[e.g.,][]{zhou_equalization_2019}. The treatment ($D$) is a binary indicator of whether the respondent graduated from college by age 29. The outcome is the percentile rank of the respondent's adult income, averaged over five survey waves between age 35 and 44, divided by the square root of family size. For the conditional decomposition, we define $\Q$ as the Armed Forces Qualification Test (AFQT) score, measured in 1980. The AFQT score is a widely used measure of academic achievement that predicts college completion. 

We measure an extensive set of confounders ($\X$) at baseline, including gender, race, parental income percentile, parental education, parental presence, the number of siblings, urban residence, educational expectation, friends' educational expectation, AFQT score, age at the baseline survey, the Rotter score of control locus, the Rosenberg self-esteem score, language spoken at home, Metropolitan Statistical Area category, separation from mother, school satisfaction, region of residence, and mother's working status.

We present four estimates each for our unconditional and conditional decompositions, using different models for the nuisance functions to assess robustness: three alternative ML methods (gradient boosting machine [GBM], neural networks, and random forests) and one set of parametric models. Depending on whether the left-hand-side variable is continuous or binary, the parametric models are linear or logit. Specifically, for $\mu(d,\X,g)$, we use all two-way interactions between $D$ and $\{\X, G\}$, along with their main effects. For $\pi(d,\X,g)$ and $p_g(\Q)$, the logit models contain only main effects. For $\E(D \mid \Q,g)$ and $\E[\hat{\delta}_d \left(Y, D, \X, G \right) \mid \Q, g ]$, we include all main effects and two-way interactions between $G$ and $\Q$. Appendix H contains model diagnostics and a sensitivity analysis for unobserved confounding. Statistical significance is assessed at the 0.05 level.

\subsection{Results}

\begin{figure}[H]
\centering

\subfigure[Unconditional decomposition]
{
\begin{tikzpicture}[scale = 0.8] 
    \node at (0,0) {};
    \node at (1,1) {};
    
    \node[anchor=center, align=center, inner sep=0pt] at (0,0) {\includegraphics[width=.7\textwidth]{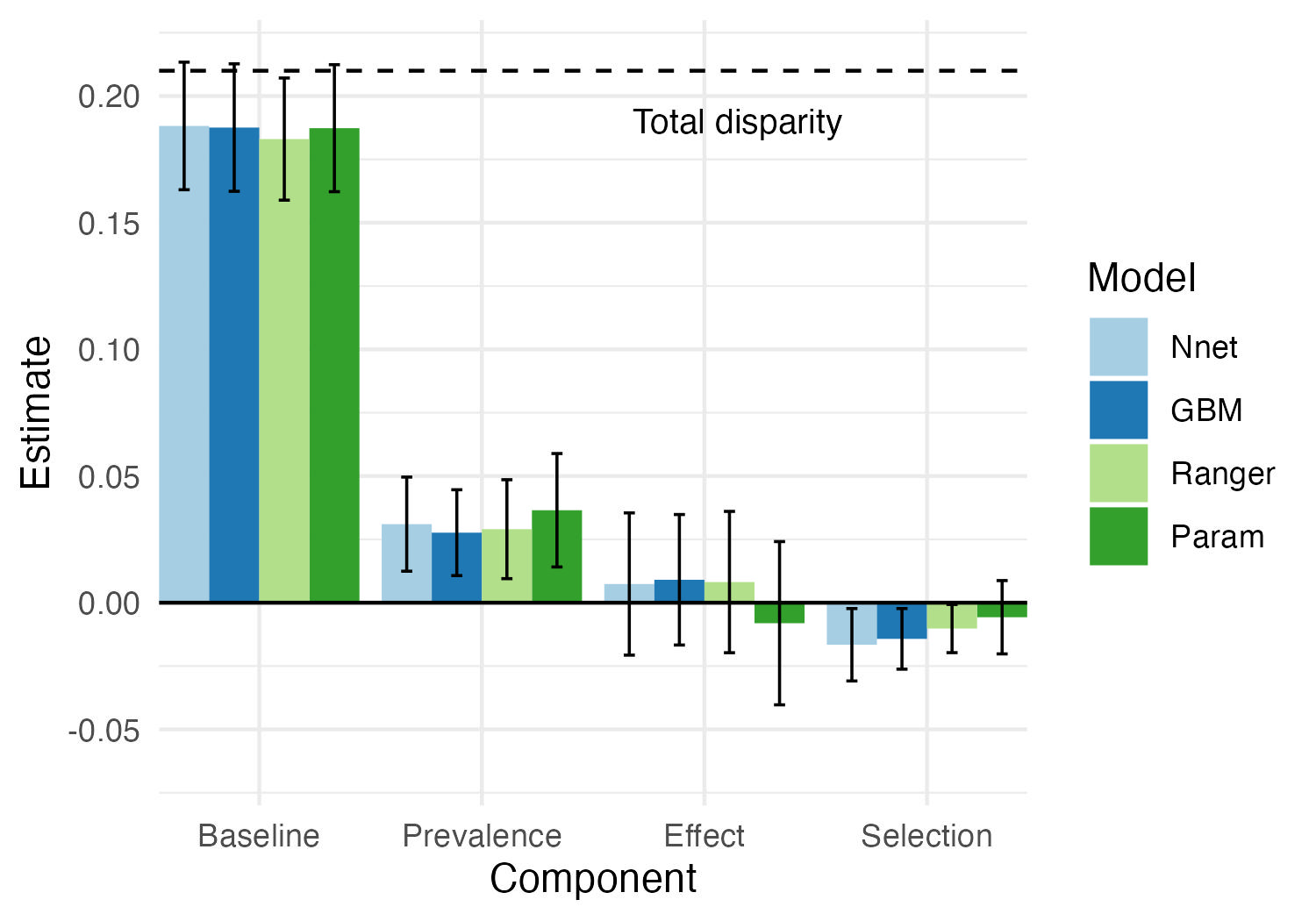}};
\end{tikzpicture}
}
\subfigure[Conditional decomposition]
{
\begin{tikzpicture}[scale = 0.8]  
    \node at (0,0) {};
    \node at (1,1) {};
    
    \node[anchor=center, align=center, inner sep=0pt] at (0,0) {\includegraphics[width=.7\textwidth]{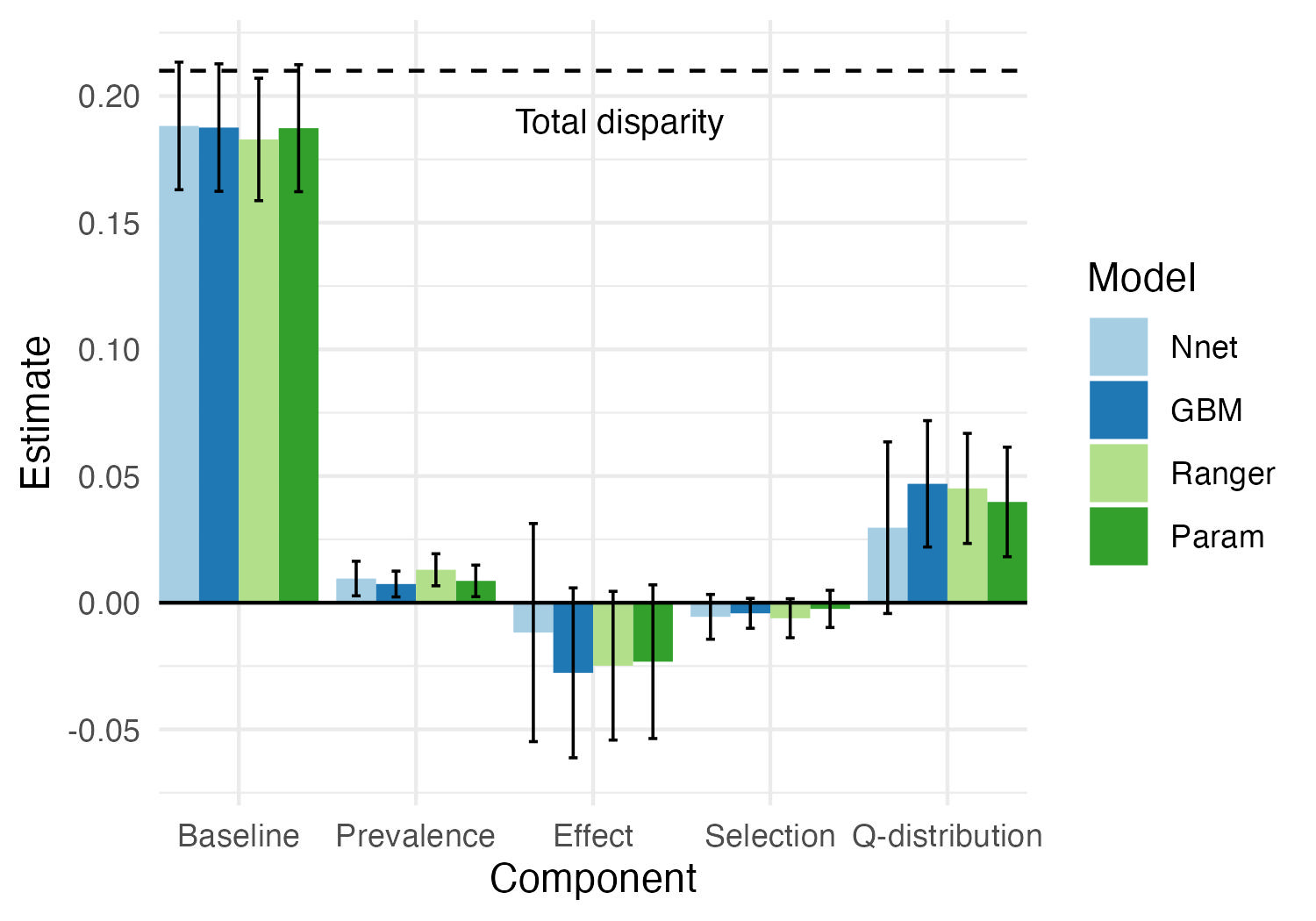}}; 
\end{tikzpicture}
}

\caption{Decomposition estimates. For nuisance function models, Nnet=neural networks, GBM=generalized boosting machines, Ranger=random forests, Para=parametric models. Error bars indicate 95\% confidence intervals, computed according to Theorems \ref{prop3} and \ref{prop6}.} \label{fig:Result}
\end{figure}

Figure \ref{fig:Result} presents our main results for the components of the unconditional and conditional decompositions across different models for the nuisance functions.\footnote{See Appendix Tables A3 and A4 for numerical details. To aid interpretation, Appendix Table A2 additionally reports estimated group-specific means of baseline potential outcome, treatment proportions, ATEs, and covariances between the treatment and the treatment effect, as well as group differences in these quantities.} 
Descriptively, we find that individuals from lower-income origins on average achieve income 21 percentiles lower in their 30s and 40s than individuals from higher-income origins. This confirms the existence of intergenerational income persistence and represents the total disparity that we decompose. Below, we present our results in the order of the three-step sequential interventions and start with the unconditional decomposition.

Our estimates for the selection component in the unconditional decomposition are consistently negative. These estimates are statistically significant for all three ML models but not for the parametric models of the nuisance functions. Consequently, randomizing college graduation within each income-origin group to remove selection (the first step of the three-step sequential intervention) would increase the total disparity in adult income by around 7\%. Thus, this intervention would increase intergenerational income persistence and decrease income mobility. Differential selection into college graduation reduces intergenerational income persistence because selection is positive in the disadvantaged group and negative in the advantaged group (Appendix Table A2). This lends support to stipulations from earlier research that obtaining a college degree is more of a rational decision in pursuit of economic returns among disadvantaged individuals, and more of an adherence to the social norm of attending college among advantaged individuals \citep{mare_social_1980, hout_social_2012}.  The selection component is the central conceptual contribution of our approach and also the most novel finding of our empirical application. 

The prevalence component in the unconditional decomposition is positive, substantively large, and statistically significant across all models for the nuisance functions, accounting for about 15\% of the total disparity in adult income. Consequently, equalizing college graduation rates across income-origin groups as in the second step of the three-step sequential intervention would reduce the total disparity in adult income by about 15\%. In other words, such an intervention would decease intergenerational income persistence and increase income mobility. Underlying the prevalence component is the striking gap in college graduation rates by parental income, as 34\% of respondents in the higher-income group, but only 9\% of the lower-income group, obtained a college degree by age 29 (see Appendix Table A2).

The effect component in the unconditional decomposition is statistically insignificant due to minimal between-group effect heterogeneity. In Appendix Table A2, we show that the group-specific ATEs of college graduation on adult income range from 11 to 15 percentiles across models, all of which are statistically significant. However, the group difference in ATEs is always statistically insignificant. 

The baseline component constitutes nearly 90\% of the total disparity in adult income across all models for the nuisance functions. This demonstrates that most of intergenerational income persistence is due to processes that do not involve college graduation or its effects.

In sum, our unconditional decomposition reveals that college graduation in the United States plays two contradictory roles in intergenerational income persistence. On the one hand, the higher college graduation rate among individuals from higher-income origins increases intergenerational income persistence. On the other hand, some part of the income persistence is offset by the positive selection into college among individuals from lower-income origins.

The conditional decomposition quantifies the contributions of college graduation within levels of the AFQT achievement score. Hence, it informs settings in which policy makers cannot, or do not want to, change the factual relationship between prior achievement and college completion, perhaps due to normative constraints or meritocratic preferences. 

In the conditional decomposition, all estimates for the conditional selection component are negative and statistically insignificant. The conditional prevalence component is positive and statistically significant, although it is much smaller than its unconditional counterpart and only accounts for 3\% to 6\% of the total disparity. Hence, equalizing chances of graduating college within levels of prior achievement (in the manner of the second step of the three-step sequential intervention in Section \ref{sec: con_intervention}) would still somewhat decrease intergenerational income persistence.

Estimates for the conditional effect component are negative and statistically insignificant. The $\Q$-distribution component is positive and statistically significant. It reflects the strong associations between the AFQT score ($Q$) on one hand and parental income ($G)$, college graduation ($D$), and the college effect on income ($\tau$) on the other (see Footnote \ref{fn: Q_interpretation}). The positive $\Q$-distribution component also shows that, net of prior achievement, college graduation plays a much more limited role in producing intergenerational income persistence.\footnote{In Appendix G, as a robustness check, Table A5 presents estimates of the conditional decomposition based on an estimation procedure combining parametric and ML models for specific nuisance functions. As explained in the last paragraph of Section \ref{sec:cond_estimation} and Appendix G, this theoretically may make the asymptotic inference more exact for the conditional prevalence, conditional effect, and $\Q$-distribution components. In practice, however, the empirical findings are qualitatively unchanged.} Finally, by definition, the baseline component is the same in the unconditional and the conditional decompositions.

In Appendix H, we conduct a sensitivity analysis for the conditional ignorability assumption (\ref{assu3}) against unobserved confounding. Leveraging results in \citet{opacic_disparity_2023}, we derive bias formulas for the unconditional and conditional prevalence components. We find that in order to reduce our point
estimates to zero, there must be an unobserved confounder that is about three times as strong as educational expectation, which is shown to be an especially strong observed confounder. This suggests that our estimates are quite robust to unobserved confounding.

Finally, we also estimate the change-in-disparity components in the URED and CRED of \citet{jackson_decomposition_2018} and \citet{jackson_meaningful_2021}. As explained above, these decompositions estimate the impacts of (marginally or conditionally) randomly equalizing the treatment and do not isolate differential prevalence of college graduation as a distinct mechanism. Consequently, both the URED and  CRED underestimate the extent to which differential college graduation rates, independent of selection, contribute to intergenerational income persistence (see Appendix Tables A3 and A4).

\section{Discussion}
The goal of causal disparity analysis is to enumerate, quantify, and disambiguate the mechanisms by which a treatment variable contributes to an observed outcome disparity between groups. We introduced a new nonparametric decomposition approach that is more appropriate for the causal explanation of descriptive disparities and differentiates more mechanisms than prior approaches. In particular, we identify differential selection into treatment as a previously overlooked mechanism and novel policy lever. We developed ML-based estimators that are semiparametrically efficient, asymptotically normal, and multiply robust. Empirically, we demonstrate that our approach provides new insights by documenting that differential prevalence of, and selection into, college graduation play important but counteracting roles in the production of intergenerational income persistence.

In future work, we plan to extend our approach in multiple directions. First, we will develop analogous decompositions for non-binary treatment and group variables, multiple treatments, non-continuous outcomes, and time-to-event outcomes, under the conceptual umbrella of sequential interventions. 
Second, as an alternative to the conditional ignorability assumption, it will be valuable to develop an instrumental variable-based identification strategy for the decompositions, possibly via the marginal treatment effect framework \citep{heckman_structural_2005}.
Third, in social science applications, period or cohort can play the role of the group variable, so longitudinal changes can also be decomposed. Fourth, in terms of estimation, one could employ targeted learning \citep{van_der_laan_targeted_2011} for improved finite-sample performance. 

\subsection*{Acknowledgments}
We thank Paul Bauer, Xavier d'Haultfoeuille, Laurent Davezies, Eric Grodsky, Jim Heckman, Aleksei Opacic, Chan Park, Stephen Raudenbush, Ben Rosche, Michael Sobel, Jiwei Zhao, and Xiang Zhou for helpful suggestions. We are also grateful for the insightful comments of three AOAS reviewers. Earlier versions of this paper have been presented at GESIS, PAA, ACIC, RC28, JSM, ICLC, CREST, and the Universities of Wisconsin, Heidelberg, and Chicago. 

\subsection*{Funding}
The authors gratefully acknowledge core grants to the Center for Demography and Ecology (P2CHD047873) and to the Center for Demography of Health and Aging (P30AG017266) at UW-Madison, a Romnes Fellowship to Felix Elwert at UW-Madison, and a Wisconsin Partnership Program grant from the UW-Madison School of Medicine and Public Health to Christie Bartels and Felix Elwert.

\section*{Appendices}
\section*{Appendix A: Proofs for Section 2}
\subsection*{A.1. Equation (1)}
By Assumption 1 (SUTVA), $Y=(1-D) Y^0 + D Y^1 = Y^0 + D\tau$. Thus, $\E_g(Y)=\E_g( Y^0 + D\tau  )=
\E_g(Y^0)+\E_g(D)\E_g(\tau)+\Cov_g(D, \tau)$.
Equation (1) then follows. 

\subsection*{A.2. Equation (2)}
Note that $R(D \mid g')$ denotes a randomly drawn value of treatment $D$ from group $g'$.
\begin{align*}
   &\phantom{{}={}}  \E_g \left(Y^{R(D \mid g') } \right) \\
   &= \E_g (Y^1  \mid  R(D  \mid g')=1)\Pro_g(R(D  \mid g')=1) + \E_g (Y^0  \mid  R(D  \mid g')=0)\Pro_g(R(D  \mid g')=0) \\
   &= \E_g (Y^1)\E_{g'}(D) + \E_g (Y^0)(1-\E_{g'}(D)) \\ 
   &= \E_g (Y^0) + \E_{g'}(D)\E_g(\tau).
\end{align*}

\subsection*{A.3. Equivalence results in Section 2.3.2}
For the CDE, 
\begin{align*}
&\phantom{{}={}} \E (Y^{a, 0})-\E(Y^{b, 0}) \\
&= \E(Y^{a, 0} \mid  G=a)-\E(Y^{b, 0} \mid  G=b) \\
&= \E_a(Y^0)-\E_b(Y^0),
\end{align*}
where the first equality is by the unconditional ignorability of $G$, and the second equality is by the SUTVA.

For the PIE, 
\begin{align*}
  &\phantom{{}={}} \E \left[ \left(Y^{b, 1} - Y^{b,0} \right) \E(D^a) \right]  \\
  &= \E \left(Y^{b, 1} - Y^{b,0} \right) \E(D^a) \\
  &= \E \left(Y^{b, 1} - Y^{b,0} \mid G=b \right) \E(D^a \mid G=a) \\
  &= \E_b \left(Y^{1} - Y^{0} \right) \E_a(D)
\end{align*}
where the first equality holds by the cross-world independence assumption. The second equality holds by the unconditional ignorability of $G$. And the third holds by the SUTVA. The equivalence result for the PAI can be similarly proved and is omitted.

\subsection*{A.4. Footnote 8}
We first note that the outcome disparity can also be decomposed as such: 
\begin{align*}
&\phantom{{}={}} \E_a(Y)-\E_b(Y) \\
&= \E_a(Y^0)-\E_b(Y^0) +  \E(D)[ \E_a(\tau) - \E_b(\tau)] +\E(\tau)[\E_a(D)-\E_b(D)]  \\
&\phantom{{}={}} + \Cov_a(D, \tau) - \Cov_b(D, \tau) - [p_a-p_b][\E_a(D) - \E_b(D)][\E_a(\tau)-\E_b(\tau)], 
\end{align*} 
where $\E(\tau)$ and $\E(D)$ are the marginal ATE and treatment prevalence, and $p_g$ is the proportion of the population in group $g$.
And the remaining disparity in Lundberg's (\citeyear{lundberg_gap-closing_2024}) unconditional decomposition can be easily rewritten as 
$\E_a(Y^0)-\E_b(Y^0) + \E(D)[ \E_a(\tau) - \E_b(\tau)] \nonumber$. It then follows that the change in disparity is what appears in Footnote 8.

\subsection*{A.5. Equation (5)}
\begin{align*}
    &\phantom{{}={}} \E_a(Y)-\E_b(Y) \\ 
    &= \E_a(Y^0) - \E_b(Y^0) + \E_a(D\tau) - \E_b(D\tau) - \E_a[\E_a(D \mid \Q) \E_a(\tau \mid \Q)] + \E_b[\E_b(D \mid \Q) \E_b(\tau \mid \Q)] \\
    &\phantom{{}={}} + \E_a[\E_a(D \mid \Q) \E_a(\tau \mid \Q)] - \E_b[\E_b(D \mid \Q) \E_b(\tau \mid \Q)]   \\
    &= \E_a(Y^0) - \E_b(Y^0) + \E_a[\Cov_a(D, \tau \mid \Q)] - \E_b[\Cov_b(D, \tau \mid \Q)] \\
    &\phantom{{}={}} + \int \E_a(D \mid \q) \E_a(\tau \mid \q) f_a(\q) \dd \q - \int \E_b(D \mid \q) \E_b(\tau \mid \q) f_b(\q) \dd \q \\
    &= \E_a(Y^0)-\E_b(Y^0) \\
    &\phantom{{}={}} + \int [\E_a(D \mid \q)-\E_b(D \mid \q)]\E_b(\tau \mid \q) f_b(\q) \dd \q + \int \E_a(D \mid \q) \E_b(\tau \mid \q) [f_a(\q)-f_b(\q)] \dd \q \\
    &\phantom{{}={}} + \int [\E_a(\tau \mid \q)-\E_b(\tau \mid \q)] \E_a(D \mid \q) f_a(\q) \dd \q + \E_a[\Cov_a(D, \tau \mid \Q)] - \E_b[\Cov_b(D, \tau \mid \Q)] .
\end{align*}
Note that the last equality uses Assumption 2 (common support). 

\subsection*{A.6. Equation (6)}
\begin{align*}
    &\phantom{{}={}} \E_g \left(Y^{R(D \mid g',\Q)} \right) \\
    &= \int \E_g \left(Y^{R(D \mid g',\q)} \mid \q \right) f_g(\q) \dd \q \\
    &= \int \E_g(Y^1 \mid \q, R(D \mid g', \q)=1) \Pro_g(R(D \mid g', \q)=1 \mid \q) f_g(\q) \dd \q \\
    &\phantom{{}={}} + \int \E_g(Y^0 \mid \q, R(D \mid g', \q)=0) \Pro_g(R(D \mid g', \q)=0 \mid \q) f_g(\q) \dd \q \\
    &= \int [\E_g(Y^1 \mid \q) \E_{g'}(D \mid \q) + \E_g(Y^0 \mid \q)(1-\E_{g'}(D \mid \q)) ] f_g(\q) \dd \q  \\
    &= \E_g(Y^0) + \int \E_g(\tau \mid \q) \E_{g'}(D \mid \q) f_g(\q) \dd \q .
\end{align*}
All expectations are taken over $\q \in \text{supp}_g(\Q)$. For $\E_g \left(Y^{R(D \mid g',\Q)} \right)$ to be well-defined, we require $\text{supp}_g(\Q) \subseteq \text{supp}_{g'}(\Q)$. 

\subsection*{A.7. Equation (7)}
\begin{align*}
    &\phantom{{}={}} \E_a(Y)-\E_b(Y)-\left[\E_a \left(Y^{R(D \mid a,\Q)} \right)-\E_b \left(Y^{R(D \mid b,\Q)} \right) \right] \\
    &= \E_a \left(Y^0 \right)-\E_b \left(Y^0 \right) + \E_a(D \tau) - \E_b(D \tau) -\left[\E_a(Y^{R(D \mid a,\Q)})-\E_b(Y^{R(D \mid b,\Q)}) \right] \\
    &= \E_a[\E_a(D \tau \mid \Q)] - \E_a[\E_a(\tau \mid \Q)\E_a(\tau \mid \Q)] - \lbrace \E_b[\E_b(D \tau \mid \Q)] - \E_b[\E_b(\tau \mid \Q)\E_b(\tau \mid \Q)] \rbrace \\
    &= \E_a[\Cov_a(D, \tau \mid \Q)]- \E_b[\Cov_b(D, \tau \mid \Q)].
\end{align*}
Other results in equation (7) follow directly from equation (6).

\subsection*{A.8. Equivalence result in Section 2.4.2}
To establish the equivalence, we invoke the following assumptions: 
\begin{enumerate}
    \item Five ignorability assumptions
    \begin{enumerate}
        \item $Y^{b,\q,d} \indep G, \forall \q,d$
        \item $Y^{b,\q,d} \indep \Q \mid b, \forall \q,d$
        \item $D^{g,\q} \indep G, \forall g,\q$
        \item $D^{g,\q} \indep \Q \mid g, \forall g,\q$
        \item $\Q^b \indep G$
    \end{enumerate}
    \item Three SUTVA-type assumptions
    \begin{enumerate}
        \item $\E_b \left(Y^{b,\q,d} \mid \q \right)=\E_b(Y^d \mid \q), \forall \q,d$
        \item $\E_g(D^{g,\q} \mid \q) = \E_g(D \mid \q), \forall g,\q$
        \item $f_{b}(\q^b)=f_{b}(\q), \forall \q$
    \end{enumerate}
    \item Two cross-world independence-type assumptions
    \begin{enumerate}
        \item $Y^{b,\q,1} \indep \left\{ \Q^b, D^{a,\q} \right\}, \forall \q$
        \item $D^{g,\q} \indep \Q^b, \forall g,\q$.
    \end{enumerate}
\end{enumerate}
Note that $f_b(\q^b) \defeq f_{\Q^b \mid b}(\q)$ is the density function of $\Q^b$ in group $b$ evaluated at $\q$, and $f_{b}(\q) \defeq f_{\Q \mid b}(\q)$ is the density function of $\Q$ in group $b$ evaluated at $\q$.

Under these assumptions,
\begin{align*}
    &\phantom{{}={}} \E \left( Y^{b,\Q^b,D^{a,\Q^b}} \right) \\
    &= \int \E\left( Y^{b,\q,D^{a,\q}} \mid \Q^b=\q \right) f_{\Q^b}(\q) \dd \q \\
    &= \int \E\left( Y^{b,\q,1} \mid \Q^b=\q, D^{a,\q}=1 \right) \E\left( D^{a,\q} \mid \Q^b=\q \right) f_{\Q^b}(\q) \dd \q \\
    &\phantom{{}={}} + \int \E\left( Y^{b,\q,0} \mid \Q^b=\q, D^{a,\q}=0 \right) \left[1-\E\left( D^{a,\q} \mid \Q^b=\q \right) \right] f_{\Q^b}(\q) \dd \q \\
    &= \int \E\left( Y^{b,\q,1} \right) \E\left( D^{a,\q} \right) f_{\Q^b}(\q) \dd \q + \int \E\left( Y^{b,\q,0} \right) \left[1-\E\left( D^{a,\q} \right) \right] f_{\Q^b}(\q) \dd \q \\
    &= \int \E_b\left( Y^{b,\q,1} \mid \q \right) \E_a\left( D^{a,\q} \mid \q \right) f_{\Q^b \mid b}(\q) \dd \q + \int \E_b\left( Y^{b,\q,0} \mid \q \right) \left[1-\E_a\left( D^{a,\q} \mid \q \right) \right] f_{\Q^b \mid b}(\q) \dd \q \\
    &= \int \E_b\left( Y^{1} \mid \q \right) \E_a\left( D \mid \q \right) f_{\Q \mid b}(\q) \dd \q + \int \E_b\left( Y^{0} \mid \q \right) \left[1-\E_a\left( D \mid \q \right) \right] f_{\Q \mid b}(\q) \dd \q.
\end{align*}
Above, the third equality is due to the cross-world independence-type assumptions. The fourth equality holds by the ignorability assumptions. The fifth (last) equality is due to the SUTVA-type assumptions. 

Similarly, 
\begin{align*}
    &\phantom{{}={}} \E \left( Y^{b,\Q^b,D^{b,\Q^b}} \right) \\
    &= \int \E_b\left( Y^{1} \mid \q \right) \E_b\left( D \mid \q \right) f_{\Q \mid b}(\q) \dd \q + \int \E_b\left( Y^{0} \mid \q \right) \left[1-\E_b\left( D \mid \q \right) \right] f_{\Q \mid b}(\q) \dd \q.
\end{align*}
The equivalence result then follows. Figure A1 illustrates the effect paths referred to in Section 2.4.2.

\begin{figure}[h]
    \centering
        \begin{tikzpicture} 
    \node at (0,0) {};
    \node at (1,1) {};
    
    \node[anchor = center, align = center] (g) at (0,0) {$G$};
    \node[anchor = center, align=center] (d) at (3,0) {$D$}; 
    \node[anchor = center, align=center] (y) at (6,0) {$Y$};   
    \node[anchor = center, align=center] (q) at (3,-1.5) {$\Q$};  
    
    \draw[->, line width=0.4mm] (g) to (d);
    \draw[->, line width=0.4mm] (d) to (y);
    \draw[->, line width=0.4mm] (g) to [bend left = 30] (y);  
    \draw[->, line width=0.4mm] (g) to (q);
    \draw[->, line width=0.4mm] (q) to (d);
    \draw[->, line width=0.4mm] (q) to (y);
\end{tikzpicture}
    \caption*{Figure A1. Illustration of effect paths in the path-specific decomposition.}
\end{figure}

\subsection*{A.9. Expression (8)}
\begin{align*}
    &\phantom{{}={}} \E_b \left(Y^{R(D \mid a, \Q)} \right)-\E_b(Y) \\
    &= \E_b \left(Y^0 \right) + \int \E_b(\tau \mid \q) \E_a(D \mid \q) f_b(\q) \dd \q - \E_b \left(Y^0 \right) - \E_b(D \tau) \\
    &= \int [\E_a(D \mid \q) - \E_b(D \mid \q) ] \E_b(\tau \mid \q) f_b(\q) \dd \q \\
    &\phantom{{}={}} + \E_b [\E_b(D \mid \Q)\E_b(\tau \mid \Q) ] - \E_b[\E_b(D \tau \mid \Q)] \\
    &= \int [\E_a(D \mid \q) - \E_b(D \mid \q) ] \E_b(\tau \mid \q) f_b(\q) \dd \q - \E_b[\Cov_b(D, \tau \mid \q)].
\end{align*}

\subsection*{A.10. Footnote 13}
In the randomized intervention notation, the change in disparity of the CRED of \citet{lundberg_gap-closing_2024} is
$$\E_a(Y)-\E_b(Y)-\left[\E_a \left(Y^{R(D \mid \Q)} \right)-\E_b \left(Y^{R(D \mid \Q)} \right) \right],$$ where $Y^{R(D \mid \Q)}$ is the potential outcome of an individual when they were given a random draw of $D$ from those in the pooled population who share the same $\Q$ values with them. Below, we start from the original form that appears in \citet{lundberg_gap-closing_2024} (see his equation [2]), which is equivalent to the form expressed in the randomized intervention notation. 

\begin{align*}
&\phantom{{}={}} \E_a(Y)-\E_b(Y) - \\ 
&\phantom{{}={}} \left\lbrace \E_a \left[\Pro(D=0 \mid \Q) Y^0 + \Pro(D=1 \mid \Q) Y^1 \right] - \E_b \left[\Pro(D=0 \mid \Q) Y^0 + \Pro(D=1 \mid \Q) Y^1 \right] \right\rbrace \\
&= \E_a(Y)-\E_b(Y) - \lbrace \E_a(Y^0) + \E_a[\E(D \mid \Q) \tau] - \E_b(Y^0) - \E_b[\E(D \mid \Q) \tau] \rbrace \\
&= \E_a(D\tau)-\E_a[\E(D \mid \Q) \E_a(\tau \mid \Q)] - \lbrace \E_b(D\tau)-\E_b[\E(D \mid \Q) \E_b(\tau \mid \Q)] \rbrace \\
&= \int [\E_a(D \tau \mid \q) - \E(D \mid \q) \E_a(\tau \mid \q)] f_a(\q) \dd \q - \int [\E_b(D \tau \mid \q) - \E(D \mid \q) \E_b(\tau \mid \q)] f_b(\q) \dd \q\\
&= \E_a[\Cov_a(D,\tau \mid \Q)] - \E_b[\Cov_b(D,\tau \mid \Q)] \\
&\phantom{{}={}} + \int [\E_a(D \mid \q) - \E(D \mid \q)] \E_a(\tau \mid \q) f_a(\q) \dd \q - \int [\E_b(D \mid \q) - \E(D \mid \q)] \E_b(\tau \mid \q) f_b(\q) \dd \q\\
&= \E_a[\Cov_a(D,\tau \mid \Q)] - \E_b[\Cov_b(D,\tau \mid \Q)] \\
&\phantom{{}={}} + \int \Pro(G=b \mid \q) [\E_a(D \mid \q) - \E_b(D \mid \q)] \E_a(\tau \mid \q) f_a(\q) \dd \q\\
&\phantom{{}={}} - \int \Pro(G=a \mid \q) [\E_b(D \mid \q) - \E_a(D \mid \q)] \E_b(\tau \mid \q) f_b(\q) \dd \q\\
&= \E_a[\Cov_a(D,\tau \mid \Q)] - \E_b[\Cov_b(D,\tau \mid \Q)] \\
&\phantom{{}={}} + \int [\E_a(D \mid \q) - \E_b(D \mid \q)][\E_a(\tau \mid \q)f_a(\q)\Pro(G=b \mid \q) + \E_b(\tau \mid \q)f_b(\q)\Pro(G=a \mid \q)] \dd \q.
\end{align*}

\subsection*{Appendix B. Graphical illustration of the unconditional decomposition}
    Using the randomized intervention notation, we present a graphical representation of our unconditional decomposition in Figure A2. The graph visualizes the distinct ways the four components contribute to the observed group disparity in outcome, i.e., one can vary the four components on the graph and obtain different outcome disparities. From this graph, it is clear that the selection component represents the contribution of differential effectiveness of treatment assignment across groups, where effectiveness is defined relative to random assignment of the treatment.
\citet{imai_experimental_2023} introduce a single-group version of Figure A2. 
But they do not note the covariance representation of the difference between an observed treatment assignment rule and the hypothetical random assignment with the same treatment prevalence.

\begin{figure}[ht]
\centering
\begin{tikzpicture}
    \node at (0,1) {};
    \node at (1,2) {};

    \draw[->, line width=0.3mm] (2,-3.5) to (10,-3.5);
    \draw[->, line width=0.3mm] (2,-3.5) to (2,2.5);
    \draw[-, line width=0.3mm] (10,-3.5) to (10,2.5);

    \draw[-, line width=0.3mm, blue, dotted] (2,-3) to (10,-1.5);
    \draw[-, line width=0.3mm, red, solid] (2,-2.3) to (10,2);

    \draw[-, line width=0.3mm, dashed] (5,-3.5) to (5,-1.5);
    \draw[-, line width=0.3mm, dashed] (7.5,-3.5) to (7.5,2.3);

    \draw[-, line width=0.3mm, dashed] (5,-1.5) to (2,-1.5);
    \draw[-, line width=0.3mm, dashed] (7.5,2.3) to (2,2.3);

    \draw[<->, line width=0.3mm, {Stealth[length=1mm,width=4mm]}-{Stealth[length=1mm,width=4mm]}] (5,-1.55) to (5,-2.45);
    \draw[<->, line width=0.3mm, {Stealth[length=1mm,width=4mm]}-{Stealth[length=1mm,width=4mm]}] (7.5,0.65) to (7.5,2.25);

    \node at (5,-3.8) {$\E_b(D)$};
    \node at (8,-3.8) {$\E_a(D)$};

    \node at (2,-3.8) {0};
    \node at (10,-3.8) {1};
    \node at (6,-4.3) {Treatment prevalence};
    \node at (0.4,-0.5) [rotate=90] {Mean outcome};

    \node at (1.35,-3) {$\E_b(Y_0)$};
    \node at (1.35,-2.3) {$\E_a(Y_0)$};

    \node at (10.7,-1.5) {$\E_b(Y_1)$};
    \node at (10.7,2) {$\E_a(Y_1)$};

    \node at (1.35,-1.5) {$\E_b(Y)$};
    \node at (1.35,2.3) {$\E_a(Y)$};

    \node at (4,-2) {$\Cov_b(D, \tau)$};
    \node at (6.5,1.5) {$\Cov_a(D, \tau)$};

    \draw (5,-1.5) [blue,fill=white] circle (2pt);
    \fill (7.5,2.3) [red] circle (2pt);
\end{tikzpicture}
\caption{Figure A2. Illustration of the unconditional decomposition. The heights of the  red (solid) dot and the blue (empty) dot respectively indicate the observed mean outcomes in group $a$ and $b$. The position of each dot on the $x$ axis corresponds to the observed treatment prevalence for the corresponding group. The red (solid) line and the blue (dotted) line respectively represent the mean potential outcomes in group $a$ and group $b$ under random assignment of the treatment, which vary by the prevalence of the assigned treatment. The lines are straight due to the hypothetical assignment being random. When nobody receives the treatment, the height of each line is $\E_g \left(Y^0 \right)$; and when everyone receives the treatment, the height of the line is $\E_g \left(Y^1 \right)$. The slope of each line is hence $\E_g(\tau)$. For a group $g$, the vertical distance between the dot and the line is $\E_g(Y)-\E_g \left(Y^{R(D \mid g)} \right)=\Cov_g(D, \tau)$.} \label{fig:Illustration}
\end{figure}

\subsection*{Appendix C. Efficient influence functions}
We use the Gateaux derivative approach to derive the EIFs \citep{ichimura_influence_2022}, which is more succinct than the approach traditionally used in the semiparametric causal inference literature \citep[e.g.,][]{hahn_role_1998}. To further simplify the derivation, we leverage some rules of calculating Gateaux derivatives in \citet{hines_demystifying_2022} and \citet{kennedy_semiparametric_2022}.

Let $\one_{\Tilde{o}}(o)$ be the point mass density at a single empirical observation,  $\Tilde{o}$. Let subscript $\mathcal{P}_t$ indicate a regular parametric submodel indexed by $t$. The subscript is omitted for the true model. By construction, $f_{\mathcal{P}_t}(o)=t \one_{\Tilde{o}}(o) + (1-t)f(o)$, i.e., the submodel is the true model perturbed in the direction of a single observation $\Tilde{o}$. Under this construction, the EIF of an estimand, $\xi$, is the Gateaux derivative at the truth, i.e., $\phi(\xi)=\frac{\partial \xi_{\mathcal{P}_t}}{\partial t} \big|_{t=0}$. For an arbitrary function $g(o)$, we denote $ \frac{\partial g_{\mathcal{P}_t}(o)}{\partial t} \big|_{t=0}$ as $\partial g(o)$. 

We derive the EIFs for the general case of  weighted estimands. Let $w(\X, G)$ be the  survey weight. Following \citet{hirano_efficient_2003}, we assume the survey weight is a known function of $\{\X, G \}$. When no survey weights are needed, $w(\X, G)$ reduces to 1 for every individual.

In this derivation, we also use the following definitions:
\begin{align*}
    h_g & \defeq \E(w(\X,  G) \mid g) \\
    h_g(\Q) & \defeq \E(w(\X,  G) \mid \Q, g).
\end{align*}

\subsubsection*{C.1. EIFs for the unconditional decomposition}
First, note that we only need to derive EIFs for two generic functions, $\xi_{dg} \defeq \E \left(Y^d \frac{ w(\X,g)}{h_g} \mid g \right)$ for an arbitrary group $g$; and $\xi_{dgg'} \defeq \E \left(Y^d \frac{ w(\X,g)}{h_g} \mid g \right)\E \left(D \frac{ w(\X,g')}{h_{g'}} \mid g' \right)$ for two arbitrary groups $g$ and $g'$, which may be the same group, and an arbitrary treatment status $d$. The EIFs for the decomposition components then follow from adding and subtracting these functions evaluated at appropriate $g$, $g'$, and $d$ values. Under conditional ignorability (Assumption 3), these estimands can be identified as the following functionals:
\begin{align*}
    \xi_{dg} &= \E \left[\mu(d,\X,g) \frac{ w(\X,g)}{h_g} \mid g \right] \\
    \xi_{dgg'} &= \E \left[\mu(d,\X,g) \frac{ w(\X,g)}{h_g} \mid g \right] \E \left[D \frac{ w(\X,g')}{h_{g'}}  \mid g' \right].
\end{align*}
We will also rely on the overlap assumption (Assumption 4) below, as $\pi(d,\X,g)$ will appear in the denominator.

We start with the EIF of $\xi_{dg}$.

\begin{align*}
    \phi(\xi_{dg}) &= \partial \E_{\mathcal{P}_t} \left[\mu_{\mathcal{P}_t}(d,\X,g) \frac{w(\X,g)}{\E_{\mathcal{P}_t}(w(\X,g) \mid g)}  \mid g \right] \\
    &= \frac{1}{h_g} \partial \E_{\mathcal{P}_t} \left[\mu_{\mathcal{P}_t}(d,\X,g) w(\X,g)  \mid g \right] +  \E[\mu(d,\X,g)w(\X,g) \mid g] \partial \frac{1}{\E_{\mathcal{P}_t}(w(\X,g)\mid g)} \\
    &= \frac{1}{h_g}\frac{\one_{\Tilde{g}}(g)}{p_g}  \{\mu(d,\Tilde{\x},g)w(\Tilde{\x},g) -\E[\mu(d,\X,g)w(\X,g) \mid g]\}  \\
    &\phantom{{}={}} + \frac{1}{h_g} \E[\partial  \mu_{\mathcal{P}_t}(d,\X,g)w(\X,g) \mid g ] - \frac{1}{(h_g)^2}\frac{\one_{\Tilde{g}}(g)}{p_g}[w(\Tilde{\x},g)-h_g]\E[\mu(d,\X,g)w(\X,g) \mid g] \\
    &= \frac{\one_{\Tilde{g}}(g)}{p_g}  \frac{w(\Tilde{\x},g)}{h_g} [\mu(d,\Tilde{\x},g)-\xi_{dg}] + \frac{1}{h_g} \E[\partial  \mu_{\mathcal{P}_t}(d,\X,g)w(\X,g) \mid g ] \\
    &= \frac{\one_{\Tilde{g}}(g)}{p_g}  \frac{w(\Tilde{\x},g)}{h_g} [\mu(d,\Tilde{\x},g)-\xi_{dg}] + \frac{1}{h_g} \E \left\{ \frac{\one_{\Tilde{d},\Tilde{\x},\Tilde{g}}(d,\X,g)}{f(d,\X,g)} [\Tilde{y}-\mu(d,\X,g)] w(\X,g) \mid g \right\} \\
        &= \frac{\one_{\Tilde{g}}(g)}{p_g}  \frac{w(\Tilde{\x},g)}{h_g} [\mu(d,\Tilde{\x},g)-\xi_{dg}] + \frac{\one_{\Tilde{g}}(g)}{p_g} \frac{w(\Tilde{\x},g)}{h_g} \frac{\one_{\Tilde{d}}(d)}{\pi(d, \Tilde{\x},g)} [\Tilde{y}-\mu(d,\Tilde{\x},g)] \\
    &= \frac{\one(G=g)}{p_g}\frac{w(\X,g)}{h_g}\left\{ \frac{\one(D=d)}{\pi(d, \X,g)} [Y-\mu(d,\X,g)] + \mu(d,\X,g) - \xi_{dg} \right\}.
\end{align*}

And without survey weights, $\phi(\xi_{dg})$ simplifies to
$$\frac{\one(G=g)}{p_g} \left\{ \frac{\one(D=d)}{\pi(d, \X,g)} [Y-\mu(d,\X,g)] + \mu(d,\X,g) - \xi_{dg} \right\}.$$

Now, for $\xi_{dgg'}$, 
\begin{equation*}
    \phi(\xi_{dgg'})=\phi(\xi_{dg})\E \left(D \frac{w(\X,g')}{h_{g'}} \mid g' \right) + \xi_{dg} \phi \left[ \E \left(D \frac{w(\X,g')}{h_{g'}} \mid g' \right) \right].
\end{equation*}
Since 
\begin{align*}
    &\phantom{{}={}} \phi \left[ \E \left(D \frac{w(\X,g')}{h_{g'}} \mid g' \right) \right] \\
    &= \partial \E_{\mathcal{P}_t} \left(D \frac{w(\X,g')}{\E_{\mathcal{P}_t}(w(\X,g')\mid g')} \mid g' \right) \\
    &= \frac{1}{h_{g'}} \frac{\one_{\Tilde{g}}(g')}{p_{g'}}\left[\Tilde{d} w(\Tilde{\x},g')-\E(Dw(\X,g')\mid g') \right] - \frac{1}{(h_{g'})^2}\frac{\one_{\Tilde{g}}(g')}{p_{g'}}[w(\Tilde{\x},g')-h_{g'}]\E(Dw(\X,g')\mid g') \\
    &= \frac{\one_{\Tilde{g}}(g')}{p_{g'}} \frac{w(\Tilde{\x},g')}{h_{g'}} \left[ \Tilde{d}-\E \left(D \frac{w(\X,g')}{h_{g'}} \mid g' \right) \right] \\
    &= \frac{\one(G=g')}{p_{g'}} \frac{w(\X, g')}{h_{g'}} \left[D - \E \left(D \frac{w(\X,g')}{h_{g'}} \mid g' \right) \right] ,
\end{align*}
we obtain the EIF for $\xi_{dgg'}$,
\begin{align*}
    \phi(\xi_{dgg'}) &= \frac{\one(G=g)}{p_g} \frac{w(\X,g)}{h_g} \left[\frac{\one(D=d)}{\pi(d,\X,g)}(Y-\mu(d,\X,g)) + \mu(d,\X,g) \right] \E \left(D \frac{w(\X,g')}{h_{g'}} \mid g' \right) \\
    &\phantom{{}={}} + \frac{\one(G=g')}{p_{g'}} \frac{w(\X, g')}{h_{g'}} \E \left[Y^d \frac{w(\X,g)}{h_g} \mid g \right]  \left[D - \E \left(D \frac{w(\X,g')}{h_{g'}} \mid g' \right) \right] \\
    &\phantom{{}={}} -  \frac{\one(G=g)}{p_g}\frac{w(\X,g)}{h_g} \xi_{dgg'} .
\end{align*}
Without survey weights, $\phi(\xi_{dgg'})$ simplifies to
\begin{align*}
    &\phantom{{}={}} \frac{\one(G=g)}{p_g}  \left[\frac{\one(D=d)}{\pi(d,\X,g)}(Y-\mu(d,\X,g)) + \mu(d,\X,g) \right] \E \left(D \mid g' \right) \\
    &\phantom{{}={}} + \frac{\one(G=g')}{p_{g'}} \E \left[Y^d \mid g \right] \left[D - \E \left(D \mid g' \right) \right] - \frac{\one(G=g)}{p_g} \xi_{dgg'}.
\end{align*}

In the main text of the paper, we write $\phi(\xi_{dg})$ and  $\phi(\xi_{dgg'})$ as $\phi_{dg}(Y,D,\X,G)$ and $\phi_{dgg'}(Y,D,\X,G)$, respectively, to highlight that they are functions of observed variables.

Also note that the EIF for the total disparity, $\E \left(Y \frac{w(\X,a)}{h_{a}} \mid a \right) - \E \left(Y \frac{w(\X,b)}{h_{b}} \mid b \right)$, is 
\begin{align*}
    &\phantom{{}={}} \phi(\text{Total}) \\
    &= \frac{\one(G=a)}{p_a}\frac{w(\X,a)}{h_a} \left[Y - \E\left(Y \frac{w(\X,a)}{h_a} \mid a \right)\right] - \frac{\one(G=b)}{p_b}\frac{w(\X,b)}{h_b}  \left[Y - \E \left(Y \frac{w(\X,b)}{h_b} \mid b \right)\right],
\end{align*}
which, without survey weights, becomes
\begin{equation*}
    \frac{\one(G=a)}{p_a} \left[Y - \E\left(Y \mid a \right)\right] - \frac{\one(G=b)}{p_b} \left[Y - \E \left(Y \mid b \right)\right].
\end{equation*}

Finally, the EIFs for the unconditional decomposition components are 

\begin{align*}
    \phi(\text{Baseline}) &= \phi(\xi_{0a})-\phi(\xi_{0b})  \\
    \phi(\text{Prevalence}) &= \phi(\xi_{1ba})-\phi(\xi_{1bb})-\phi(\xi_{0ba})+\phi(\xi_{0bb}) \\
    \phi(\text{Effect}) &= \phi(\xi_{1aa})-\phi(\xi_{0aa}) - \phi(\xi_{1ba})+\phi(\xi_{0ba}) \\
    \phi(\text{Selection}) &= \phi(\text{Total}) - \phi(\text{Baseline}) - \phi(\text{Prevalence}) - \phi(\text{Effect}) .
\end{align*}

\subsubsection*{C.2. EIFs for the conditional decomposition}
Similar to the unconditional case, we focus on the generic function
$$\xi_{dgg'g''} \defeq \E \left[\E \left(Y^d \frac{w(\X,g)}{h_{g}(\Q)}  \mid \Q, g \right) \E \left(D \frac{w(\X,g')}{h_{g'}(\Q)} \mid \Q, g' \right) \frac{w(\X,g'')}{h_{g''}}  \mid g'' \right],$$
where $(g,g',g'')$ is an arbitrary combination of group memberships out of the 8 possible combinations. We maintain the conditional ignorability and overlap assumptions.  
\begin{align*}
    &\phantom{{}={}} \phi(\xi_{dgg'g''}) \\
    &= \partial \E_{\mathcal{P}_t} \left[\E_{\mathcal{P}_t} \left(Y^d \frac{w(\X,g)}{\E_{\mathcal{P}_t}(w(\X,g) \mid \Q,g)}  \mid \Q, g \right) \E_{\mathcal{P}_t} \left(D \frac{w(\X,g')}{\E_{\mathcal{P}_t}(w(\X,g') \mid \Q,g')} \mid \Q, g' \right) \right. \\ 
    &\phantom{{}={}} \left. \vphantom{\pheight}
    w(\X,g'') \mid g'' \right] \frac{1}{\E_{\mathcal{P}_t}(w(\X,g) \mid g'')} \\
    &= \partial \E_{\mathcal{P}_t} \left[\E_{\mathcal{P}_t} \left(Y^d \frac{w(\X,g)}{\E_{\mathcal{P}_t}(w(\X,g) \mid \Q,g)}  \mid \Q, g \right) \E_{\mathcal{P}_t} \left(D \frac{w(\X,g')}{\E_{\mathcal{P}_t}(w(\X,g') \mid \Q,g')} \mid \Q, g' \right) w(\X,g'') \mid g'' \right] \frac{1}{h_{g''}} \\
    &\phantom{{}={}} + \E \left[\E \left(Y^d \frac{w(\X,g)}{\E(w(\X,g) \mid \Q,g)}  \mid \Q, g \right) \E \left(D \frac{w(\X,g')}{\E(w(\X,g') \mid \Q,g')} \mid \Q, g' \right) 
    w(\X,g'') \mid g'' \right] \\
    &\phantom{{}={}} \partial \frac{1}{\E_{\mathcal{P}_t}(w(\X,g) \mid g'')} \\
    &=\frac{\one_{\Tilde{g}(g'')}}{p_{g''}}\frac{w(\X,g'')}{h_{g''}} \E \left(Y^d \frac{w(\X,g)}{h_{g}(\Q)}  \mid \Q, g \right) \E \left(D \frac{w(\X,g')}{h_{g'}(\Q)} \mid \Q, g' \right) - \frac{\one_{\Tilde{g}}(g'')}{p_{g''}}\xi_{dgg'g''} \\
    &\phantom{{}={}} + \E \left[\partial \E_{\mathcal{P}_t} \left(Y^d \frac{w(\X,g)}{\E_{\mathcal{P}_t}(w(\X,g) \mid \Q,g)}  \mid \Q, g \right) \E_{\mathcal{P}_t} \left(D \frac{w(\X,g')}{\E_{\mathcal{P}_t}(w(\X,g') \mid \Q,g')} \mid \Q, g' \right) w(\X,g'') \mid g'' \right] \frac{1}{h_{g''}} \\
    &\phantom{{}={}} - \xi_{dgg'g''}h_{g''}\frac{1}{(h_{g''})^2}\frac{\one_{\Tilde{g}(g'')}}{p_{g''}}\left[w(\X,g'')-h_{g''}\right] \\
    &= \frac{\one(G=g'')}{p_{g''}}\frac{w(\X,g'')}{h_{g''}} \left[ \E \left(Y^d \frac{w(\X,g)}{h_{g}(\Q)}  \mid \Q, g \right) \E \left(D \frac{w(\X,g')}{h_{g'}(\Q)} \mid \Q, g' \right) - \xi_{dgg'g''} \right] \\
    &\phantom{{}={}} + \E \left[\partial \E_{\mathcal{P}_t} \left(Y^d \frac{w(\X,g)}{\E_{\mathcal{P}_t}(w(\X,g) \mid \Q,g)}  \mid \Q, g \right) \E_{\mathcal{P}_t} \left(D \frac{w(\X,g')}{\E_{\mathcal{P}_t}(w(\X,g') \mid \Q,g')} \mid \Q, g' \right) w(\X,g'') \mid g'' \right] \frac{1}{h_{g''}}.
\end{align*}

And
\begin{align*}
    &\phantom{{}={}} \E \left[\partial \E_{\mathcal{P}_t} \left(Y^d \frac{w(\X,g)}{\E_{\mathcal{P}_t}(w(\X,g) \mid \Q,g)}  \mid \Q, g \right) \E_{\mathcal{P}_t} \left(D \frac{w(\X,g')}{\E_{\mathcal{P}_t}(w(\X,g') \mid \Q,g')} \mid \Q, g' \right) 
    w(\X,g'') \mid g'' \right] \frac{1}{h_{g''}} \\
    &= \E \left[\partial \E_{\mathcal{P}_t} \left(Y^d \frac{w(\X,g)}{\E_{\mathcal{P}_t}(w(\X,g) \mid \Q,g)}  \mid \Q, g \right) \E \left(D \frac{w(\X,g')}{h_{g'}(\Q)} \mid \Q, g' \right) w(\X,g'') \mid g'' \right]\frac{1}{h_{g''}} \\
    &\phantom{{}={}} + \E \left[\E \left(Y^d \frac{w(\X,g)}{h_g(\Q)}  \mid \Q, g \right) \partial  \E_{\mathcal{P}_t} \left(D \frac{w(\X,g')}{\E_{\mathcal{P}_t}(\X,g')} \mid \Q, g' \right) w(\X,g'') \mid g'' \right]\frac{1}{h_{g''}} \\
    &= \frac{\one(G=g) p_{g''}(\Q)}{p_g(\Q)p_{g''}} \frac{w(\X,g)}{h_g(\Q)} \frac{w(\X,g'')}{h_{g''}} \left\{ \frac{\one(D=d)}{\pi(d,\X,g)} [Y-\mu(d,\X,g)]+\mu(d,\X,g) \right. \\
    &\phantom{{}={}} \left. - \E \left(Y^d \frac{w(X,Q,g)}{h_g(\Q)} \mid \Q,g \right) \right\} \E \left(D \frac{w(\X,Q,g')}{h_{g'}(\Q)} \mid \Q,g' \right) \\
    &\phantom{{}={}} + \frac{\one(G=g')p_{g''}(\Q)}{p_{g'}(\Q)p_{g''}} \frac{w(\X,g')}{h_{g'}(\Q)} \frac{w(\X,g'')}{h_{g''}} \left[ D-\E \left(D \frac{w(\X,g')}{h_{g'}(\Q)} \mid \Q, g' \right) \right]\E\left( Y^d \frac{w(\X,g)}{h_g(\Q)} \mid \Q, g \right).
\end{align*}

Hence,
\begin{align*}
    &\phantom{{}={}} \phi(\xi_{dgg'g''}) \\
    &= \frac{\one(G=g'')}{p_{g''}}\frac{w(\X,g'')}{h_{g''}} \left[ \E \left(Y^d \frac{w(\X,g)}{h_{g}(\Q)}  \mid \Q, g \right) \E \left(D \frac{w(\X,g')}{h_{g'}(\Q)} \mid \Q, g' \right) - \xi_{dgg'g''} \right] \\
    &\phantom{{}={}} + \frac{w(\X,g'')}{h_{g''}}  \frac{\one(G=g) p_{g''}(\Q)}{p_g(\Q)p_{g''}} \frac{w(\X,g)}{h_g(\Q)} \left\{ \frac{\one(D=d)}{\pi(d,\X,g)} [Y-\mu(d,\X,g)]+\mu(d,\X,g) \right. \\
    &\phantom{{}={}} \left. - \E \left(Y^d \frac{w(\X,g)}{h_g(\Q)} \mid \Q,g \right) \right\} \E \left(D \frac{w(\X,g')}{h_{g'}(\Q)} \mid \Q,g' \right) \\
    &\phantom{{}={}} + \frac{w(\X,g'')}{h_{g''}}  \frac{\one(G=g')p_{g''}(\Q)}{p_{g'}(\Q)p_{g''}} \frac{w(\X,g')}{h_{g'}(\Q)} \left[ D-\E \left(D \frac{w(\X,g')}{h_{g'}(\Q)} \mid \Q, g' \right) \right] \E\left( Y^d \frac{w(\X,g)}{h_g(\Q)} \mid \Q, g \right), 
\end{align*}
which, in the absence of survey weights, simplifies to 
\begin{align*}
    &\phantom{{}={}} \frac{\one(G=g'')}{p_{g''}} \left[ \E \left(Y^d \mid \Q, g \right) \E \left(D \mid \Q, g' \right) - \xi_{dgg'g''} \right] \\
    &\phantom{{}={}} + \frac{\one(G=g) p_{g''}(\Q)}{p_g(\Q)p_{g''}} \left\{ \frac{\one(D=d)}{\pi(d,\X,g)} [Y-\mu(d,\X,g)]+\mu(d,\X,g) - \E \left(Y^d \mid \Q,g \right) \right\} \E \left(D \mid \Q,g' \right) \\
    &\phantom{{}={}} + \frac{\one(G=g')p_{g''}(\Q)}{p_{g'}(\Q)p_{g''}} \left[ D-\E \left(D \mid \Q, g' \right) \right] \E\left( Y^d \mid \Q, g \right).
\end{align*}
Recall that $\E\left( Y^d \mid \Q, g \right)$ is identified as $\omega(d, \Q, g) \defeq \E[\mu(d,\X,g) \mid \Q,g] $. Finally, in the main text of the paper, we write $\phi(\xi_{dgg'g''})$ as $\phi_{dgg'g''}(Y,D,\X,G)$ to highlight that it is a function of observed variables.

\subsection*{Appendix D. Asymptotic distributions}

We follow the procedure of using the von Mises expansion to prove asymptotic properties of cross-fitted EIF-based one-step estimators \citep{hines_demystifying_2022, kennedy_semiparametric_2022, fisher_visually_2021}. In order for the cross-fitted one-step estimator to be  $\sqrt{n}-$consistent, asymptotically normal, and semiparametrically efficient, we just need two conditions to hold. That is, both the empirical process term and the ``remainder term" in the von Mises expansion are $o_p(n^{-1/2})$.

We use the notation $\P(f(O)) \defeq \int f(O)d\P(O)$, and $\P_n$ denotes the corresponding sample average. Also let $\| \cdot \|$ denote the $L_2$-norm, such that $\| f(O) \|^2=\P(f(O)^2)$. And $\hat{\xi}$ is defined to be a substitution estimator for $\xi$.
Formally, for all $d,g$ and $\xi$, we need $(\P-\P_n) \left[ \hat{\phi}(Y,d,\X,g)-\phi(Y,d,\X,g) \right]=o_p(n^{-1/2})$, 
and $ \hat{\xi} + \P[\hat{\phi}(Y,d,\X,g)] - \xi =o_p(n^{-1/2})$. In this appendix, we prove that the assumptions specified in the main text are sufficient for the unconditional and conditional decompositions to attain this convergence result. By cross-fitting, all of $\hat{\mu}(d,\X,g), \hat{\pi}(d,\X,g), \hat{p}_g(\Q)$, $\hat{\omega}(d,\Q,g)$, and $\hat{\E}(D \mid \Q,g)$ are fitted using data not in the current subsample, which we implicitly condition on throughout.

\subsubsection*{D.1. Inference for the unconditional decomposition}

First, for $\xi_{dg}$, the remainder term is 
\begin{align*}
    R_{2,dg} &= \hat{\xi}_{dg} + \P\left\{ \frac{\one(G=g)}{\hat{p}_g} \left[ \frac{\one(D=d)}{\hat{\pi}(d, \X,g)} (Y-\hat{\mu}(d,\X,g)) + \hat{\mu}(d,\X,g) - \hat{\xi}_{dg} \right] \right\} - \xi_{dg} 
    \\
    &= \hat{\xi}_{dg} + \P\left\{ \frac{\one(G=g)}{\hat{p}_g} \left[ \frac{\one(D=d)}{\hat{\pi}(d, \X,g)} (Y-\hat{\mu}(d,\X,g)) + \hat{\mu}(d,\X,g) - \mu(d,\X,g) \right] \right\}
    \\
    &\phantom{{}={}} + \P\left[\frac{\one(G=g)}{\hat{p}_g} \mu(d,\X,G)\right] - \P\left[\frac{\one(G=g)}{\hat{p}_g} \hat{\xi}_{dg} \right] - \xi_{dg} 
    \\
    &= \hat{\xi}_{dg} +  \P\left\{ \frac{\one(G=g)}{\hat{p}_g}  \left[1- \frac{\pi(d,\X,g)}{\hat{\pi}(d, \X,g)}\right] \left(\hat{\mu}(d,\X,g) - \mu(d,\X,g)\right) \right\} + \frac{p_g}{\hat{p}_g}\xi_{dg} - \frac{p_g}{\hat{p}_g}\hat{\xi}_{dg} - \xi_{dg} \\
    &= \left(1- \frac{p_g}{\hat{p}_g}\right) \left(\hat{\xi}_{dg} - \xi_{dg}\right) + \P\left\{ \frac{\one(G=g)}{\hat{p}_g}  \left[1- \frac{\pi(d,\X,g)}{\hat{\pi}(d, \X,g)}\right] \left[\hat{\mu}(d,\X,g) - \mu(d,\X,g)\right] \right\}.
\end{align*}
The first term is a lower order term \citep[see the proof of Theorem 1 in][]{vansteelandt_assumption-lean_2022}. For the second term,

\begin{align*}
    &\phantom{{}={}} \left|\P\left\{ \frac{\one(G=g)}{\hat{p}_g}  \left[1- \frac{\pi(d,\X,g)}{\hat{\pi}(d, \X,g)}\right] \left[\hat{\mu}(d,\X,g) - \mu(d,\X,g)\right] \right\} \right| \\
    & \leq \frac{1}{\eta \hat{p}_g} | \P \left\{ \left[ \hat{\pi}(d, \X,g)- \pi(d,\X,g)\right]   \left[\hat{\mu}(d,\X,g) - \mu(d,\X,g) \right] \right\} | \\
    & \leq \frac{1}{\eta \hat{p}_g}  \| \hat{\pi}(d, \X,g)- \pi(d,\X,g) \| \|\hat{\mu}(d,\X,g) - \mu(d,\X,g) \| \\
    &= o_p(n^{-1/2}),
\end{align*}
where the second inequality uses the Cauchy–Schwarz inequality. 

For the empirical process term,
\begin{align*}
    &\phantom{{}={}} \hat{\phi}_{dg}(Y,\X)- \phi_{dg}(Y,\X) \\
    &= \frac{\one(G=g)}{\hat{p}_{g}} \left[ 1-\frac{\one(D=d)}{\pi(d,\X,g)} \right] \left[\hat{\mu}(d,\X,g)-\mu(d,\X,g)\right]  \\
    &\phantom{{}={}} + \frac{\one(G=g)}{\hat{p}_{g}} \frac{\one(D=d)[Y-\hat{\mu}(d,\X,g)]}{\hat{\pi}(d,\X,g)\pi(d,\X,g)}\left[\pi(d,\X,g)-\hat{\pi}(d,\X,g)\right] \\
    &\phantom{{}={}} + \one(G=g) \frac{p_g-\hat{p}_g}{\hat{p}_g p_g} \left\{ \frac{\one(D=d)}{\pi(d, \X,g)} [Y-\mu(d,\X,g)] + \mu(d,\X,g) \right\} \\
    &\phantom{{}={}} + \frac{\one(G=g)}{\hat{p}_{g}} \left(\xi_{dg}-\hat{\xi}_{dg} \right) + \one(G=g) \frac{\hat{p}_g-p_g}{\hat{p}_g p_g}\xi_{dg}.
\end{align*}
Note that $(\P_n-\P) \left[ \frac{\one(G=g)}{\hat{p}_{g}} \left(\xi_{dg}-\hat{\xi}_{dg} \right) \right]$ is a lower order term. Then, using the Chebyshev's inequality argument commonly used in the double ML literature \citep{chernozhukov_doubledebiasedneyman_2017}, the empirical process term is $o_p(n^{-1/2})$ under stated conditions. 

Second, for $\xi_{dgg'}$,
\begin{align*}
    R_{2,dgg'} &= \hat{\xi}_{dgg'}+\P \left\{\frac{\one(G=g)}{\hat{p}_g}  \left[\frac{\one(D=d)}{\hat{\pi}(d,\X,g)}(Y-\hat{\mu}(d,\X,g)) + \hat{\mu}(d,\X,g) \right] \hat{\E} \left(D \mid g' \right) \right. \\
    &\phantom{{}={}} + \left. \frac{\one(G=g')}{\hat{p}_{g'}} \hat{\E} \left(Y^d \mid g \right) \left[D - \hat{\E} \left(D \mid g' \right) \right] - \frac{\one(G=g)}{\hat{p}_g} \hat{\xi}_{dgg'} \right\} - \xi_{dgg'} \\
    &= \P \left\{ \frac{\one(G=g)}{\hat{p}_g} \left[1-\frac{\pi(d,\X,g)}{\hat{\pi}(d,\X,g)} \right]\left[\hat{\mu}(d,\X,g)-\mu(d,\X,g)\right] \right\} \hat{\E}(D\mid g') \\
    &\phantom{{}={}} + \left[\frac{p_g}{\hat{p}_g} \E(Y^d \mid g) - \frac{p_{g'}}{\hat{p}_{g'}}\hat{\E}(Y^d \mid g) \right] \left[\hat{\E}(D \mid g') - \E(D\mid g') \right] + \left(1-\frac{p_g}{\hat{p}_g}\right) \left(\hat{\xi}_{dgg'}-\xi_{dgg'} \right),
\end{align*}
where, under stated conditions, the first term is $o_p(n^{-1/2})$, the second term is $o_p(1)O_p(n^{-1/2})=o_p(n^{-1/2})$, and the last term is again a lower order term. 

Also,
\begin{align*}
    &\phantom{{}={}} \hat{\phi}_{dgg'}(Y,\X)- \phi_{dgg'}(Y,\X) \\
    &= \frac{\one(G=g)}{\hat{p}_{g}} \hat{\E}(D \mid g')  \left[ 1-\frac{\one(D=d)}{\pi(d,\X,g)} \right] \left[\hat{\mu}(d,\X,g)-\mu(d,\X,g)\right]  \\
    &\phantom{{}={}} + \frac{\one(G=g)}{\hat{p}_{g}} \hat{\E}(D \mid g')  \frac{\one(D=d)[Y-\hat{\mu}(d,\X,g)]}{\hat{\pi}(d,\X,g)\pi(d,\X,g)}\left[\pi(d,\X,g)-\hat{\pi}(d,\X,g)\right] \\
    &\phantom{{}={}} + \frac{\one(G=g)}{\hat{p}_{g}} \left[\hat{\E}(D \mid g')-\E(D \mid g')\right] \left\{ \frac{\one(D=d)}{\pi(d,\X,g)} [Y-\mu(d,\X,g)] + \mu(d,\X,g) \right\} \\
    &\phantom{{}={}} + \one(G=g)\frac{p_g-\hat{p}_g}{\hat{p}_g p_g} \E(D \mid g') \left\{ \frac{\one(D=d)}{\pi(d,\X,g)} [Y-\mu(d,\X,g)] + \mu(d,\X,g) \right\} \\
    &\phantom{{}={}} + \frac{\one(G=g')}{\hat{p}_{g'}} \left[\hat{\E}(Y^d \mid g)-\E(Y^d \mid g)\right]D + \one(G=g') \frac{p_g-\hat{p}_g}{\hat{p}_g p_g} \E(Y^d \mid g)D \\
    &\phantom{{}={}} + \frac{\one(G=g')}{p_{g'}} \E(Y^d \mid g) \left[\hat{\E}(D \mid g')-\E(D \mid g') \right] + \frac{\one(G=g')}{\hat{p}_{g'}} \left[\hat{\E}(Y^d \mid g)-\E(Y^d \mid g)\right] \hat{\E}(D \mid g') \\
    &\phantom{{}={}} + \one(G=g')\frac{p_{g'}-\hat{p}_{g'}}{\hat{p}_{g'}p_{g'}} \E(Y^d \mid g)\hat{\E}(D \mid g') + \one(G=g)\frac{\hat{p}_{g}-p_{g}}{\hat{p}_{g}p_{g}}\xi_{dgg'} + \frac{\one(G=g)}{\hat{p}_{g}}(\xi_{dgg'}-\hat{\xi}_{dgg'}).
\end{align*}
Thus, the empirical process term, $(\P_n-\P)\left[\hat{\phi}_{dgg'}(Y,\X)- \phi_{dgg'}(Y,\X) \right]$, is $o_p(n^{-1/2})$ under stated conditions. Note that 
$$\hat{\E}(Y^d \mid g)=\P_n\left\{ \frac{\one(D=d)}{\hat{\pi}(d,\X,g)} [Y-\hat{\mu}(d,\X,g)] + \hat{\mu}(d,\X,g) \right\},$$
hence consistent estimation of $\pi(d,\X,g)$ and $\mu(d,\X,g)$ ensures that $\hat{\E}(Y^d \mid g)-\E(Y^d \mid g)=o_p(1)$.

\subsubsection*{D.2. Inference for the conditional decomposition}

For the components of our conditional decomposition, either $g=g''$ or $g'=g''$. In what follows, we first show that the empirical process term, $(\P_n-\P) \left[ \hat{\phi}_{dgg'g''}(Y,\X)- \phi_{dgg'g''}(Y,\X) \right]$, is $o_p(n^{-1/2})$. Then, we show that the remainder term is also $o_p(n^{-1/2})$ in both cases relevant to us, i.e., when $g=g''$ and when $g'=g''$.

For the empirical process term, 
\begin{align*}
    &\phantom{{}={}} \hat{\phi}_{dgg'g''}(Y,\X)-\phi_{dgg'g''}(Y,\X) \\
    &= \frac{\one(G=g'')}{\hat{p}_{g''}}\left(\xi_{dgg'g''}-\hat{\xi}_{dgg'g''}\right) + \one(G=g'')\frac{\hat{p}_{g''}-p_{g''}}{\hat{p}_{g''}p_{g''}} \xi_{dgg'g''} \\
    &\phantom{{}={}} + \one(G=g'')\frac{p_{g''}-\hat{p}_{g''}}{p_{g''}\hat{p}_{g''}}\omega(d,\Q,g)\E(D \mid \Q,g') \\
    &\phantom{{}={}} + \frac{\one(G=g'')}{\hat{p}_{g''}} \omega(d,\Q,g)\left[ \hat{\E}(D \mid \Q,g')-\E(D \mid \Q,g')\right] \\
    &\phantom{{}={}} + \frac{\one(G=g'')}{\hat{p}_{g''}} \hat{\E}(D \mid \Q,g')\left[\hat{\omega}(d,\Q,g)-\omega(d,\Q,g) \right] \\
    &\phantom{{}={}} + \frac{\one(G=g) \hat{p}_{g''}(\Q)}{\hat{p}_{g}(\Q) \hat{p}_{g''}} \omega(d,\Q,g) \left[ \E(D \mid \Q,g')-\hat{\E}(D \mid \Q,g') \right] \\
    &\phantom{{}={}} + \frac{\one(G=g) \hat{p}_{g''}(\Q)}{\hat{p}_{g}(\Q) \hat{p}_{g''}} \hat{\E}(D \mid \Q,g')\left[\omega(d,\Q,g)-\hat{\omega}(d,\Q,g) \right] \\
    &\phantom{{}={}} + \one(G=g)\frac{p_{g''}(\Q)}{p_g(\Q)}\frac{\hat{p}_{g''}-p_{g''}}{\hat{p}_{g''}p_{g''}} \omega(d,\Q,g)\E(D \mid \Q,g') \\
    &\phantom{{}={}} + \frac{\one(G=g)}{\hat{p}_{g''}}\frac{1}{\hat{p}_g(\Q)}\left[p_{g''}(\Q)-\hat{p}_{g''}(\Q) \right] \omega(d,\Q,g)\E(D \mid \Q,g') \\
    &\phantom{{}={}} + \frac{\one(G=g)}{\hat{p}_{g''}} \frac{\hat{p}_g(\Q)-p_g(\Q)}{\hat{p}_g(\Q)p_g(\Q)} p_{g''}(\Q)\omega(d,\Q,g)\E(D \mid \Q,g') \\
    &\phantom{{}={}} + \frac{\one(G=g)}{\hat{p}_{g''}}\frac{\hat{p}_{g''}(\Q)}{\hat{p}_{g}(\Q)} \left[ 1-\frac{\one(D=d)}{\pi(d,\X,g)} \right] \left[\hat{\mu}(d,\X,g)-\mu(d,\X,g)\right] \\
    &\phantom{{}={}} + \frac{\one(G=g)}{\hat{p}_{g''}}\frac{\hat{p}_{g''}(\Q)}{\hat{p}_{g}(\Q)} \frac{\one(D=d)[Y-\hat{\mu}(d,\X,g)]}{\hat{\pi}(d,\X,g)\pi(d,\X,g)}\left[\pi(d,\X,g)-\hat{\pi}(d,\X,g)\right] \\
    &\phantom{{}={}} + \one(G=g)\frac{p_{g''}-\hat{p}_{g''}}{p_{g''}\hat{p}_{g''}} \frac{p_{g''}(\Q)}{p_g(\Q)} \left\{ \frac{\one(D=d)}{\pi(d,\X,g)} [Y-\mu(d,\X,g)] + \mu(d,\X,g) \right\} \\
    &\phantom{{}={}} + \frac{\one(G=g)}{\hat{p}_{g''}} \frac{p_g(\Q)-\hat{p}_g(\Q)}{p_g(\Q)\hat{p}_g(\Q)}p_{g''}(\Q) \left\{ \frac{\one(D=d)}{\pi(d,\X,g)} [Y-\mu(d,\X,g)] + \mu(d,\X,g) \right\} \\
    &\phantom{{}={}} + \frac{\one(G=g')\hat{p}_{g''}(\Q)}{\hat{p}_{g'}(\Q)\hat{p}_{g''}} D \left[ \hat{\omega}(d,\Q,g)-\omega(d,\Q,g) \right] \\
    &\phantom{{}={}} + \one(G=g')\frac{p_{g''}-\hat{p}_{g''}}{p_{g''}\hat{p}_{g''}} \frac{p_{g''}(\Q)}{p_{g'}(\Q)}D\omega(d,\Q,g) \\
    &\phantom{{}={}} + \frac{\one(G=g')}{\hat{p}_{g''}} \frac{1}{\hat{p}_{g'}(\Q)} \left[ \hat{p}_{g''}(\Q)-p_{g''}(\Q) \right] D \omega(d,\Q,g) \\
    &\phantom{{}={}} + \frac{\one(G=g')}{\hat{p}_{g''}} \frac{p_{g'}(\Q)-\hat{p}_{g'}(\Q)}{p_{g'}(\Q)\hat{p}_{g'}(\Q)} p_{g''}(\Q) D \omega(d,\Q,g) \\
    &\phantom{{}={}} + \frac{\one(G=g')\hat{p}_{g''}(\Q)}{\hat{p}_{g'}(\Q) \hat{p}_{g''}} \omega(d,\Q,g) \left[ \E(D \mid \Q,g')-\hat{\E}(D \mid \Q,g') \right] \\
    &\phantom{{}={}} + \frac{\one(G=g')\hat{p}_{g''}(\Q)}{\hat{p}_{g'}(\Q) \hat{p}_{g''}} \hat{\E}(D \mid \Q,g')\left[ \omega(d,\Q,g)-\hat{\omega}(d,\Q,g) \right] \\
    &\phantom{{}={}} + \one(G=g')\frac{\hat{p}_{g''}-p_{g''}}{\hat{p}_{g''}p_{g''}} \frac{p_{g''}(\Q)}{p_{g'}(\Q)} \E(D \mid \Q,g') \omega(d,\Q,g) \\
    &\phantom{{}={}} + \frac{\one(G=g')}{\hat{p}_{g''}} \frac{1}{\hat{p}_{g'}(\Q)} \left[ p_{g''}(\Q)-\hat{p}_{g''}(\Q) \right] \E(D \mid \Q,g')\omega(d,\Q,g) \\
    &\phantom{{}={}} + \frac{\one(G=g')}{\hat{p}_{g''}} \frac{\hat{p}_{g'}(\Q)-p_{g'}(\Q)}{\hat{p}_{g'}(\Q)p_{g'}(\Q)} p_{g''}(\Q) \E(D \mid \Q,g')\omega(d,\Q,g).
\end{align*}
Using arguments similar to above, we can show that the empirical process term is indeed asymptotically negligible under stated conditions. Next, we turn to the remainder term.

\subsubsection*{D.2.1 When $g=g''$}
Note that for the conditional prevalence component, all $\xi$ terms satisfy $g=g''$.
\begin{align*}
&\phantom{{}={}} R_{2,dgg'g''} \\
&= \hat{\xi}_{dgg'g} - \xi_{dgg'g} \\
    &\phantom{{}={}} + \P \left\{- \frac{\one(G=g)}{\hat{p}_{g}}  \hat{\xi}_{dgg'g} \right\} \\
    &\phantom{{}={}} + \P \left\{ \frac{\one(G=g) }{\hat{p}_{g}} \left\{ \frac{\one(D=d)}{\hat{\pi}(d,\X,g)} [Y-\hat{\mu}(d,\X,g)]+\hat{\mu}(d,\X,g) \right\} \hat{\E} \left(D \mid \Q,g' \right) \right\} \\
    &\phantom{{}={}} + \P \left\{ \frac{\one(G=g')\hat{p}_{g}(\Q)}{\hat{p}_{g'}(\Q)\hat{p}_{g}} \left[ D-\hat{\E} \left(D \mid \Q, g' \right) \right] \hat{\omega}(d,\Q,g) \right\} \\
    &=  \left(1-\frac{p_g}{\hat{p}_g}\right)\hat{\xi}_{dgg'g} - \xi_{dgg'g} \\
    &\phantom{{}={}} + \P \left\{ \frac{\one(G=g) }{\hat{p}_{g}}  \left[ 1-\frac{\pi(d,\X,g)}{\hat{\pi}(d,\X,g)} \right]  [\hat{\mu}(d,\X,g)-\mu(d,\X,g)]  \hat{\E} \left(D \mid \Q,g' \right) \right\} \\
    &\phantom{{}={}}  + \P \left\{ \frac{\one(G=g) }{\hat{p}_{g}} \omega(d,\Q,g) \hat{\E}(D \mid \Q, g') \right\}\\
    &\phantom{{}={}} + \P \left\{ \frac{\one(G=g')\hat{p}_{g}(\Q)}{\hat{p}_{g'}(\Q)\hat{p}_{g}} \left[ \E(D \mid \Q, g')-\hat{\E} \left(D \mid \Q, g' \right) \right] \hat{\omega}(d,\Q,g) \right\} \\
    &= \left(1-\frac{p_g}{\hat{p}_g}\right)\left(\hat{\xi}_{dgg'g} - \xi_{dgg'g} \right) \\
    &\phantom{{}={}} + \P \left\{ \frac{\one(G=g) }{\hat{p}_{g}} \left[ 1-\frac{\pi(d,\X,g)}{\hat{\pi}(d,\X,g)} \right]  [\hat{\mu}(d,\X,g)-\mu(d,\X,g)]  \hat{\E} \left(D \mid \Q,g' \right) \right\} \\
    &\phantom{{}={}} + \P \left\{ \left[ \frac{\one(G=g) }{\hat{p}_{g}} \omega(d,\Q,g) - \frac{\one(G=g')\hat{p}_{g}(\Q)}{\hat{p}_{g'}(\Q)\hat{p}_{g}} \hat{\omega}(d,\Q,g) \right] \left[ \hat{\E}(D \mid \Q, g') - \E(D \mid \Q, g') \right] \right\}.
\end{align*}
Then it follows from similar arguments as above that $R_{2,dgg'g''}=o_p(n^{-1/2})$ under stated conditions. In particular, for the last line, note that if $g=g'=g''$,
\begin{align*}
    &\phantom{{}={}} \P\left\{ \left[ \one(G=g) \omega(d,\Q,g) - \one(G=g') \frac{\hat{p}_{g}(\Q)}{\hat{p}_{g'}(\Q)} \hat{\omega}(d,\Q,g) \right] \left[ \hat{\E}(D \mid \Q, g') - \E(D \mid \Q, g') \right] \right\} \\
    &=  \P\left\{ \one(G=g) \left[ \omega(d,\Q,g) - \hat{\omega}(d,\Q,g) \right] \left[ \hat{\E}(D \mid \Q, g') - \E(D \mid \Q, g') \right] \right\},
\end{align*}
so $\left\| \omega(d,\Q,g) - \hat{\omega}(d,\Q,g)  \right\| \left\| \hat{\E}(D \mid \Q, g) - \E(D \mid \Q, g)  \right\| = o_p(n^{-1/2}), \forall d,g$, is sufficient for the last line. 

If $g= g'' \neq g'$,
\begin{align*}
    &\phantom{{}={}} \P\left\{ \left[ \one(G=g) \omega(d,\Q,g) - \one(G=g') \frac{\hat{p}_{g}(\Q)}{\hat{p}_{g'}(\Q)} \hat{\omega}(d,\Q,g) \right] \left[ \hat{\E}(D \mid \Q, g') - \E(D \mid \Q, g') \right] \right\} \\
    &= \P \left\{ \one(G=g')\frac{\hat{p}_g(\Q)}{\hat{p}_{g'}(\Q)} \left[ \omega(d,\Q,g) - \hat{\omega}(d,\Q,g) \right] \left[ \hat{\E}(D \mid \Q, g') - \E(D \mid \Q, g') \right] \right\} \\
    &\phantom{{}={}} +  
    \P \left\{ \frac{\hat{p}_g(\Q)}{\hat{p}_{g'}(\Q)} \left[\one(G=g)-\one(G=g')\right] \omega(d,\Q,g) \left[ \hat{\E}(D \mid \Q, g') - \E(D \mid \Q, g') \right] \right\} \\
    &\phantom{{}={}} + \P \left\{ \one(G=g) \left[ 1-\frac{\hat{p}_g(\Q)}{\hat{p}_{g'}(\Q)} \right] \omega(d,\Q,g) \left[ \hat{\E}(D \mid \Q, g') - \E(D \mid \Q, g') \right] \right\} ,
\end{align*}
so the following conditions are sufficient for the last line: for some $\zeta < \infty$, $\left| \frac{\hat{p}_g(\Q)}{\hat{p}_{g'}(\Q)} \right| \leq \zeta$ with probability 1, $\left\| \hat{\E}(D \mid \Q, g) - \E(D \mid \Q, g) \right\|=o_p(n^{-1/2})$, and $\left\| \omega(d,\Q,g) - \hat{\omega}(d,\Q,g) \right\|=o_p(1)$, $\forall d,g,g'$. 

\subsubsection*{D.2.2 When $g'=g''$}
Note that all $\xi$ terms satisfy $g'=g''$ for the conditional effect component.

\begin{align*}
    &\phantom{{}={}} R_{2,dgg'g''} \\
    &= \hat{\xi}_{dgg'g'} - \xi_{dgg'g'} \\
    &\phantom{{}={}} + \P \left\{- \frac{\one(G=g')}{\hat{p}_{g'}} \hat{\xi}_{dgg'g'} \right\} \\
    &\phantom{{}={}} + \P \left\{ \frac{\one(G=g) \hat{p}_{g'}(\Q)}{\hat{p}_g(\Q)\hat{p}_{g'}} \left\{ \frac{\one(D=d)}{\hat{\pi}(d,\X,g)} [Y-\hat{\mu}(d,\X,g)]+\hat{\mu}(d,\X,g) - \hat{\omega}(d,\Q,g) \right\} \hat{\E} \left(D \mid \Q,g' \right) \right\} \\
    &\phantom{{}={}} + \P \left\{ \frac{\one(G=g')}{\hat{p}_{g'}} D \cdot \hat{\omega}(d,\Q,g) \right\} \\
    &= \left(1-\frac{p_g}{\hat{p}_g}\right)\left(\hat{\xi}_{dgg'g'} - \xi_{dgg'g'} \right) \\
    &\phantom{{}={}} + \P\left\{ \frac{\one(G=g) \hat{p}_{g'}(\Q) }{\hat{p}_g(\Q) \hat{p}_{g} } \left[ 1-\frac{\pi(d,\X,g)}{\hat{\pi}(d,\X,g)} \right]  [\hat{\mu}(d,\X,g)-\mu(d,\X,g)]  \hat{\E} \left(D \mid \Q,g' \right) \right\} \\
    &\phantom{{}={}} + \P\left\{ \left[ \frac{\one(G=g') }{\hat{p}_{g'}}\E(D \mid \Q,g') - \frac{\one(G=g) \hat{p}_{g'}(\Q) }{\hat{p}_g(\Q) \hat{p}_{g'} } \hat{\E}(D \mid \Q,g') \right] \left[ \hat{\omega}(d,\Q,g) - \omega(d,\Q,g) \right] \right\}.
\end{align*}
Under stated conditions, $R_{2,dgg'g''}=o_p(n^{-1/2})$. In particular, for the last line, the following conditions are sufficient: for some $\zeta < \infty$, $\left| \frac{\hat{p}_{g'}(\Q)}{\hat{p}_{g}(\Q)}\right| \leq \zeta$ with probability 1, $\left\| \hat{\E}(D \mid \Q, g) - \E(D \mid \Q, g) \right\|=o_p(1)$, and $\left\| \omega(d,\Q,g) - \hat{\omega}(d,\Q,g) \right\|=o_p(n^{-1/2})$, $\forall d,g,g'$.

\subsubsection*{D.2.3 Summary of convergence rate conditions}
When $g=g'=g''$, which holds for the conditional selection component, $\left\| \hat{\omega}(d,\Q,g) - \omega(d,\Q,g)  \right\| \left\| \hat{\E}(D \mid \Q, g) - \E(D \mid \Q, g)  \right\| = o_p(n^{-1/2})$ is sufficient. Hence, we obtain a form of rate double robustness with respect to $\omega(d,\Q,g)$ and $\E(D \mid \Q, g)$. Second, when $g= g'' \neq g'$, which holds for the conditional prevalence component, the following condition is sufficient\footnote{Recall that by Assumption 6a, for some $\zeta<\infty$, $\hat{p}_g(\Q)/\hat{p}_{g'}(\Q) \leq \zeta$ with probability 1, $\forall g, g'$. And by Assumption 6b, $ \left\| \hat{\omega}(d,\Q,g) - \omega(d,\Q,g) \right\|=o_p(1)$, $\left\| \hat{\E}(D \mid \Q, g) - \E(D \mid \Q, g) \right\|=o_p(1)$, $\forall d,g$.}: $\left\| \hat{\E}(D \mid \Q, g) - \E(D \mid \Q, g) \right\|=o_p(n^{-1/2})$. Third, when $g'=g'' \neq g$, which holds for the conditional effect component, a sufficient condition is $\left\| \hat{\omega}(d,\Q,g) - \omega(d,\Q,g)  \right\|=o_p(n^{-1/2})$. Therefore, the assumption is weaker for conditional selection than other components, which is also the case for Theorem 5.

\subsection*{Appendix E. Multiple robustness}
The proof for Theorem 2 is omitted, as it is similar to the proof of the double robustness of the AIPW estimator for the ATE \citep{robins_estimation_1994}.

Below, we prove the multiple robustness of $\hat{\xi}_{dgg'g''}$, which is stated in Theorem 5. 
We use a tilde to denote the probability limit of a nuisance estimator. For example, $\hat{\mu}(d,\X,g)$ converges to $\Tilde{\mu}(d,\X,g)$. 
Under consistent estimation of $p_g$, the one-step estimator $\hat{\xi}_{dgg'g''}$ converges in probability to 
\begin{align*}
    &\phantom{{}={}} \E \left\{ \frac{\one(G=g'')}{p_{g''}} \Tilde{\omega}(d,\Q,g)  \Tilde{\E} \left(D \mid \Q, g' \right) \right\} \\
    &+ \E \left\{ \frac{\one(G=g) \Tilde{p}_{g''}(\Q)}{\Tilde{p}_g(\Q)p_{g''}} \left\{ \frac{\one(D=d)}{\Tilde{\pi}(d,\X,g)} [Y-\Tilde{\mu}(d,\X,g)]+\Tilde{\mu}(d,\X,g) - \Tilde{\omega}(d,\Q,g) \right\} \Tilde{\E} \left(D \mid \Q,g' \right) \right\} \\
    &+ \E \left\{ \frac{\one(G=g')\Tilde{p}_{g''}(\Q)}{\Tilde{p}_{g'}(\Q)p_{g''}} \left[ D-\Tilde{\E} \left(D \mid \Q, g' \right) \right] \Tilde{\omega}(d,\Q,g) \right\} 
    \\
    &= \E \left\{ \frac{\one(G=g'')}{p_{g''}} \Tilde{\omega}(d,\Q,g)  \Tilde{\E} \left(D \mid \Q, g' \right) \right\} \\
    &+ \E \left\{ \frac{\one(G=g) \Tilde{p}_{g''}(\Q)}{\Tilde{p}_g(\Q)p_{g''}} \left\{ \frac{\pi(d,\X,g)}{\Tilde{\pi}(d,\X,g)} [\mu(d,\X,g)-\Tilde{\mu}(d,\X,g)] + \Tilde{\mu}(d,\X,g) \right\} \Tilde{\E} \left(D \mid \Q,g' \right) \right\} \\
    &- \E  \left\{ \frac{\one(G=g) \Tilde{p}_{g''}(\Q)}{\Tilde{p}_g(\Q)p_{g''}}  \Tilde{\omega}(d,\Q,g) \Tilde{\E} \left(D \mid \Q,g' \right) \right\} \\
    &+ \E \left\{ \frac{\one(G=g')\Tilde{p}_{g''}(\Q)}{\Tilde{p}_{g'}(\Q)p_{g''}} \left[ \E(D \mid \Q,g')-\Tilde{\E} \left( D \mid \Q, g' \right) \right] \Tilde{\omega}(d,\Q,g) \right\} \\
    &= \E \left\{ \frac{p_{g''}(\Q)}{p_{g''}} \Tilde{\omega}(d,\Q,g)  \Tilde{\E} \left(D \mid \Q, g' \right) \right\} \\
    &+ \E \left\{ \frac{ p_g(\Q) \Tilde{p}_{g''}(\Q)}{\Tilde{p}_g(\Q)p_{g''}} \E \left\{ \frac{\pi(d,\X,g)}{\Tilde{\pi}(d,\X,g)} [\mu(d,\X,g)-\Tilde{\mu}(d,\X,g)] + \Tilde{\mu}(d,\X,g) \mid \Q, g \right\}  \Tilde{\E} \left(D \mid \Q,g' \right) \right\} \\
    &- \E  \left\{ \frac{p_g(\Q) \Tilde{p}_{g''}(\Q)}{\Tilde{p}_g(\Q)p_{g''}}  \Tilde{\omega}(d,\Q,g)  \Tilde{\E} \left(D \mid \Q,g' \right) \right\} \\
    &+ \E \left\{ \frac{p_{g'}(\Q) \Tilde{p}_{g''}(\Q)}{\Tilde{p}_{g'}(\Q)p_{g''}} \left[ \E(D \mid \Q,g')-\Tilde{\E} \left( D \mid \Q, g' \right) \right] \Tilde{\omega}(d,\Q,g) \right\}. 
\end{align*}
Now, when $g=g''$, this probability limit becomes 
\begin{align*}
    &\phantom{{}+{}} \E \left\{ \frac{ p_g(\Q) }{p_{g}} \E \left\{ \frac{\pi(d,\X,g)}{\Tilde{\pi}(d,\X,g)} [\mu(d,\X,g)-\Tilde{\mu}(d,\X,g)] + \Tilde{\mu}(d,\X,g) \mid \Q, g \right\}  \Tilde{\E} \left(D \mid \Q,g' \right) \right\} \\
    &+ \E \left\{ \frac{p_{g'}(\Q) \Tilde{p}_{g}(\Q)}{\Tilde{p}_{g'}(\Q)p_{g}} \left[ \E(D \mid \Q,g')-\Tilde{\E} \left( D \mid \Q, g' \right) \right] \Tilde{\omega}(d,\Q,g) \right\}.
\end{align*}
Next, when $g'=g''$, the probability limit becomes
\begin{align*}
    &\phantom{{}+{}} \E \left\{ \frac{ p_g(\Q) \Tilde{p}_{g'}(\Q)}{\Tilde{p}_g(\Q)p_{g'}} \E \left\{ \frac{\pi(d,\X,g)}{\Tilde{\pi}(d,\X,g)} [\mu(d,\X,g)-\Tilde{\mu}(d,\X,g)] + \Tilde{\mu}(d,\X,g) \mid \Q, g \right\}  \Tilde{\E} \left(D \mid \Q,g' \right) \right\} \\
    &- \E  \left\{ \frac{p_g(\Q) \Tilde{p}_{g'}(\Q)}{\Tilde{p}_g(\Q)p_{g'}}  \Tilde{\omega}(d,\Q,g)  \Tilde{\E} \left(D \mid \Q,g' \right) \right\} + \E \left\{ \frac{p_{g'}(\Q)}{p_{g'}} \E(D \mid \Q,g') \Tilde{\omega}(d,\Q,g) \right\}.
\end{align*}

Finally, when $g=g'=g''$, it becomes
\begin{align*}
    &\phantom{{}+{}} \E \left\{ \frac{ p_g(\Q) }{p_{g}} \E \left\{ \frac{\pi(d,\X,g)}{\Tilde{\pi}(d,\X,g)} [\mu(d,\X,g)-\Tilde{\mu}(d,\X,g)] + \Tilde{\mu}(d,\X,g) \mid \Q, g \right\}  \Tilde{\E} \left(D \mid \Q,g \right) \right\} \\
    &+ \E \left\{ \frac{p_{g}(\Q)}{p_{g}} \left[ \E(D \mid \Q,g)-\Tilde{\E} \left( D \mid \Q, g \right) \right] \Tilde{\omega}(d,\Q,g) \right\}.
\end{align*}
Then it is relatively easy to verify the results in Theorem 5. Note that the condition for the conditional selection component to be consistently estimated is the weakest compared with other components. In particular, for the conditional selection component, it is not necessary to consistently estimate $p_g(\Q)$.

\subsection*{Appendix F. Relation between selection concepts}
Various concepts of ``selection'', ``sorting'', and ``selectivity'' have appeared in the social science literature on the effects of education on later-life outcomes. Below, we clarify the relationship between our selection concept and those appearing in prior work. In our framework, group-specific selection is defined as 
\begin{align*}
 &\phantom{{}={}} \Cov_g(D,\tau) = [\E_g(\tau \mid  D=1)-\E_g(\tau)]\E_g(D).
\end{align*}

First, our definition is closely related to the ``sorting on gains'' in economics \citep{heckman_structural_2005, heckman_returns_2018}, which is defined as the difference between the average treatment effect on the treated and the ATE, i.e., 
\begin{equation*}
    \E_g(\tau \mid  D=1)-\E_g(\tau).
\end{equation*}
Since $\E_g(D)$ is always non-negative, the sign of our group-specific selection is always the same as that of sorting on gains.

Second, akin to our framework, recent works in the ``great equalizer'' literature  \citep{zhou_equalization_2019, fiel_great_2020, karlson_college_2019} have discussed \emph{differential} selection into college completion by parental income. Implicitly, these authors define group-specific selectivity as 
\begin{equation*}
    \E_g(Y \mid  D=1) - \E_g(Y \mid  D=0) - \E_g(\tau).
\end{equation*}
This implicit definition can be validated by rewriting the group difference in this selectivity term.
\begin{align*}
    \phantom{{}={}}& \E_a(Y \mid  D=1) - \E_a(Y  \mid  D=0) - \E_a(\tau) - [\E_b(Y \mid  D=1) - \E_b(Y  \mid  D=0) - \E_b(\tau)] \\
    =& \underbrace{\E_a(Y \mid  D=1)-\E_b(Y \mid  D=1) - [ \E_a(Y \mid  D=0)-\E_b(Y \mid  D=0)]}_{\text{The descriptive test of the great equalizer thesis}}  \\
    \phantom{{}={}}&- \underbrace{\lbrace \E_a(Y^1)-\E_b(Y^1) - [\E_a(Y^0)-\E_b(Y^0)] \rbrace}_{\text{The causal test of the great equalizer thesis}} 
\end{align*}
If the descriptive test returns a more negative value than the causal test, these authors will conclude that members of the advantaged group are less selected than their disadvantaged peers in the transition to college completion.\footnote{This conclusion, however, corresponds more closely to $\E_a(Y \mid D=1)-\E_a(Y^1)-[\E_b(Y \mid D=1)-\E_b(Y^1)]$ being negative. For example, if $\E_a(Y \mid D=1)-\E_a(Y^1)=0$, while $\E_b(Y \mid D=1)-\E_b(Y^1) >0$, then the treated individuals in group $b$ (and group $b$ only) have higher $Y_1$ than the untreated individuals in the same group, which means that the treated individuals in group $b$ (and group $b$ only) are positively selected on $Y_1$ \citep[see][]{yu2024counter}.}  

Taking the difference between the selectivity term in the great equalizer literature and sorting on gains, we obtain $\E_g(Y^0 \mid  D=1)-\E_g(Y^0 \mid  D=0)$, which is selection on baseline potential outcome.\footnote{In \citet{heckman_characterizing_1998}, \citet{heckman_structural_2005} and \citet{heckman_returns_2018}, selection on baseline outcome is referred to as the ``selection bias''.}
Therefore, the selectivity term in the great equalizer literature captures  selection on baseline outcome, on top of selection on treatment effect captured in sorting on gains and our selection term \citep[see][p.58-9]{morgan_counterfactuals_2014}. 

Third, a large literature on effect heterogeneity by propensity score \citep{brand_who_2010, xie_estimating_2012, brand_uncovering_2021} is  closely aligned with our framework in its conceptualization of selection into treatment. Originally developed in the context of education effects, this framework has been applied in a wide array of topics.
Recall that $\pi(1,\X,g)$ denotes the propensity score, then under Assumption 3 (conditional ignorability),
\begin{align*}
&\phantom{{}={}} \Cov_g(D,\tau)  \\
&= \E_g[\Cov_g(D,\tau \mid \X)] + \Cov_g[\E_g(D \mid \X),\E_g(\tau \mid \X)] \\
&= \Cov_g[\pi(1,\X,g), \E_g(\tau \mid \X)] \\
&= \E_g [\pi(1,\X,g) \E_g(\tau \mid \X)] - \E_g(D)\E_g(\tau) \\
&= \E_g \lbrace \E_g [\pi(1,\X,g) \E_g(\tau \mid \X) \mid \pi(1,\X,g)] \rbrace - \E_g(D)\E_g(\tau) \\
&= \E_g \lbrace \pi(1,\X,g) \E_g[\tau \mid \pi(1,\X,g)] \rbrace - \E_g[\pi(1,\X,g)]\E_g\lbrace\E_g[\tau \mid \pi(1,\X,g)] \rbrace \\
&= \Cov_g [\pi(1,\X,g), \E_g(\tau \mid \pi(1,\X,g)].
\end{align*}
Hence, under conditional ignorability, our selectivity term equals the covariance between the propensity score and the conditional treatment effect given the propensity score. Due to this relationship, the estimators we developed for the generic functions underlying our unconditional decomposition can also be used to nonparametrically estimate effect heterogeneity by propensity score. 

\subsection*{Appendix G. Supplemental tables}

{\centering
\begin{longtable}{lccc}
\multicolumn{4}{l}{Table A1. Relationship between baseline covariates and outcome missingness } \\
\hline
 Covariate & & \multirow{2}{*}{\makecell{Missing percentage or \\ correlation coefficient}} & P value \\
\\
\hline
Parental income &  & -0.079$^\#$ & $<0.001$ \\
Race & White & 25.27 & \multirow{3}{*}{$<0.001$} \\
 & Black & 14.45 \\
 & Hispanic & 12.38 \\
Gender & Male & 21.09 & \multirow{2}{*}{$0.385$} \\
 & Female & 19.63 \\
Mother's year of schooling & & 0.003$^\#$ & 0.870 \\
Parental presence & Yes & 19.31 & \multirow{2}{*}{$0.089$} \\
 & No & 22.21 \\
Number of siblings & & 0.002$^\#$ & 0.900 \\
Urban residence & Yes & 19.65 & \multirow{2}{*}{$0.108$} \\
 & No & 22.86 \\
\multirow{2}{*}{\makecell{Expecting bachelor \\ degree or higher}} & Yes & 19.28 & \multirow{2}{*}{$0.250$} \\
 & No & 21.20 \\
AFQT score & & 0.021$^\#$ & 0.294 \\
Age & & -0.015$^\#$ & 0.453 \\
\multirow{2}{*}{\makecell{Friends expecting \\ bachelor degree or higher}} & Yes & 19.15 & \multirow{2}{*}{$0.190$} \\
& No & 21.31 \\
Rotter score & & -0.002$^\#$ & 0.914 \\
Rosenberg score & & -0.013$^\#$ & 0.517 \\
School satisfaction & & 0.003 & 0.888 \\
\multirow{2}{*}{\makecell{Speak foreign language \\ at home}} & Yes & 18.07 & \multirow{2}{*}{$0.174$} \\
 & No & 20.91 \\
\multirow{4}{*}{\makecell{Metropolitan Statistical \\ Area category}} & Not in SMSA & 22.73 & \multirow{4}{*}{$0.057$} \\
& In SMSA, not central city & 18.00 \\
& In SMSA, in central city & 22.32 \\
& In SMSA, central city unknown & 18.59 \\
Separate from mother & Yes & 25.83 & \multirow{2}{*}{$0.158$} \\
 & No & 20.08 \\
Mother working & Yes & 18.27 & \multirow{2}{*}{$0.004$} \\
 & No & 23.00 \\
\multirow{4}{*}{SMSA} & Northeast & 23.21 & \multirow{4}{*}{$0.031$} \\
 & North central & 22.42 \\
 & South & 18.75 \\
 & West & 16.95 \\
 \hline
   \multicolumn{4}{l} {\makecell{Note: N=2580. The sample is individuals with no missing values in any baseline covariates. \\ In the column for missing percentage or correlation coefficient, values with $\#$ are correlation \\ coefficients between covariates and the outcome missingness indicator. P values for missing  \\ percentages are based on Chi-squared tests.}} 
\end{longtable}
}

{\centering
\begin{longtable}{lccc}
\multicolumn{4}{l}{Table A2. Group-specific estimates for the unconditional decomposition } \\
  \hline
   &  Top 40\% income & Bottom 40\% income  & Top-Bottom \\ 
  \hline
  Treatment proportion & 0.337 & 0.086 & 0.251 \\ 
   & (0.310, 0.364) & (0.067, 0.105) & (0.218, 0.284) \\
  & \multicolumn{3}{l}{Neural networks}  \\
  Baseline outcome & 0.599 & 0.411 & 0.188 \\ 
  & (0.582, 0.617) & (0.393, 0.430) & (0.163,0.213) \\
  ATE & 0.145 & 0.123 & 0.022 \\ 
  & (0.103, 0.188) & (0.052, 0.195) & (-0.061, 0.105) \\
  $\Cov(D, \tau)$ & -0.014 & 0.002 & -0.017 \\ 
  & (-0.027, -0.002) & (-0.004, 0.009) & (-0.031,-0.002) \\
  & \multicolumn{3}{l}{Gradient boosted machines}  \\
  Baseline outcome & 0.599 & 0.411 & 0.188 \\ 
  & (0.581, 0.616) & (0.393, 0.429) & (0.162, 0.213) \\
  ATE & 0.137 & 0.110 & 0.027 \\ 
  & (0.100, 0.174) & (0.044, 0.177) & (-0.049, 0.103) \\
  $\Cov(D, \tau)$ & -0.011 & 0.004 & -0.014 \\ 
  & (-0.021, 0.000) & (-0.002, 0.009) & (-0.026,-0.002) \\
  & \multicolumn{3}{l}{Random forests}  \\
  Baseline outcome & 0.593 & 0.410 & 0.183 \\ 
  & (0.578,0.609) & (0.392,0.429) & (0.159,0.207) \\
  ATE & 0.140 & 0.116 & 0.024 \\ 
  & (0.107, 0.172) & (0.040, 0.192) & (-0.058, 0.107) \\
  $\Cov(D, \tau)$ & -0.006 & 0.004 & -0.010 \\ 
  & (-0.014, 0.001) & (-0.002, 0.010) & (-0.020,-0.001) \\
  & \multicolumn{3}{l}{Parametric models}  \\
  Baseline outcome & 0.598 & 0.410 & 0.187 \\ 
  & (0.580, 0.615) & (0.393, 0.428) & (0.162,0.212) \\
  ATE & 0.121 & 0.145 & -0.024 \\ 
  & (0.080, 0.162) & (0.059, 0.232) & (-0.120, 0.072) \\
  $\Cov(D, \tau)$ & -0.005 & 0.001 & -0.006 \\ 
  & (-0.017, 0.008) & (-0.006, 0.008) & (-0.020,0.009) \\
   \hline
   \multicolumn{4}{l} {\makecell{Note: 95\% confidence intervals are in parentheses and are computed according to Theorem 3. \\ Weight stabilization is used. For ML models, cross-fitting is used.}} 
\end{longtable}
}

{\centering
\begin{longtable}{lllll}
\multicolumn{5}{l}{Table A3. Unconditional decomposition estimates } \\
  \hline
   & Neural networks & GBM & Random forests & Parametric \\ 
  \hline
Total & 0.210 & 0.210 & 0.210 & 0.210 \\ 
& (0.188,0.232) & (0.188,0.232) & (0.188,0.232) & (0.188,0.232) \\ 
Baseline & 0.188 & 0.188 & 0.183 & 0.187 \\ 
& (0.163,0.213) & (0.162,0.213) & (0.159,0.207) & (0.162,0.212) \\ 
Prevalence & 0.031 & 0.028 & 0.029 & 0.036 \\ 
& (0.012,0.050) & (0.011,0.045) & (0.009,0.049) & (0.014,0.059) \\ 
Effect & 0.007 & 0.009 & 0.008 & -0.008 \\ 
& (-0.021,0.035) & (-0.017,0.035) & (-0.020,0.036) & (-0.040,0.024) \\ 
Selection & -0.017 & -0.014 & -0.010 & -0.006 \\ 
& (-0.031,-0.002) & (-0.026,-0.002) & (-0.020,-0.001) & (-0.020,0.009) \\ 
Change in disparity & 0.029 & 0.024 & 0.025 & 0.035 \\ 
& (0.007,0.050) & (0.005,0.043) & (0.002,0.048) & (0.008,0.063) \\ 
   \hline
  \multicolumn{5}{l} {\makecell{Note: 95\% confidence intervals are in parentheses and are computed according to Theorem 3. \\ Weight stabilization is used. For ML models, cross-fitting is used. The change in disparity uses the \\ definition in Jackson and Vanderweele's (\citeyear{jackson_decomposition_2018}) URED. }}
\end{longtable}
}

{\centering
\begin{tabular}{lllll}
\multicolumn{5}{l}{Table A4. Conditional decomposition estimates} \\
  \hline
 & Neural networks & GBM & Random forests & Parametric \\ 
  \hline
Total & 0.210 & 0.210 & 0.210 & 0.210 \\ 
& (0.188,0.232) & (0.188,0.232) & (0.188,0.232) & (0.188,0.232) \\ 
Baseline & 0.188 & 0.188 & 0.183 & 0.187 \\ 
& (0.163,0.213) & (0.162,0.213) & (0.159,0.207) & (0.162,0.212) \\ 
Conditional prevalence & 0.010 & 0.007 & 0.013 & 0.009 \\ 
& (0.003,0.016) & (0.002,0.012) & (0.007,0.019) & (0.002,0.015) \\ 
Conditional effect & -0.012 & -0.028 & -0.025 & -0.023 \\ 
& (-0.055,0.031) & (-0.061,0.006) & (-0.054,0.004) & (-0.054,0.007) \\ 
Conditional selection & -0.006 & -0.004 & -0.006 & -0.002 \\ 
& (-0.014,0.003) & (-0.010,0.002) & (-0.014,0.002) & (-0.010,0.005) \\ 
$\Q$-distribution & 0.030 & 0.047 & 0.045 & 0.040 \\ 
& (-0.004,0.063) & (0.022,0.072) & (0.023,0.067) & (0.018,0.061) \\ 
Change in disparity & 0.006 & 0.006 & 0.007 & 0.006 \\ 
& (-0.003,0.016) & (-0.001,0.013) & (-0.002,0.017) & (-0.003,0.016) \\ 
\hline
  \multicolumn{5}{l} {\makecell{Note: 95\% confidence intervals are in parentheses and are computed according to Theorem 6. \\ Weight stabilization is used. For ML models, cross-fitting is used. The change in disparity uses \\ the definition in Jackson's (\citeyear{jackson_meaningful_2021}) CRED. }}
\end{tabular}
}

In addition, in Table A5, we present a set of estimates for the conditional decomposition as a robustness check. This is motivated by Assumption 6c, which requires the parametric convergence rate for $\hat{\E}(D \mid \Q,g)$ for the conditional prevalence component, the parametric convergence rate for $\hat{\omega}(d,\Q,g)$ for the conditional effect component, and the same rate for both nuisance function models for the $\Q$-distribution component. Thus, to make asymptotic inference more exact, we implement the following procedure. 

We estimate $\mu(d,\X,g)$, $\pi(d,\X,g)$, and $p_g(\Q)$ using ML with cross-fitting. Then we estimate $\E(D \mid \Q,g)$ and $\omega(d,\Q,g)$ parametrically using linear or logistic regressions without cross-fitting. For the parametric models, we include the group indicator, the AFQT score, the squared AFQT score, and the interactions between the group indicator and the AFQT variables. For $\omega(d,\Q,g)$, we apply the pseudo-outcome approach detailed in the main text, but with a different cross-fitting procedure. Here, we use cross-fitted estimates of $\mu(d,\X,g)$ and $\pi(d,\X,g)$ to construct the pseudo-outcomes, then we estimate $\omega(d,\Q,g)$ without cross-fitting. The results are quantitatively similar and qualitatively identical across Tables A4 and A5.

{\centering
\begin{longtable}{lllll}
\multicolumn{5}{l}{Table A5. Conditional decomposition estimates with mixed nonparametric and parametric models} \\
  \hline
 & Neural networks & GBM & Random forests \\ 
  \hline
Total & 0.210 & 0.210 & 0.210 \\ 
 & (0.188,0.232) & (0.188,0.232) & (0.188,0.232) \\ 
Baseline & 0.188 & 0.182 & 0.183 \\ 
 & (0.163,0.213) & (0.157,0.208) & (0.159,0.207) \\ 
Conditional prevalence & 0.008 & 0.008 & 0.008 \\ 
 & (0.002,0.014) & (0.002,0.013) & (0.002,0.014) \\ 
Conditional effect & -0.018 & -0.019 & -0.024 \\ 
 & (-0.049,0.013) & (-0.048,0.010) & (-0.054,0.006) \\ 
Conditional selection & -0.006 & -0.001 & -0.004 \\ 
 & (-0.014,0.003) & (-0.008,0.005) & (-0.010,0.002) \\ 
$\Q$-distribution & 0.037 & 0.040 & 0.047 \\ 
 & (0.016,0.058) & (0.022,0.059) & (0.025,0.069) \\ 
Change in disparity & 0.006 & 0.007 & 0.006 \\ 
 & (-0.002,0.015) & (-0.000,0.014) & (-0.003,0.014) \\ 
\hline
  \multicolumn{4}{l} {\makecell{Note: 95\% confidence intervals are in parentheses and are computed according to Theorem 6. \\ Weight stabilization is used. $\mu(d,\X,g)$, $\pi(d,\X,g)$, and $p_g(\Q)$ are estimated using ML with cross-fitting. \\ $\E(D \mid \Q,g)$ and $\omega(d,\Q,g)$ are estimated using parametric  models without cross-fitting. \\ The change in disparity uses the definition in Jackson's (\citeyear{jackson_meaningful_2021}) CRED. }}
\end{longtable}
}

\subsection*{Appendix H. Diagnostic checks and sensitivity analysis for the empirical application}
\subsubsection*{H.1. Covariate overlap}
First, we assess covariate overlap in the data. Our Assumption 4 requires that at any level of the covariates, there are both individuals receiving the treatment and individuals receiving the control. We empirically assess this assumption by plotting the histograms of estimated propensity scores in Figure A3. 

Across the four alternative models for the propensity score, the basic pattern is the same. There appear to be a large number of individuals whose probability of completing college is close to 0. However, there is no exact 0 or 1 in the estimated propensity scores, which means that Assumption 4 is minimally satisfied. 
\begin{figure}[htbp]
    \centering
    \includegraphics[width=1\linewidth]{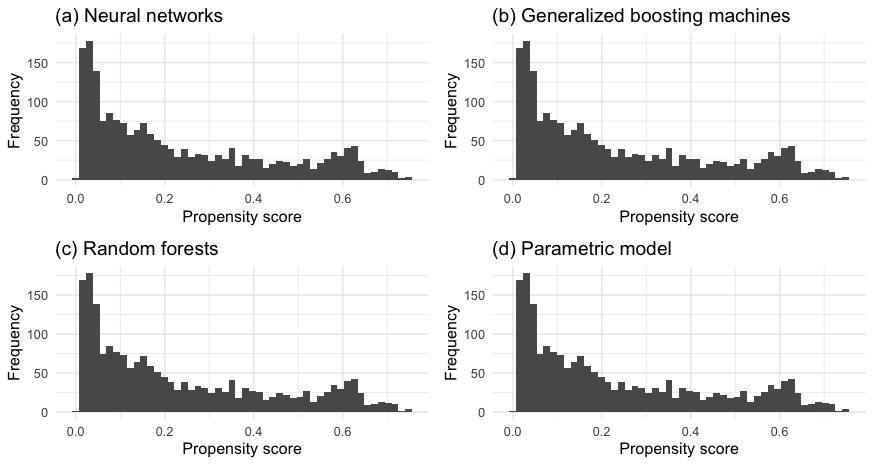}
    \caption*{Figure A3. Covariate overlap}
    \label{fig:overlap}
\end{figure}

\subsubsection*{H.2. Covariate balance after weighting}
We perform a diagnostic check for the propensity score models. The propensity score function is $\pi(d,\X,G) \defeq \Pro(D=d \mid \X,G)$, which is a nuisance function shared by both the unconditional and conditional decompositions. Our estimators can be viewed as involving a weight based on the propensity score, $\frac{\one(D=d)}{\pi(d,\X,G)}$.  Also recall that $\hat{\pi}(d,\X,G)$ is the estimated counterpart of $\pi(d,\X,G)$. In this subsection, covariates refer to $\C \defeq \{\X,G\}$, and $C_j$ is the $j$- covariate. 

Theoretically, all covariates should have the same distribution across treatment and control observations if the true propensity score is used to weight the data \citep{li_balancing_2018}. Thus, covariate balance after weighting has been used as a diagnostic check for estimators of the propensity score. 
There is a variety of covariate balance measures in the weighting literature \citep{austin_moving_2015,li_balancing_2018,cannas_comparison_2019}. We focus on the absolute standardized difference (ASD) as defined in \citet[][p.3666]{austin_moving_2015}: 

\begin{equation*}
    100 \times \left| \frac{\sum C_j \frac{D}{\hat{\pi}(1,\X,G)} }{\sum \frac{D}{\hat{\pi}(1,\X,G)}} - \frac{\sum C_j \frac{1-D}{\hat{\pi}(0,\X,G)} }{\sum \frac{1-D}{\hat{\pi}(0,\X,G)}} \right| \bigslash \sqrt{\left(s_1^2 + s_0^2 \right)/2},
\end{equation*}
where
\begin{equation*}
    s_d^2 \defeq \frac{\sum \frac{\one(D=d)}{\hat{\pi}(d,\X,G)}}{ \left( \sum \frac{\one(D=d)}{\hat{\pi}(d,\X,G)} \right)^2 - \sum \left( \frac{\one(D=d)}{\hat{\pi}(d,\X,G)}\right)^2 } \sum \frac{\one(D=d)}{\hat{\pi}(d,\X,G)} \left(C_j - \frac{\sum C_j \frac{\one(D=d)}{\hat{\pi}(d,\X,G)} }{\sum \frac{\one(D=d)}{\hat{\pi}(d,\X,G)}} \right)^2.
\end{equation*}
The unweighted ASD is defined by replacing $\hat{\pi}(1,\X,G)$ and $\hat{\pi}(0,\X,G)$ with 1 \citep[see][p.310-311]{imbens_causal_2015}. Note that the way $s_d^2$ is computed treats $\frac{\one(D=d)}{\hat{\pi}(d,\X,G)}$ as a ``reliability weight'' and corrects the weighted sample variance accordingly.  

\begin{figure}[htb]
    \centering
        \begin{minipage}{\textwidth}
        \hspace{-1.5em}
        \includegraphics[width=1.04\linewidth]{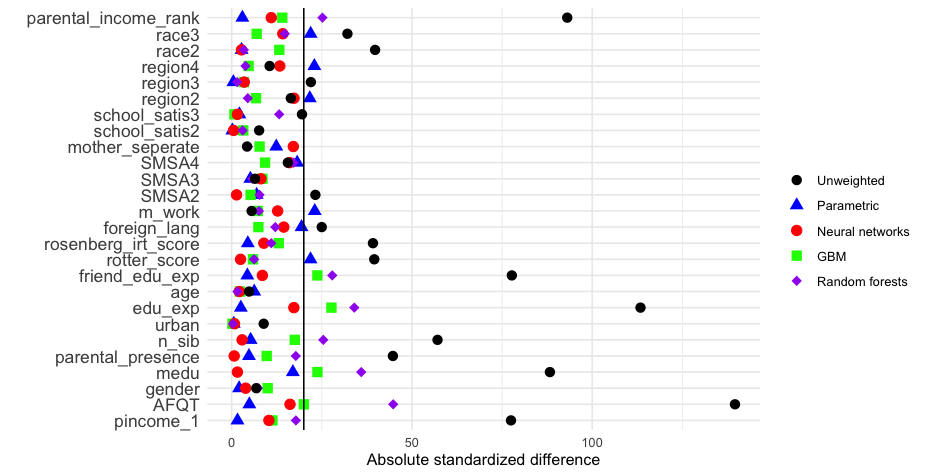}
    \end{minipage}
    \caption*{Figure A4. Covariate balance before and after weighting by covariate and propensity score model. The vertical line indicates the rule-of-thumb threshold of 20 \citep{cannas_comparison_2019}.}
    \label{fig:balance}
\end{figure}

We present the ASDs in Figure A4. Two observations are noteworthy. First, before weighting, several covariates are very imbalanced across treatment and control groups. After weighting, extreme imbalances almost all disappeared, regardless of the model. Using the rule-of-thumb threshold of 20 \citep{cannas_comparison_2019}, all models achieve good enough balance for most covariates. In particular, neural networks keep the ASDs for all covariates under 20. On the other hand, random forests leave a few covariates quite unbalanced after weighting. It is also important to recall that our estimators are multiply robust, such that even if the propensity score function is not consistently estimated, the decomposition components may still be consistently estimated.

\subsubsection*{H.3. Outcome model fit}
We also perform a diagnostic check for the outcome regression models. The outcome regression function is $\mu(d,\X,g) \defeq \E(Y \mid d,\X,g)$, and $\hat{\mu}(d,\X,g)$ is its estimated counterpart. It is another nuisance function shared by both the unconditional and conditional decompositions.

Here, we measure model fit using the mean squared error (MSE): $\frac{1}{n}\sum \left[Y- \hat{\mu}(D,\X,G) \right]^2$. We compare the ML models with the intercept-only model and the parametric model used in the empirical application in terms of the MSE as a sanity check.
The ML models should have lower MSEs than the intercept-only model and the parametric model. Otherwise, hyper-parameters may not have been sufficiently tuned. To make comparison across models fair, we use cross-fitting for all models, as opposed to only the ML models like we did in the main text. In other words, the MSEs are all calculated out of sample.

\begin{table}[h]
    \centering
    \begin{tabular}{l c}
    \multicolumn{2}{l}{Table A6. Mean squared errors of outcome models} \\
  \hline
        Model & MSE \\
        \hline
        Intercept model & 0.0680 \\
        Parametric model & 0.0484 \\
        Neural networks & 0.0480 \\
        Generalized boosted machines & 0.0474 \\
        Random forests & 0.0480 \\
        \hline
        \multicolumn{2}{l}{\makecell{Note: To ensure fair comparison, cross-fitting is used for all models. \\ As specified in Section 4.2 of the main text, the parametric model contains \\ all two-way interactions between $D$ and $\{\X, G\}$ as well as the main effects.}}
    \end{tabular}
    \label{tab:mse}
\end{table}

Results are shown in Table A6. All models used in the empirical application obtain much lower MSEs than the intercept-only model. Furthermore, all ML models have lower MSEs than the parametric model. These suggest that the hyper-parameters of the ML models are satisfactorily tuned via cross-validation. Again, due to the multiple robustness of our estimators, consistent estimation of the outcome regression function is not necessary for consistent estimation of the decomposition components. 

\subsubsection*{H.4. Sensitivity analysis}
We conduct a sensitivity analysis for our decompositions using results in \citet{opacic_disparity_2023}, who provide bias formulas for a class of estimands, including $\E_g\left( Y^{R(D \mid g')} \right)$ and $\E_g\left( Y^{R(D \mid g',\Q)} \right)$.\footnote{There is an emerging literature on sensitivity analysis for random equalization decompositions \citep{park_sensitivity_2023,park_estimation_2024,shen2024calibratedsensitivityanalysisweighted}.} They do so in a setting where there is a binary unobserved confounder $U \in \{0,1\}$, and they make two simplifying assumptions. In particular,
\begin{itemize}[leftmargin=0em]
    \item[] Sensitivity assumption 1. $Y^d \indep D \mid \x,g,u, \forall d,\x,g,u$.
    \item[] Sensitivity assumption 2. $\rho_{1g} \defeq \E(U \mid D=1,\x,g) - \E(U \mid D=0,\x,g)$ does not depend on $\x$.
    \item[] Sensitivity assumption 3. $\rho_{2g} \defeq \E(Y \mid d,\x,g,U=1) - \E(Y \mid d,\x,g,U=0)$ does not depend on $d$ or $\x$.
\end{itemize}
Note that $\rho_{1g}$ is the conditional association between $U$ and the treatment, and $\rho_{2g}$ is the conditional association between $U$ and the outcome.
The bias is defined as the identified functional derived under Assumption 3 in the main text minus the one derived under Sensitivity assumption 1. Based on the results of \citet{opacic_disparity_2023}, it is easy to show that, under Sensitivity Assumptions 1, 2, and 3, 
\begin{align*}
    \text{bias}\left[ \E_g\left( Y^{R(D \mid g')} \right) \right] &= \rho_{1g} \rho_{2g} [ \E(D \mid g') - \E(D \mid g) ] \\
    \text{bias}\left[ \E_g\left( Y^{R(D \mid g',\Q)} \right) \right] &= \rho_{1g} \rho_{2g} \{ \E[ \E(D \mid \Q, g') \mid g] - \E(D \mid g) \}.
\end{align*}

In our decompositions, there are five components that are estimated to be significantly different from zero by all ML models: baseline, prevalence, selection, conditional prevalence, and $\Q$-distribution.  Baseline and $\Q$-distribution components are not part of the unconditional and conditional total contributions of the treatment, and their biases cannot be directly evaluated using the bias formulas above. For the selection component, under Sensitivity Assumptions 1, 2, and 3, the bias is necessarily 0.\footnote{The bias for the conditional selection component is also constrained to be 0 under these assumptions. Echoing Theorems 5 and 6 in the main text, the conditional selection component is thus more robust to inconsistencies and lower convergence rates in nuisance function estimation as well as to unobserved confounding relative to other components of the conditional contribution of the treatment.} Therefore, we focus on the possible biases of the prevalence and conditional prevalence components. In particular, using results from Sections 2.2 and 2.4.1 in the main text, we have
\begin{align*}
    \text{bias}(\text{prevalence}) &= \rho_{1b} \rho_{2b} [ \E(D \mid a) - \E(D \mid b) ] \\
    \text{bias}(\text{conditional prevalence}) &= \rho_{1b} \rho_{2b} \{ \E[ \E(D \mid \Q, a) \mid b] - \E(D \mid b) \}.
\end{align*}

Therefore, in order to conduct sensitivity analysis for the prevalence and conditional prevalence components, we need to estimate $\E[ \E(D \mid \Q, a) \mid b]$ and specify $\{\rho_{1b}, \rho_{2b}\}$. For the conditional prevalence estimate based on each nuisance function model, we use the same nuisance function model to estimate $\E[ \E(D \mid \Q, a) \mid b]$. For $\rho_{1b}$ and $\rho_{2b}$, we benchmark their values using the observed confounder, educational expectation (binary indicator for ``bachelor degree or higher''), which we denote as $\Tilde{U}$. In Figure A4, this is the binary baseline covariate that has the strongest imbalance in the unweighted data. Hence, it is likely an exceptionally strong confounder. We estimate what $\rho_{1b}$ and $\rho_{2b}$ would be had educational expectation been $U$ by fitting two models in group $b$ with only main effects: 
\begin{align*}
    \E(\Tilde{U} \mid D, \X \setminus \Tilde{U}, b) &= \kappa_{10} + \kappa_{11} D + \boldsymbol{\kappa}_{12}^\intercal \{ \X \setminus \Tilde{U} \} \\
    \E(Y \mid D,\X \setminus \Tilde{U}, \Tilde{U}, b) &= \kappa_{20} + \kappa_{21} D + \kappa_{22}^\intercal \{ \X \setminus \Tilde{U} \} + \kappa_{23}\Tilde{U}.
\end{align*}
Then, $\kappa_{11}$ and $\kappa_{23}$ respectively serve as benchmarks for $\rho_{1b}$ and $\rho_{2b}$. These benchmarks suggest the magnitude of the biases if the unobserved confounder confounds the prevalence and conditional prevalence components as much as the observed educational expectation does. In our application, $\kappa_{11}$ and $\kappa_{23}$ are estimated to be 0.27 and 0.05, respectively.

    \begin{figure}[H]
    \centering
    \includegraphics[width=1.1\linewidth]{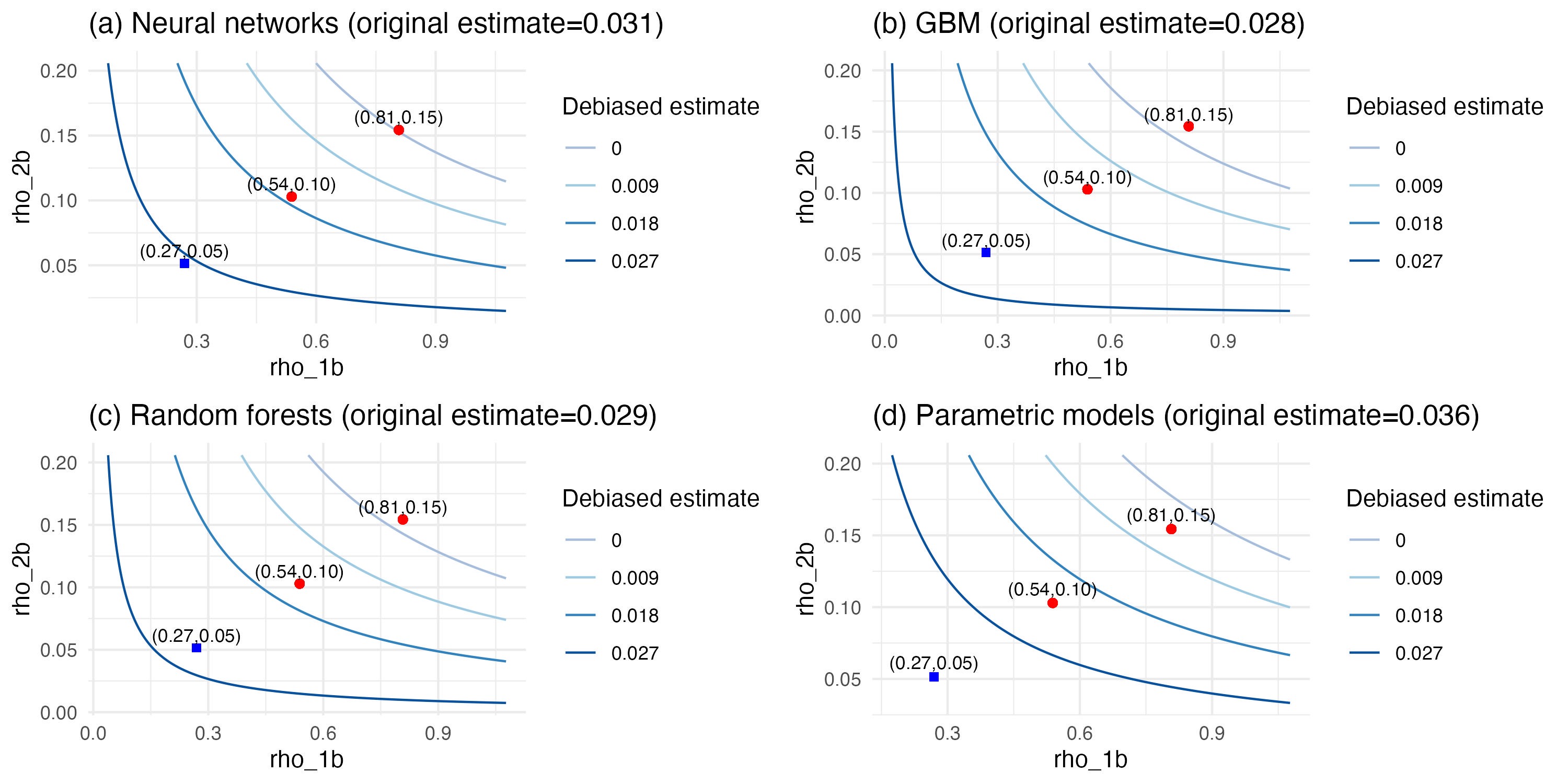}
    \caption*{Figure A5. Contour plot for debiased estimates of the prevalence component. The blue (square) dot indicates the benchmark values of $\rho_{1b}$ and $\rho_{2b}$ based on treating educational expectation as an unobserved confounder. The two red (circle) dots show twice and thrice the benchmark values, respectively.}
\end{figure}

\begin{figure}[H]
    \centering
    \includegraphics[width=1.1\linewidth]{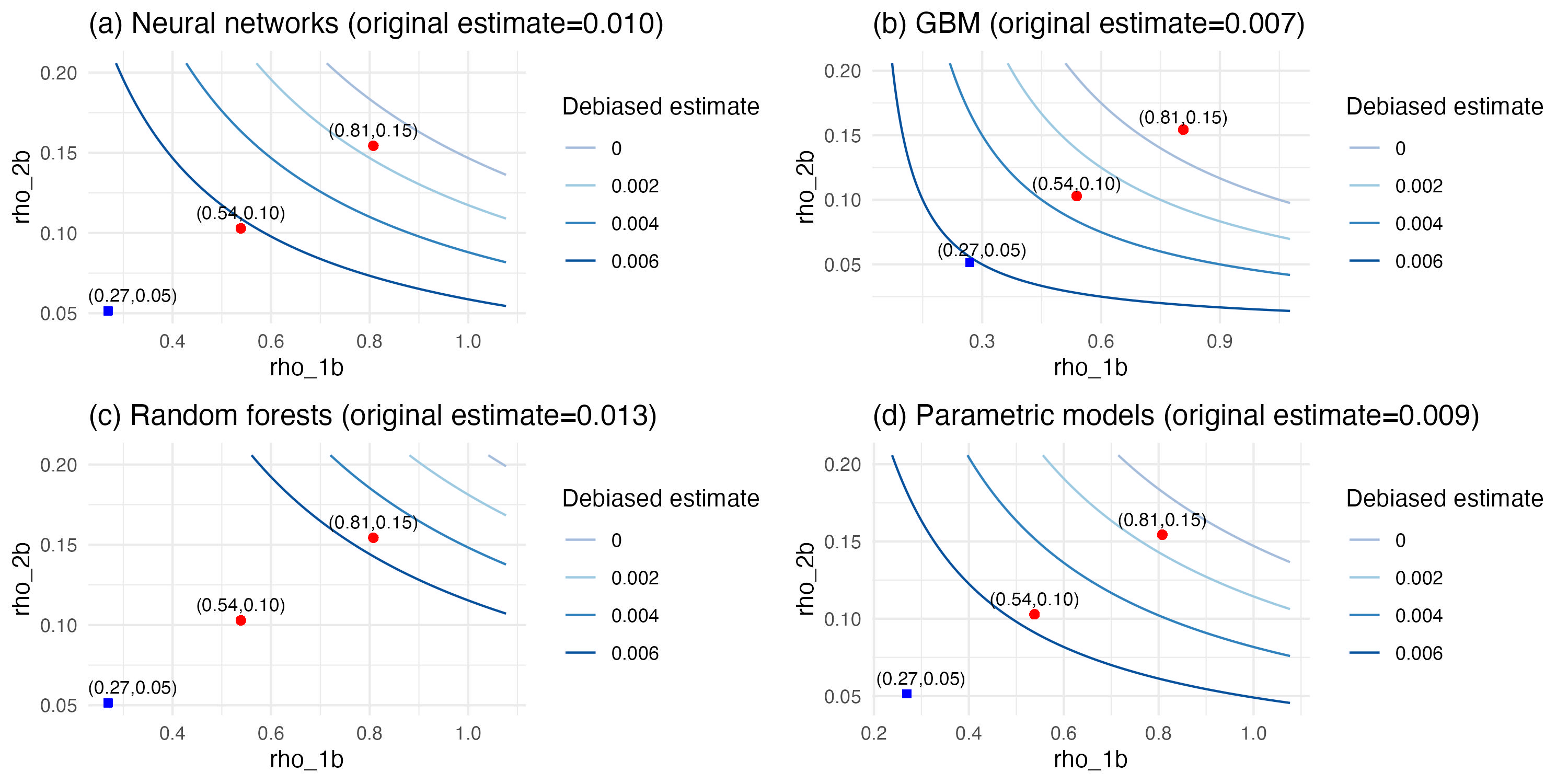}
    \caption*{Figure A6. Contour plot for debiased estimates of the conditional prevalence component. The blue (square) dot indicates the benchmark values of $\rho_{1b}$ and $\rho_{2b}$ based on treating educational expectation as an unobserved confounder. The two red (circle) dots show twice and thrice the benchmark values, respectively.}
\end{figure}

In Figures A5 and A6, we show debiased estimates of the prevalence and the conditional prevalence components, respectively. The debiased estimates are obtained by subtracting the biases under certain $\rho_{1b}$ and $\rho_{2b}$ values from the original estimates. For both unconditional and conditional prevalence components and across nuisance function models, a confounder whose conditional associations with the treatment and the outcome are twice as large as those of educational expectation would not bring the estimates to zero. In fact, to reduce the original estimates to zero, there must be a confounder that is at least around three times as strong as educational expectation. Based on these results, our empirical estimates appear reasonably robust to unobserved confounding. We leave the development of sensitivity analysis for other components of our decompositions to future work.

\FloatBarrier

\bibliographystyle{asr.bst} 
\bibliography{Bibliography.bib}       

\begin{thebibliography}{}
\newcommand{\enquote}[1]{``#1''}

\bibitem[\protect\citeauthoryear{Austin and Stuart}{Austin and
  Stuart}{2015}]{austin_moving_2015}
Austin, Peter~C. and Elizabeth~A. Stuart. 2015.
\newblock \enquote{Moving towards best practice when using inverse probability
  of treatment weighting ({IPTW}) using the propensity score to estimate causal
  treatment effects in observational studies.}
\newblock {\em Statistics in Medicine\/} 34:3661--3679.

\bibitem[\protect\citeauthoryear{Avin, Shpitser, and Pearl}{Avin
  et~al.}{2005}]{avin_identiability_2005}
Avin, Chen, Ilya Shpitser, and Judea Pearl. 2005.
\newblock \enquote{Identiﬁability of {Path}-{Speciﬁc} {Effects}.}
\newblock In {\em Proceedings of {International} {Joint} {Conference} on
  {Artificial} {Intelligence}\/}, pp. 357--363, Edinburgh, Schotland.

\bibitem[\protect\citeauthoryear{Bailey and Dynarski}{Bailey and
  Dynarski}{2011}]{bailey_gains_2011}
Bailey, Martha and Susan Dynarski. 2011.
\newblock \enquote{Gains and {Gaps}: {Changing} {Inequality} in {U}.{S}.
  {College} {Entry} and {Completion}.}
\newblock Technical Report w17633, National Bureau of Economic Research,
  Cambridge, MA.

\bibitem[\protect\citeauthoryear{Bickel, Klaassen, Ritov, and Wellner}{Bickel
  et~al.}{1998}]{bickel_efficient_1998}
Bickel, Peter~J, Chris~A.J Klaassen, Ya'acov Ritov, and Jon~A Wellner. 1998.
\newblock {\em Efficient and {Adaptive} {Estimation} for {Semiparametric}
  {Models}\/}.
\newblock New York, NY: Springer.

\bibitem[\protect\citeauthoryear{Blau and Duncan}{Blau and
  Duncan}{1967}]{blau_american_1978}
Blau, Peter~Michael and Otis~Dudley Duncan. 1967.
\newblock {\em The {American} {Occupational} {Structure}\/}.
\newblock Free Press.

\bibitem[\protect\citeauthoryear{Blinder}{Blinder}{1973}]{blinder_wage_1973}
Blinder, Alan~S. 1973.
\newblock \enquote{Wage {Discrimination}: {Reduced} {Form} and {Structural}
  {Estimates}.}
\newblock {\em The Journal of Human Resources\/} 8:436.

\bibitem[\protect\citeauthoryear{Bloome, Dyer, and Zhou}{Bloome
  et~al.}{2018}]{bloome_educational_2018}
Bloome, Deirdre, Shauna Dyer, and Xiang Zhou. 2018.
\newblock \enquote{Educational {Inequality}, {Educational} {Expansion}, and
  {Intergenerational} {Income} {Persistence} in the {United} {States}.}
\newblock {\em American Sociological Review\/} 83:1215--1253.

\bibitem[\protect\citeauthoryear{Brady, Finnigan, and Hübgen}{Brady
  et~al.}{2017}]{brady_rethinking_2017}
Brady, David, Ryan~M. Finnigan, and Sabine Hübgen. 2017.
\newblock \enquote{Rethinking the {Risks} of {Poverty}: {A} {Framework} for
  {Analyzing} {Prevalences} and {Penalties}.}
\newblock {\em American Journal of Sociology\/} 123:740--786.

\bibitem[\protect\citeauthoryear{Brand and Xie}{Brand and
  Xie}{2010}]{brand_who_2010}
Brand, Jennie~E. and Yu~Xie. 2010.
\newblock \enquote{Who {Benefits} {Most} from {College}?: {Evidence} for
  {Negative} {Selection} in {Heterogeneous} {Economic} {Returns} to {Higher}
  {Education}.}
\newblock {\em American Sociological Review\/} 75:273--302.

\bibitem[\protect\citeauthoryear{Brand, Xu, Koch, and Geraldo}{Brand
  et~al.}{2021}]{brand_uncovering_2021}
Brand, Jennie~E., Jiahui Xu, Bernard Koch, and Pablo Geraldo. 2021.
\newblock \enquote{Uncovering {Sociological} {Effect} {Heterogeneity} {Using}
  {Tree}-{Based} {Machine} {Learning}.}
\newblock {\em Sociological Methodology\/} 51:189--223.

\bibitem[\protect\citeauthoryear{Breen}{Breen}{2004}]{breen_social_2004}
Breen, Richard (ed.). 2004.
\newblock {\em Social {Mobility} in {Europe}\/}.
\newblock Oxford University Press.

\bibitem[\protect\citeauthoryear{Cannas and Arpino}{Cannas and
  Arpino}{2019}]{cannas_comparison_2019}
Cannas, Massimo and Bruno Arpino. 2019.
\newblock \enquote{A comparison of machine learning algorithms and covariate
  balance measures for propensity score matching and weighting.}
\newblock {\em Biometrical Journal\/} 61:1049--1072.

\bibitem[\protect\citeauthoryear{Cha, Weeden, and Schnabel}{Cha
  et~al.}{2023}]{cha_is_2023}
Cha, Youngjoo, Kim~A. Weeden, and Landon Schnabel. 2023.
\newblock \enquote{Is the {Gender} {Wage} {Gap} {Really} a {Family} {Wage}
  {Gap} in {Disguise}?}
\newblock {\em American Sociological Review\/} 88:972--1001.

\bibitem[\protect\citeauthoryear{Cheng, Brand, Zhou, Xie, and Hout}{Cheng
  et~al.}{2021}]{cheng_heterogeneous_2021}
Cheng, Siwei, Jennie~E. Brand, Xiang Zhou, Yu~Xie, and Michael Hout. 2021.
\newblock \enquote{Heterogeneous returns to college over the life course.}
\newblock {\em Science Advances\/} 7:eabg7641.

\bibitem[\protect\citeauthoryear{Chernozhukov, Chetverikov, Demirer, Duflo,
  Hansen, and Newey}{Chernozhukov
  et~al.}{2017}]{chernozhukov_doubledebiasedneyman_2017}
Chernozhukov, Victor, Denis Chetverikov, Mert Demirer, Esther Duflo, Christian
  Hansen, and Whitney Newey. 2017.
\newblock \enquote{Double/{Debiased}/{Neyman} {Machine} {Learning} of
  {Treatment} {Effects}.}
\newblock {\em American Economic Review\/} 107:261--265.

\bibitem[\protect\citeauthoryear{Chernozhukov, Chetverikov, Demirer, Duflo,
  Hansen, Newey, and Robins}{Chernozhukov
  et~al.}{2018a}]{chernozhukov_double/debiased_2018}
Chernozhukov, Victor, Denis Chetverikov, Mert Demirer, Esther Duflo, Christian
  Hansen, Whitney Newey, and James Robins. 2018a.
\newblock \enquote{Double/debiased machine learning for treatment and
  structural parameters.}
\newblock {\em The Econometrics Journal\/} 21:C1--C68.

\bibitem[\protect\citeauthoryear{Chernozhukov, Fernández-Val, and
  Luo}{Chernozhukov et~al.}{2018b}]{chernozhukov_sorted_2018}
Chernozhukov, Victor, Iván Fernández-Val, and Ye~Luo. 2018b.
\newblock \enquote{The {Sorted} {Effects} {Method}: {Discovering}
  {Heterogeneous} {Effects} {Beyond} {Their} {Averages}.}
\newblock {\em Econometrica\/} 86:1911--1938.

\bibitem[\protect\citeauthoryear{Didelez, Dawid, and Geneletti}{Didelez
  et~al.}{2006}]{didelez_direct_2006}
Didelez, Vanessa, A~Philip Dawid, and Sara Geneletti. 2006.
\newblock \enquote{Direct and {Indirect} {Eﬀects} of {Sequential}
  {Treatments}.}
\newblock In {\em Proceedings of the {Twenty}-{Second} {Conference} on
  {Uncertainty} in {Artificial} {Intelligence}\/}, edited by  R~Dechter and
  T.~S Richardson, pp. 138--146, Arlington, Virginia. AUAI Press.

\bibitem[\protect\citeauthoryear{Diderichsen, Hallqvist, and
  Whitehead}{Diderichsen et~al.}{2019}]{diderichsen_differential_2019}
Diderichsen, Finn, Johan Hallqvist, and Margaret Whitehead. 2019.
\newblock \enquote{Differential vulnerability and susceptibility: how to make
  use of recent development in our understanding of mediation and interaction
  to tackle health inequalities.}
\newblock {\em International Journal of Epidemiology\/} 48:268--274.

\bibitem[\protect\citeauthoryear{Díaz, Hejazi, Rudolph, and Van
  Der~Laan}{Díaz et~al.}{2021}]{diaz_nonparametric_2021}
Díaz, I, N~S Hejazi, K~E Rudolph, and M~J Van Der~Laan. 2021.
\newblock \enquote{Nonparametric efficient causal mediation with intermediate
  confounders.}
\newblock {\em Biometrika\/} 108:627--641.

\bibitem[\protect\citeauthoryear{Featherman and Hauser}{Featherman and
  Hauser}{1978}]{featherman_opportunity_1978}
Featherman, David~L. and Robert~M. Hauser. 1978.
\newblock {\em Opportunity and {Change}\/}.
\newblock Studies in population. Academic Press.

\bibitem[\protect\citeauthoryear{Fiel}{Fiel}{2020}]{fiel_great_2020}
Fiel, Jeremy~E. 2020.
\newblock \enquote{Great {Equalizer} or {Great} {Selector}? {Reconsidering}
  {Education} as a {Moderator} of {Intergenerational} {Transmissions}.}
\newblock {\em Sociology of Education\/} 93:353--371.

\bibitem[\protect\citeauthoryear{Fisher and Kennedy}{Fisher and
  Kennedy}{2021}]{fisher_visually_2021}
Fisher, Aaron and Edward~H. Kennedy. 2021.
\newblock \enquote{Visually {Communicating} and {Teaching} {Intuition} for
  {Influence} {Functions}.}
\newblock {\em The American Statistician\/} 75:162--172.

\bibitem[\protect\citeauthoryear{Fortin, Lemieux, and Firpo}{Fortin
  et~al.}{2011}]{fortin_decomposition_2011}
Fortin, Nicole, Thomas Lemieux, and Sergio Firpo. 2011.
\newblock \enquote{Decomposition {Methods} in {Economics}.}
\newblock In {\em Handbook of {Labor} {Economics}\/}, volume~4, pp. 1--102.
  Elsevier.

\bibitem[\protect\citeauthoryear{Geneletti}{Geneletti}{2007}]{geneletti_identifying_2007}
Geneletti, Sara. 2007.
\newblock \enquote{Identifying {Direct} and {Indirect} {Effects} in a
  {Non}-{Counterfactual} {Framework}.}
\newblock {\em Journal of the Royal Statistical Society Series B: Statistical
  Methodology\/} 69:199--215.

\bibitem[\protect\citeauthoryear{Hahn}{Hahn}{1998}]{hahn_role_1998}
Hahn, Jinyong. 1998.
\newblock \enquote{On the {Role} of the {Propensity} {Score} in {Efficient}
  {Semiparametric} {Estimation} of {Average} {Treatment} {Effects}.}
\newblock {\em Econometrica\/} 66:315.

\bibitem[\protect\citeauthoryear{Heckman, Ichimura, Smith, and Todd}{Heckman
  et~al.}{1998}]{heckman_characterizing_1998}
Heckman, James, Hidehiko Ichimura, Jeffrey Smith, and Petra Todd. 1998.
\newblock \enquote{Characterizing {Selection} {Bias} {Using} {Experimental}
  {Data}.}
\newblock {\em Econometrica\/} 66:1017.

\bibitem[\protect\citeauthoryear{Heckman, Humphries, and Veramendi}{Heckman
  et~al.}{2018}]{heckman_returns_2018}
Heckman, James~J, John~Eric Humphries, and Gregory Veramendi. 2018.
\newblock \enquote{Returns to {Education}: {The} {Causal} {Effects} of
  {Education} on {Earnings}, {Health}, and {Smoking}.}
\newblock {\em Journal of Political Economy\/} 126:50.

\bibitem[\protect\citeauthoryear{Heckman, Stixrud, and Urzua}{Heckman
  et~al.}{2006}]{heckman_effects_2006}
Heckman, James~J., Jora Stixrud, and Sergio Urzua. 2006.
\newblock \enquote{The {Effects} of {Cognitive} and {Noncognitive} {Abilities}
  on {Labor} {Market} {Outcomes} and {Social} {Behavior}.}
\newblock {\em Journal of Labor Economics\/} 24:411--482.

\bibitem[\protect\citeauthoryear{Heckman and Vytlacil}{Heckman and
  Vytlacil}{2005}]{heckman_structural_2005}
Heckman, James~J. and Edward Vytlacil. 2005.
\newblock \enquote{Structural {Equations}, {Treatment} {Effects}, and
  {Econometric} {Policy} {Evaluation}.}
\newblock {\em Econometrica\/} 73:669--738.

\bibitem[\protect\citeauthoryear{Hines, Dukes, Diaz-Ordaz, and
  Vansteelandt}{Hines et~al.}{2022}]{hines_demystifying_2022}
Hines, Oliver, Oliver Dukes, Karla Diaz-Ordaz, and Stijn Vansteelandt. 2022.
\newblock \enquote{Demystifying {Statistical} {Learning} {Based} on {Efficient}
  {Influence} {Functions}.}
\newblock {\em The American Statistician\/} 76:292--304.

\bibitem[\protect\citeauthoryear{Hirano, Imbens, and Ridder}{Hirano
  et~al.}{2003}]{hirano_efficient_2003}
Hirano, Keisuke, Guido~W Imbens, and Geert Ridder. 2003.
\newblock \enquote{Efficient {Estimation} of {Average} {Treatment} {Effects}
  {Using} the {Estimated} {Propensity} {Score}.}
\newblock {\em Econometrica\/} 71:1161--1189.

\bibitem[\protect\citeauthoryear{Holland}{Holland}{1986}]{holland_statistics_1986}
Holland, Paul~W. 1986.
\newblock \enquote{Statistics and {Causal} {Inference}.}
\newblock {\em Journal of the American Statistical Association\/} 81:945--960.

\bibitem[\protect\citeauthoryear{Hout}{Hout}{2012}]{hout_social_2012}
Hout, Michael. 2012.
\newblock \enquote{Social and {Economic} {Returns} to {College} {Education} in
  the {United} {States}.}
\newblock {\em Annual Review of Sociology\/} 38:379--400.

\bibitem[\protect\citeauthoryear{Howe, Napravnik, Cole, Kaufman, Adimora,
  Elston, Eron, and Mugavero}{Howe et~al.}{2014}]{howe_african_2014}
Howe, Chanelle~J., Sonia Napravnik, Stephen~R. Cole, Jay~S. Kaufman, Adaora~A.
  Adimora, Beth Elston, Joseph~J. Eron, and Michael~J. Mugavero. 2014.
\newblock \enquote{African {American} {Race} and {HIV} {Virological}
  {Suppression}: {Beyond} {Disparities} in {Clinic} {Attendance}.}
\newblock {\em American Journal of Epidemiology\/} 179:1484--1492.

\bibitem[\protect\citeauthoryear{Huber}{Huber}{2015}]{huber_causal_2015}
Huber, Martin. 2015.
\newblock \enquote{Causal {Pitfalls} in the {Decomposition} of {Wage} {Gaps}.}
\newblock {\em Journal of Business \& Economic Statistics\/} 33:179--191.

\bibitem[\protect\citeauthoryear{Ichimura and Newey}{Ichimura and
  Newey}{2022}]{ichimura_influence_2022}
Ichimura, Hidehiko and Whitney~K. Newey. 2022.
\newblock \enquote{The influence function of semiparametric estimators.}
\newblock {\em Quantitative Economics\/} 13:29--61.

\bibitem[\protect\citeauthoryear{Imai and Li}{Imai and
  Li}{2023}]{imai_experimental_2023}
Imai, Kosuke and Michael~Lingzhi Li. 2023.
\newblock \enquote{Experimental {Evaluation} of {Individualized} {Treatment}
  {Rules}.}
\newblock {\em Journal of the American Statistical Association\/} 118:242--256.

\bibitem[\protect\citeauthoryear{Imbens and Rubin}{Imbens and
  Rubin}{2015}]{imbens_causal_2015}
Imbens, Guido~W and Donald~B Rubin. 2015.
\newblock {\em Causal inference in statistics, social, and biomedical
  sciences\/}.
\newblock Cambridge University Press.

\bibitem[\protect\citeauthoryear{Ishida, Muller, and Ridge}{Ishida
  et~al.}{1995}]{ishida_class_1995}
Ishida, Hiroshi, Walter Muller, and John~M. Ridge. 1995.
\newblock \enquote{Class {Origin}, {Class} {Destination}, and {Education}: {A}
  {Cross}-{National} {Study} of {Ten} {Industrial} {Nations}.}
\newblock {\em American Journal of Sociology\/} 101:145--193.

\bibitem[\protect\citeauthoryear{Jackson}{Jackson}{2021}]{jackson_meaningful_2021}
Jackson, John~W. 2021.
\newblock \enquote{Meaningful {Causal} {Decompositions} in {Health} {Equity}
  {Research}: {Definition}, {Identification}, and {Estimation} {Through} a
  {Weighting} {Framework}.}
\newblock {\em Epidemiology\/} 32:282--290.

\bibitem[\protect\citeauthoryear{Jackson and VanderWeele}{Jackson and
  VanderWeele}{2018}]{jackson_decomposition_2018}
Jackson, John~W. and Tyler~J. VanderWeele. 2018.
\newblock \enquote{Decomposition {Analysis} to {Identify} {Intervention}
  {Targets} for {Reducing} {Disparities}:.}
\newblock {\em Epidemiology\/} 29:825--835.

\bibitem[\protect\citeauthoryear{Karlson}{Karlson}{2019}]{karlson_college_2019}
Karlson, Kristian~Bernt. 2019.
\newblock \enquote{College as equalizer? {Testing} the selectivity hypothesis.}
\newblock {\em Social Science Research\/} 80:216--229.

\bibitem[\protect\citeauthoryear{Kennedy}{Kennedy}{2022}]{kennedy_semiparametric_2022}
Kennedy, Edward~H. 2022.
\newblock \enquote{Semiparametric doubly robust targeted double machine
  learning: a review.}
\newblock Number: arXiv:2203.06469 [stat].

\bibitem[\protect\citeauthoryear{Kitagawa}{Kitagawa}{1955}]{kitagawa_components_1955}
Kitagawa, Evelyn~M. 1955.
\newblock \enquote{Components of a {Difference} {Between} {Two} {Rates}.}
\newblock {\em Journal of the American Statistical Association\/}
  50:1168--1194.

\bibitem[\protect\citeauthoryear{Kline}{Kline}{2011}]{kline_oaxaca-blinder_2011}
Kline, Patrick. 2011.
\newblock \enquote{Oaxaca-{Blinder} as a {Reweighting} {Estimator}.}
\newblock {\em American Economic Review\/} 101:532--537.

\bibitem[\protect\citeauthoryear{Li, Morgan, and Zaslavsky}{Li
  et~al.}{2018}]{li_balancing_2018}
Li, Fan, Kari~Lock Morgan, and Alan~M. Zaslavsky. 2018.
\newblock \enquote{Balancing {Covariates} via {Propensity} {Score}
  {Weighting}.}
\newblock {\em Journal of the American Statistical Association\/} 113:390--400.

\bibitem[\protect\citeauthoryear{Li, Zaslavsky, and Landrum}{Li
  et~al.}{2013}]{li_propensity_2013}
Li, Fan, Alan~M. Zaslavsky, and Mary~Beth Landrum. 2013.
\newblock \enquote{Propensity score weighting with multilevel data.}
\newblock {\em Statistics in Medicine\/} 32:3373--3387.

\bibitem[\protect\citeauthoryear{Lundberg}{Lundberg}{2024}]{lundberg_gap-closing_2024}
Lundberg, Ian. 2024.
\newblock \enquote{The {Gap}-{Closing} {Estimand}: {A} {Causal} {Approach} to
  {Study} {Interventions} {That} {Close} {Disparities} {Across} {Social}
  {Categories}.}
\newblock {\em Sociological Methods \& Research\/} 53:507--570.

\bibitem[\protect\citeauthoryear{Lundberg, Johnson, and Stewart}{Lundberg
  et~al.}{2021}]{lundberg_what_2021}
Lundberg, Ian, Rebecca Johnson, and Brandon~M Stewart. 2021.
\newblock \enquote{What {Is} {Your} {Estimand}? {Defining} the {Target}
  {Quantity} {Connects} {Statistical} {Evidence} to {Theory}.}
\newblock {\em American Sociological Review\/} 86:532--565.

\bibitem[\protect\citeauthoryear{Malinsky}{Malinsky}{2018}]{malinsky_intervening_2018}
Malinsky, Daniel. 2018.
\newblock \enquote{Intervening on structure.}
\newblock {\em Synthese\/} 195:2295--2312.

\bibitem[\protect\citeauthoryear{Mare}{Mare}{1980}]{mare_social_1980}
Mare, Robert~D. 1980.
\newblock \enquote{Social {Background} and {School} {Continuation}
  {Decisions}.}
\newblock {\em Journal of the American Statistical Association\/} 75:295--305.

\bibitem[\protect\citeauthoryear{Morgan and Winship}{Morgan and
  Winship}{2014}]{morgan_counterfactuals_2014}
Morgan, Stephen~L. and Christopher Winship. 2014.
\newblock {\em Counterfactuals and {Causal} {Inference}: {Methods} {And}
  {Principles} {For} {Social} {Research}\/}.
\newblock Analytical {Methods} for {Social} {Research}. Cambridge, UK:
  Cambridge University Press, 2nd edition edition.

\bibitem[\protect\citeauthoryear{Naimi, Schnitzer, Moodie, and Bodnar}{Naimi
  et~al.}{2016}]{naimi_mediation_2016}
Naimi, Ashley~I., Mireille~E. Schnitzer, Erica E.~M. Moodie, and Lisa~M.
  Bodnar. 2016.
\newblock \enquote{Mediation {Analysis} for {Health} {Disparities} {Research}.}
\newblock {\em American Journal of Epidemiology\/} 184:315--324.

\bibitem[\protect\citeauthoryear{Oaxaca}{Oaxaca}{1973}]{oaxaca_male-female_1973}
Oaxaca, Ronald. 1973.
\newblock \enquote{Male-{Female} {Wage} {Differentials} in {Urban} {Labor}
  {Markets}.}
\newblock {\em International Economic Review\/} 14:693.

\bibitem[\protect\citeauthoryear{Opacic, Wei, and Zhou}{Opacic
  et~al.}{2023}]{opacic_disparity_2023}
Opacic, Aleksei, Lai Wei, and Xiang Zhou. 2023.
\newblock \enquote{Disparity {Analysis}: {A} {Tale} of {Two} {Approaches}.}
\newblock {\em Unpublished manuscript\/} .

\bibitem[\protect\citeauthoryear{Park and Kang}{Park and
  Kang}{2024}]{park_groupwise_2024}
Park, Chan and Hyunseung Kang. 2024.
\newblock \enquote{A groupwise approach for inferring heterogeneous treatment
  effects in causal inference.}
\newblock {\em Journal of the Royal Statistical Society Series A: Statistics in
  Society\/} 187:374--392.

\bibitem[\protect\citeauthoryear{Park, Kang, Lee, and Ma}{Park
  et~al.}{2023}]{park_sensitivity_2023}
Park, Soojin, Suyeon Kang, Chioun Lee, and Shujie Ma. 2023.
\newblock \enquote{Sensitivity analysis for causal decomposition analysis:
  {Assessing} robustness toward omitted variable bias.}
\newblock {\em Journal of Causal Inference\/} 11:20220031.

\bibitem[\protect\citeauthoryear{Park, Qin, and Lee}{Park
  et~al.}{2024}]{park_estimation_2024}
Park, Soojin, Xu~Qin, and Chioun Lee. 2024.
\newblock \enquote{Estimation and sensitivity analysis for causal decomposition
  in health disparity research.}
\newblock {\em Sociological Methods \& Research\/} 53:571--602.

\bibitem[\protect\citeauthoryear{Pearl}{Pearl}{2001}]{pearl_direct_2001}
Pearl, Judea. 2001.
\newblock \enquote{Direct and {Indirect} {Effects}.}
\newblock In {\em Proceedings of the {Seventeenth} {Conference} on {Uncertainy}
  in {Artificial} {Intel} ligence\/}, pp. 411--20, San Francisco, CA. Morgan
  Kaufmann.

\bibitem[\protect\citeauthoryear{Robins, Rotnitzky, and Zhao}{Robins
  et~al.}{1994}]{robins_estimation_1994}
Robins, James~M, Andrea Rotnitzky, and Lue~Ping Zhao. 1994.
\newblock \enquote{Estimation of {Regression} {Coefficients} {When} {Some}
  {Regressors} {Are} {Not} {Always} {Observed}.}
\newblock {\em Journal of the American statistical Association\/} 89:846--866.

\bibitem[\protect\citeauthoryear{Rubin}{Rubin}{1974}]{rubin_estimating_1974}
Rubin, Donald~B. 1974.
\newblock \enquote{Estimating causal effects of treatments in randomized and
  nonrandomized studies.}
\newblock {\em Journal of Educational Psychology\/} 66:688--701.

\bibitem[\protect\citeauthoryear{Rubin}{Rubin}{1980}]{rubin_randomization_1980}
Rubin, Donald~B. 1980.
\newblock \enquote{Randomization {Analysis} of {Experimental} {Data}: {The}
  {Fisher} {Randomization} {Test} {Comment}.}
\newblock {\em Journal of the American Statistical Association\/} 75:591--593.

\bibitem[\protect\citeauthoryear{Rubin}{Rubin}{1986}]{rubin_comment_1986}
Rubin, Donald~B. 1986.
\newblock \enquote{Comment: {Which} {Ifs} {Have} {Causal} {Answers}.}
\newblock {\em Journal of the American Statistical Association\/} 81:961--962.

\bibitem[\protect\citeauthoryear{Semenova and Chernozhukov}{Semenova and
  Chernozhukov}{2021}]{semenova_debiased_2021}
Semenova, Vira and Victor Chernozhukov. 2021.
\newblock \enquote{Debiased machine learning of conditional average treatment
  effects and other causal functions.}
\newblock {\em The Econometrics Journal\/} 24:264--289.

\bibitem[\protect\citeauthoryear{Shen, Visoki, Barzilay, and Pimentel}{Shen
  et~al.}{2024}]{shen2024calibratedsensitivityanalysisweighted}
Shen, Andy, Elina Visoki, Ran Barzilay, and Samuel~D. Pimentel. 2024.
\newblock \enquote{A Calibrated Sensitivity Analysis for Weighted Causal
  Decompositions.}

\bibitem[\protect\citeauthoryear{Sudharsanan and Bijlsma}{Sudharsanan and
  Bijlsma}{2021}]{sudharsanan_educational_2021}
Sudharsanan, Nikkil and Maarten~J Bijlsma. 2021.
\newblock \enquote{Educational note: causal decomposition of population health
  differences using {Monte} {Carlo} integration and the g-formula.}
\newblock {\em International Journal of Epidemiology\/} 50:2098--2107.

\bibitem[\protect\citeauthoryear{van~der Laan}{van~der
  Laan}{2006}]{van_der_laan_statistical_2006}
van~der Laan, Mark~J. 2006.
\newblock \enquote{Statistical {Inference} for {Variable} {Importance}.}
\newblock {\em The International Journal of Biostatistics\/} 2.

\bibitem[\protect\citeauthoryear{Van~der Laan and Rose}{Van~der Laan and
  Rose}{2011}]{van_der_laan_targeted_2011}
Van~der Laan, Mark~J and Sherri Rose. 2011.
\newblock {\em Targeted learning: Causal Inference for Observational and
  Experimental Data\/}.
\newblock Springer {Series} in {Statistics}. New York: Springer.

\bibitem[\protect\citeauthoryear{Van~der Vaart}{Van~der
  Vaart}{2000}]{van_der_vaart_asymptotic_2000}
Van~der Vaart, Aad~W. 2000.
\newblock {\em Asymptotic {Statistics}\/}.
\newblock Cambridge {Series} on {Statistical} and {Probablistic} {Mathematics}.
  Cambridge University Press.

\bibitem[\protect\citeauthoryear{VanderWeele}{VanderWeele}{2015}]{vanderweele_explanation_2015}
VanderWeele, Tyler. 2015.
\newblock {\em Explanation in Causal Inference: Methods for Mediation and
  Interaction\/}.
\newblock Oxford University Press.

\bibitem[\protect\citeauthoryear{VanderWeele}{VanderWeele}{2014}]{vanderweele_unification_2014}
VanderWeele, Tyler~J. 2014.
\newblock \enquote{A {Unification} of {Mediation} and {Interaction}: {A}
  4-{Way} {Decomposition}.}
\newblock {\em Epidemiology\/} 25:749--761.

\bibitem[\protect\citeauthoryear{VanderWeele and Robinson}{VanderWeele and
  Robinson}{2014}]{vanderweele_causal_2014}
VanderWeele, Tyler~J. and Whitney~R. Robinson. 2014.
\newblock \enquote{On the {Causal} {Interpretation} of {Race} in {Regressions}
  {Adjusting} for {Confounding} and {Mediating} {Variables}.}
\newblock {\em Epidemiology\/} 25:473--484.

\bibitem[\protect\citeauthoryear{VanderWeele and Tchetgen~Tchetgen}{VanderWeele
  and Tchetgen~Tchetgen}{2014}]{vanderweele_attributing_2014}
VanderWeele, Tyler~J. and Eric~J. Tchetgen~Tchetgen. 2014.
\newblock \enquote{Attributing {Effects} to {Interactions}.}
\newblock {\em Epidemiology\/} 25:711--722.

\bibitem[\protect\citeauthoryear{Vansteelandt and Dukes}{Vansteelandt and
  Dukes}{2022}]{vansteelandt_assumption-lean_2022}
Vansteelandt, Stijn and Oliver Dukes. 2022.
\newblock \enquote{Assumption-lean {Inference} for {Generalised} {Linear}
  {Model} {Parameters}.}
\newblock {\em Journal of the Royal Statistical Society Series B: Statistical
  Methodology\/} 84:657--685.

\bibitem[\protect\citeauthoryear{Ward, Gartner, Keyes, Fliss, McClure, and
  Robinson}{Ward et~al.}{2019}]{ward_how_2019}
Ward, Julia~B., Danielle~R. Gartner, Katherine~M. Keyes, Mike~D. Fliss,
  Elizabeth~S. McClure, and Whitney~R. Robinson. 2019.
\newblock \enquote{How do we assess a racial disparity in health?
  {Distribution}, interaction, and interpretation in epidemiological studies.}
\newblock {\em Annals of Epidemiology\/} 29:1--7.

\bibitem[\protect\citeauthoryear{Xie, Brand, and Jann}{Xie
  et~al.}{2012}]{xie_estimating_2012}
Xie, Yu, Jennie~E. Brand, and Ben Jann. 2012.
\newblock \enquote{Estimating {Heterogeneous} {Treatment} {Effects} with
  {Observational} {Data}.}
\newblock {\em Sociological Methodology\/} 42:314--347.

\bibitem[\protect\citeauthoryear{Yamaguchi}{Yamaguchi}{2015}]{yamaguchi_decomposition_2015}
Yamaguchi, Kazuo. 2015.
\newblock \enquote{Decomposition of {Gender} or {Racial} {Inequality} with
  {Endogenous} {Intervening} {Covariates}: {An} {Extension} of the
  {DiNardo}-{Fortin}-{Lemieux} {Method}.}
\newblock {\em Sociological Methodology\/} 45:388--428.

\bibitem[\protect\citeauthoryear{Yu}{Yu}{2024}]{yu_2023}
Yu, Ang. 2024.
\newblock {\em cdgd: Causal Decomposition of Group Disparities\/}.
\newblock R package version 0.3.5.

\bibitem[\protect\citeauthoryear{Yu, Ge, and Elwert}{Yu
  et~al.}{2024}]{yu2024naturalmediationeffectsdiffer}
Yu, Ang, Li~Ge, and Felix Elwert. 2024.
\newblock \enquote{When Do Natural Mediation Effects Differ from Their
  Randomized Interventional Analogues: Test and Theory.}
\newblock Number: arXiv:2407.02671 [stat].

\bibitem[\protect\citeauthoryear{Yu, Park, Kang, and Fletcher}{Yu
  et~al.}{2021}]{yu_leveraging_2021}
Yu, Ang, Chan Park, Hyunseng Kang, and Jason Fletcher. 2021.
\newblock \enquote{Leveraging {Machine} {Learning} to {Estimate} {Effect}
  {Modification}.}
\newblock preprint, SocArXiv.

\bibitem[\protect\citeauthoryear{Yu and Zhao}{Yu and
  Zhao}{2024}]{yu2024counter}
Yu, Ang and Jiwei Zhao. 2024.
\newblock \enquote{Counterfactual Slopes and Their Applications in Social
  Stratification.}
\newblock Number: arXiv:2401.07000 [stat].

\bibitem[\protect\citeauthoryear{Zhou}{Zhou}{2019}]{zhou_equalization_2019}
Zhou, Xiang. 2019.
\newblock \enquote{Equalization or {Selection}? {Reassessing} the
  “{Meritocratic} {Power}” of a {College} {Degree} in {Intergenerational}
  {Income} {Mobility}.}
\newblock {\em American Sociological Review\/} 84:459--485.

\bibitem[\protect\citeauthoryear{Zhou}{Zhou}{2024}]{zhou_attendance_2024}
Zhou, Xiang. 2024.
\newblock \enquote{Attendance, Completion, and Heterogeneous Returns to
  College: A Causal Mediation Approach.}
\newblock {\em Sociological Methods \& Research\/} 53:1136--1166.

\bibitem[\protect\citeauthoryear{Ziol-Guest and Lee}{Ziol-Guest and
  Lee}{2016}]{ziol-guest_parent_2016}
Ziol-Guest, Kathleen~M. and Kenneth T.~H. Lee. 2016.
\newblock \enquote{Parent {Income}–{Based} {Gaps} in {Schooling}:
  {Cross}-{Cohort} {Trends} in the {NLSYs} and the {PSID}.}
\newblock {\em AERA Open\/} 2:233285841664583.

\end{thebibliography}

\end{document}